\RequirePackage{ifpdf}
\documentclass[12pt,aps,amsmath,latexsym,amsfonts]{article}
\usepackage{graphicx}
\usepackage{epsfig}
\usepackage{amsfonts,amssymb,amsmath,bm}
\usepackage{enumitem}
\usepackage[hang]{subfigure}
\usepackage{epstopdf}
\usepackage{jheppub}
\usepackage{mathtools}

\usepackage{color}

\newcommand{\be}{\begin{equation}}
\newcommand{\ee}{\end{equation}}
\newcommand{\ba}{\begin{eqnarray}}
\newcommand{\ea}{\end{eqnarray}}

\newcommand{\ket}{\rangle}
\newcommand{\qbra}[1]{\left< #1 \right|}
\newcommand{\qket}[1]{\left| #1 \right>}

\DeclarePairedDelimiter\floor{\lfloor}{\rfloor}

\def\half {\frac{1}{2}}

\def\p1{|{1}\ket}

\title{Black hole chemistry: thermodynamics with Lambda}

\author[1,2]{David Kubiz{\v n}{\'a}k}
\author[2,1]{Robert B.~Mann}
\author[3]{Mae Teo}

\affiliation[1]{Perimeter Institute for Theoretical Physics, 31 Caroline St. N., Waterloo, Ontario N2L 2Y5, Canada}
\affiliation[2]{Department of Physics and Astronomy, University of Waterloo, Waterloo, Ontario N2L 3G1, Canada}
\affiliation[3]{Stanford Institute for Theoretical Physics, Stanford University, Stanford, CA 94305, USA}

\emailAdd{dkubiznak@perimeterinstitute.ca}
\emailAdd{rbmann@uwaterloo.ca}
\emailAdd{maehwee@stanford.edu}

\abstract{
We review recent developments on the thermodynamics of black holes in extended phase space, where the cosmological constant is interpreted as thermodynamic pressure and treated as a thermodynamic variable in its own right. In this approach, the mass of the black hole is no longer
regarded as internal energy, rather it is identified with the chemical enthalpy. This leads to an extended dictionary for   black hole thermodynamic quantities, in particular a notion of   thermodynamic volume emerges for a given black hole spacetime. This volume is conjectured to satisfy the reverse isoperimetric inequality---an inequality imposing a bound on the amount of entropy black hole can carry for a fixed thermodynamic volume.
New thermodynamic phase transitions naturally emerge from these identifications. Namely, we show that   black holes
can be understood  from the viewpoint of chemistry, in terms of concepts such as Van der Waals fluids, reentrant phase transitions, and triple points. We also review the recent attempts at extending the AdS/CFT dictionary in this setting, discuss the connections with horizon thermodynamics, applications to Lifshitz spacetimes, and outline possible future directions in this field.
}

\date{} 

\begin{document}
\maketitle
\flushbottom

\section{Overview}\label{Sec:1}
Over the past four decades a preponderance of evidence has accumulated suggesting a fundamental relationship between gravitation, thermodynamics, and quantum theory.   This evidence is rooted in our understanding of black holes and their relationship to quantum physics, and developed into the sub-discipline of {\it black hole thermodynamics}.

This subject was originally quite counter-intuitive \cite{Bardeen:1973gs}.  Classically black holes were nature's ultimate
sponges, absorbing all matter and emitting nothing.  Superficially they had neither temperature nor entropy, and were characterized by only a few basic parameters:  mass, angular momentum, and  charge (if any) \cite{israel}.   However the advent of quantum field theory in curved spacetime changed all of this, leading to the famous results that the area of a black hole corresponds to its entropy \cite{Bekenstein:1973ur} and its
surface gravity corresponds to its temperature \cite{Hawking:1974sw}.  All theoretical evidence indicated that black holes radiate heat, analogous to  black body radiation, and the subject of black hole thermodynamics was born.

Black hole thermodynamics stimulated a whole new set of techniques for analyzing the behavior of black holes and gave rise to some deep insights concerning the relationship between gravity and quantum physics.
 Further investigation led to one of
 the most perplexing conundrums in physics, namely that the process of black hole radiation leads to a loss of information that is incompatible with the basic foundations of quantum physics \cite{Giddings:1995gd,Mathur:2009hf}  that have yet to be resolved   \cite{Mathur:2005zp,Almheiri:2012rt, Hawking:2016msc}.
Black hole entropy turned out to be the Noether charge associated with diffeomorphism symmetry \cite{Wald:1993nt}.
 The laws of gravitation were posited to be intimately connected with the laws of thermodynamics \cite{Jacobson:1995ab, Padmanabhan:2009vy}.  The introduction of a negative cosmological constant implied that black holes could exhibit phase behaviour \cite{Hawking:1982dh}, and somewhat later led to the holographic deployment of black holes as systems dual to those in conformal field theories \cite{Maldacena:1997re, Witten:1998qj, Witten:1998zw}, quantum chromodynamics \cite{Kovtun:2004de}, and condensed matter physics \cite{Hartnoll:2007ih, Hartnoll:2008vx}.
 Deep connections were discovered between the quantum information concept of entanglement entropy  \cite{Ryu:2006bv} and the `architecture of spacetime' \cite{Bianchi:2012ev};  the linearized Einstein equations were later shown to follow from the first law of entanglement entropy  \cite{Faulkner:2013ica}.  Geometric approaches to the thermodynamics of black holes were summarized in \cite{Gruber:2016mqb}.

Somewhat more recently reconsideration of the role of the cosmological constant, $\Lambda$, has led to the realization that  black hole thermodynamics is a much richer subject than previously thought. It led to the introduction of {\it pressure}, and with it a concept of {\it volume for a black hole}.
 New phase behaviour, analogous to that seen in gels and polymers, was found to be present.  Triple points for black holes, analogous to those in water, were discovered. Black holes could further be understood as heat engines.   In general black holes were found quite analogous to Van der Waals fluids, and in general exhibited the diverse behavior of different substances we encounter in everyday life.  This burgeoning subfield was hence given the name {\it black hole chemistry} \cite{Kubiznak:2014zwa,Mann:2016trh}.

This is the subject of this topical review.
While more focussed overviews of this subject have appeared \cite{Dolan:2012jh,Altamirano:2014tva, Kubiznak:2014zwa,Dolan:2014jva,Mann:2015luq,Mann:2016trh}, our aim here is to be comprehensive, reviewing the subject from its historical roots to its modern developments.

We begin in Sec.~\ref{Sec:2} by reviewing the laws of black hole mechanics and their relationship to black hole thermodynamics. After discussing early attempts at incorporating $\Lambda$ into the thermodynamic laws, we shall investigate how a proper treatment suggests that this quantity should be interpreted as thermodynamic pressure, thereby completing the parallel between the laws of black hole mechanics and thermodynamics.  The conjugate quantity, volume, naturally emerges and its properties are discussed in Sec.~\ref{Sec:3}.
Once these basic notions of pressure and volume for  black hole systems are defined and understood it becomes possible to analyze the {\it extended thermodynamic phase space}, one that includes these variables along with the more established quantities of   temperature, entropy, potential, charge, angular velocity, angular momentum, and energy. It is in this context that  the rich panoply of  chemical behavior of black holes is manifest, a subject we review in Sec.~\ref{Sec:4}.
We then turn to more recent developments that endeavour to understand black hole chemistry from a holographic viewpoint in Sec.~\ref{Sec:5}.  We then consider in Sec.~\ref{Sec:6} how  the concept of thermodynamic pressure can be applied and re-interpreted in more general situations, such as cosmological expanding spacetimes, or Lifshitz spacetimes that are conjectured to be dual to certain condensed matter systems. We relegate supplementary technical material into appendices and conclude our review with a discussion of what has been accomplished in this subject and what remains to be done.

By introducing $\Lambda$ as a thermodynamic variable, black hole thermodynamics has been given a new life. Let us begin exploring.

\section{Thermodynamics with Lambda}\label{Sec:2}
\subsection{A brief review of standard black hole thermodynamics}

Since black holes are classical solutions to Einstein's equations  there is no a-priori reason to  expect them to exhibit thermodynamic behaviour.  The first indication linking these two subjects came from   \textit{Hawking's area theorem} \cite{hawking}, which states that the area of the event horizon  of a black hole can
never decrease.\footnote{%
Hawking's proof applies to black hole spacetimes that satisfy certain reasonable assumptions, namely, that the spacetime on and outside the future event horizon is a regular predictable space, and that the stress-energy tensor satisfies the null energy condition, $T_{ab}k^a k^b \geq 0$, for arbitrary {null vector} $k^a$.}
Bekenstein subsequently noticed the resemblance between this area law and the second law of thermodynamics. By applying thermodynamic considerations in a set of Gedanken (thought) experiments, he proposed \cite{Bekenstein:1973ur} that each black hole should be assigned an entropy proportional to the area of its event horizon. Pursuing this analogy further, the ``four laws of {\it black hole mechanics}'' \cite{Bardeen:1973gs} were formulated by Bardeen, Carter, and Hawking under the assumption that the event horizon of the black hole is a Killing horizon, which is a null hypersurface generated by a corresponding Killing vector field. The four laws are:
\begin{enumerate}[start=0]
\item The surface gravity $\kappa$ is constant over the event horizon of a stationary black hole.
\item For a rotating charged black hole with a mass $M$, an angular momentum $J$, and a charge $Q$,
\begin{equation} \label{eqn:sfl}
\delta M = \frac{\kappa}{ 8\pi G}\delta A + \Omega \delta J + \Phi \delta Q\,,
\end{equation}\\
where $\kappa$ is its surface gravity, $\Omega$ its angular velocity, and $\Phi$ its  electric potential.
\item Hawking's area theorem: $\delta A \geq 0$, i.e. the area $A$ of a black hole's event horizon can never decrease.
\item It is impossible to reduce the surface gravity $\kappa$ to zero in a finite number of steps.
\end{enumerate}
The surface gravity $\kappa$ is defined in the presence of a Killing horizon via
\begin{equation}\label{surfgrav}
\xi^a \nabla_a \xi^b = \kappa \xi^b
\end{equation}
 for a suitably normalized Killing vector $\xi^a$ that generates the horizon.
For a static black hole, such as the Schwarzschild black hole, the surface gravity is the
force  exerted at infinity that is required to keep an object of  unit mass at the horizon.

If we only consider black holes classically, these laws  are merely a formal analogy between black hole mechanics and thermodynamics, where comparison to the first law of ordinary thermodynamics\footnote{ Here, $\delta N_j$ describes changing number of particles of a given species and $\mu_j$ is the corresponding chemical potential. Similarly the term $\Phi \delta Q$ allows for the possibility of variable electrostatic energy. }
\be\label{flawTD}
\delta U= T \delta S - P \delta V + \sum_j \mu_j \delta N_j + \Phi \delta Q
\ee
is made  with $\kappa$ playing the role of temperature and event horizon area playing the role of entropy.
In fact  classical black holes have zero temperature. 
They never emit anything;  a classical black hole immersed in a radiation bath at any finite temperature will
always absorb the radiation.

By taking quantum effects into account, Hawking discovered \cite{Hawking:1974sw} that black holes do
emit radiation with a blackbody spectrum at a characteristic temperature
\begin{equation} \label{temp}
k_B T= \frac{\hbar \kappa}{2\pi c}\,,
\end{equation}
inserting the factors of Boltzmann's constant $k_B$, the speed of light $c$, and Planck's  constant $\hbar$, the latter quantity underscoring the intrinsically quantum-mechanical nature of black hole temperature.
This discovery,  verified by many subsequent derivations, led to a paradigm shift: black holes are actual physical thermodynamic systems that have temperature and entropy; they are no longer systems that are simply described by a convenient analogy with thermodynamics.

With the relation \eqref{temp} between $T$ and $\kappa$ established, we can compare the $T \delta S$ term in the first law of thermodynamics with the $\kappa \delta A$ term for black holes to infer that the entropy is directly related to the area by
\begin{equation} \label{ent}
S=\frac{A c^3}{4\hbar G}\,,
\end{equation}
a relation confirmed, for example, by the Euclidean path integral approach \cite{Gibbons:1977mu}.
Accordingly, equation (\ref{eqn:sfl}) is nothing else but the standard {\it first law of black hole thermodynamics}:
\begin{equation}\label{flaw}
\delta M = T\delta S + \Omega \delta J + \Phi \delta Q\,
\end{equation}
for a black hole of mass $M$, charge $Q$, and angular momentum $J$, upon setting
\be
G=c=k_B=1\,,
\ee
a convention that implies  that $\hbar$ has units of [length]$^2$. Keeping this in mind, we shall henceforth suppress the explicit appearance of these quantities, restoring them on an as-needed basis.\footnote{Note that in $d$ dimensions the gravitational constant $G_d$ has units
of  [length]$^{d-4}$, which will be relevant when we consider  the AdS/CFT correspondence in Sec.~\ref{Sec:5}.}

The black hole mass $M$ is identified with the energy of the system, $U$, and the chemical potential
and electromagnetic work terms in \eqref{flawTD} play roles analogous to the rotational and electromagnetic work terms in \eqref{flaw}.

The thermodynamic variables in \eqref{flaw} are related by a useful {\it Smarr--Gibbs--Duhem relation}, which (in four dimensions) reads\footnote{ In thermodynamics this relation is known as the Gibbs--Duhem relation;  in the context of black hole physics this relation was originally derived by Smarr \cite{Smarr:1972kt}, and is often simply referred to as a {\it Smarr formula}.}

\begin{equation}\label{oldSmarr}
M=2 (TS+ \Omega J) + \Phi Q\,,
\end{equation}
expressing a relationship between the extensive $(M,J,Q)$ and intensive $(T,\Omega,\Phi)$ thermodynamic variables.

 For references on standard black hole thermodynamics, a variety of reviews are available \cite{Wald:1999vt, Traschen:1999zr, Grumiller:2014qma, Carlip:2014pma}.

\subsection{History of variable $\Lambda$}

One of the noteworthy features of the first law \eqref{flaw} is the omission of a { pressure-volume term}
$P \delta V$.  This quantity is commonplace in everyday thermodynamics, but there is no obvious notion
of {\it pressure} or {\it volume} associated with a black hole.
In the last few years a new perspective  has emerged that incorporates these notions
into black hole thermodynamics.  The basic idea is that pressure can be associated with a  negative cosmological constant $\Lambda$, a form of energy whose (positive) pressure is equal in magnitude to its (negative)  energy density.\footnote{{While one might, a-priori,  consider associating black hole pressure with a positive cosmological constant, there are a number of subtleties and unresolved issues in black hole thermodynamics in such spacetimes.  We shall consider this topic in Sec.~\ref{Sec:6}.}}. In what follows we consider black holes `immersed' in the environment of a negative cosmological constant.

An asymptotically anti de Sitter (AdS) black hole in $d$  spacetime dimensions is a solution to the Einstein equations
\be\label{EinEq}
R_{ab} - \frac{1}{2} g_{ab}R + \Lambda g_{ab} = T_{ab}\,,
\ee
where $\Lambda<0$ is often parameterized by the {\it AdS radius $l$} according to
\be
\Lambda = -\frac{(d-1)(d-2)}{2l^2} < 0\,,
\ee
and $T_{ab}$ is the matter stress-energy tensor (that vanishes sufficiently quickly as we approach the asymptotic region).

For example, let us consider a vacuum ($T_{ab}=0$) static spherically symmetric black hole solution, generalizing the asymptotically flat higher-dimensional Schwarzschild--Tangherlini solution \cite{Tangherlini:1963bw}. The metric reads
\be\label{Schw-Ads}
ds^2  = - f dt^2+\frac{dr^2}{f}+r^2  d\Omega_k^2\,,
\ee
where the metric function
\be\label{FSchw}
f=k-\left(\frac{r_0}{r}\right)^{d-3}+\frac{r^2}{l^2}\,,
\ee
and
\be
d\Omega_k^2 = d\theta^2 +\frac{\sin^2(\sqrt{k}\theta)}{k}d\Omega_{d-3}
\ee
is the metric on a compact $(d-2)$-dimensional space $\Sigma_k$ of constant curvature with
sign $k$, with $k=1$ being the $(d-2)$-sphere, $k=0$ being a torus, and $k=-1$ being
a compact hyperbolic space \cite{Aminneborg:1996iz,Mann:1996gj}. The object $d\Omega_{d-3}$ is the metric of a $(d-3)$-sphere.
 The parameter $r_0$ is related to the black hole mass. The appearance of the parameter $k$
with its three distinct values is specific to AdS spacetimes; appropriate identifications can render any constant
$(t,r)$ section of the spacetime compact \cite{Mann:1997iz}.
The presence of $r^2/l^2$ in the metric function $f$ ensures the proper AdS asymptotic behavior.

Concentrating on the spherical $(k=1)$ case, sufficiently large (as compared to the AdS radius $l$) black holes  \eqref{Schw-Ads}
have positive specific heat (unlike their asymptotically flat counterparts) and can be in stable equilibrium at a fixed temperature (with AdS space acting like a gravitational box). They can also undergo a {\it phase transition} to pure radiation depending on the temperature  \cite{Hawking:1982dh}.  In the context of the AdS/CFT correspondence, this transition, known as the Hawking--Page transition, was later understood as a confinement/deconfinement phase transition in the boundary Conformal Field Theory (CFT) \cite{Witten:1998zw}; we shall further discuss this transition in Sec.~\ref{Sec:4}.

 The notion that $\Lambda$ itself might be a dynamical variable was proposed by Teitelboim and Brown
\cite{Teitelboim:1985dp, Brown:1988kg}, and the corresponding  thermodynamic term was formally incorporated into the first law somewhat later \cite{Creighton:1995au}, though no interpretation of the conjugate variable  was considered.
The idea of associating
$\Lambda$ with pressure was subsequently explored from several perspectives \cite{Caldarelli:1999xj, Padmanabhan:2002sha}, but its proper association along with the notion of a conjugate  {\it black hole volume}
was achieved once the laws of black hole mechanics were generalized to include\footnote{See also \cite{Papadimitriou:2005ii}
for a treatment of the Smarr formula and first law with fixed $\Lambda$ and the corresponding AdS/CFT interpretation.} $\Lambda \neq 0$
\cite{Kastor:2009wy}.
The resultant  generalized  first law of black hole thermodynamics is
\be\label{law1mech}
\delta M =T\delta S  + V \delta P+\Omega \delta J+\Phi\delta Q\,,
 \ee
whose derivation is reproduced in App.~\ref{FirstLaw}.  Here
\be\label{eq:press}
P = -\frac{\Lambda}{8\pi} = \frac{(d-1)(d-2)}{16\pi l^2}
\ee
is interpreted as thermodynamic pressure,  
and the quantity $V$, given by
\be\label{Vdef}
V\equiv \left(\frac{\partial M}{\partial P}\right)_{S,Q,J}
\ee
is  its conjugate {\it thermodynamic volume} \cite{Dolan:2010ha, Cvetic:2010jb}, whose interpretation we postpone until  Sec.~\ref{Sec:3}.
The quantities $M$ and $J$ are the conserved charges  respectively associated with the
time-translation and rotational Killing vectors of the spacetime. As before, the area of the black hole event horizon is related to the entropy according to $A=4S$ and temperature   $T=\kappa/2\pi$ with $\kappa$ its surface gravity. 

The interpretation of $P$ as thermodynamic pressure naturally follows from the realization that $\Lambda<0$
induces a positive vacuum pressure in spacetime.
Comparing \eqref{law1mech} with \eqref{flawTD}
 we see that in the presence of the cosmological constant, the mass $M$ has no longer meaning of  internal energy. Rather, $M$ can be interpreted as a gravitational version of chemical {\it enthalpy}
\cite{Kastor:2009wy}, which is the total energy of a system including both its internal energy $E$ and the energy $PV$ required to  displace the vacuum energy of its environment:
\be
M=E+PV\,,
\ee
with the two quantities related by standard Legendre transformation.
In other words, $M$ is the total energy required to ``create a black hole and place it in a cosmological  (negative $\Lambda$) environment''.

Hence by permitting $\Lambda$ to be a variable quantity we recover the familiar
 pressure-volume term from chemical thermodynamics. Extending to cases with
multiple rotations and $U(1)$ charges, the {\it generalized first law of black hole thermodynamics} reads \cite{Kastor:2009wy, Cvetic:2010jb, Dolan:2013ft,Altamirano:2014tva}
\be
\delta M= T\delta S +\sum_i^{N} \Omega^i\delta J^i\, {+  V \delta P }
+\sum_j \Phi^j \delta Q^j\,,
\label{firstBH}
\ee
where $ N=\floor{\frac{d-1}{2}}$ is the  largest integer less than or equal to $\frac{d-1}{2}$ and  represents an upper bound on possible number of independent rotations in $d$ dimensions \cite{Myers:1986un}. (higher-dimensional rotating black hole solutions with $\Lambda$ were constructed in \cite{Hawking:1998kw, Gibbons:2004uw, Gibbons:2004js} and are reviewed in App.~\ref{AppB}.)
Here the $\Phi^j$  are the conjugate (gauge independent) potentials for the electric (and magnetic) $U(1)$ charges, $\Phi^j=\Phi^j_+-\Phi^j_{\infty}$, allowing for both: a non-trivial potential on the horizon $\Phi^j_+$ and at infinity $\Phi^j_\infty$.
 Similarly, $\Omega^i=\Omega^i_{+}-\Omega^i_{\infty}$\,,
where the quantities $\Omega_\infty^i$ allow for the possibility of a rotating frame at infinity \cite{Gibbons:2004ai}.
The thermodynamic volume $V$ may therefore be interpreted as the change in the mass under variations in $\Lambda$, with the black hole entropy, angular momenta, and charges held fixed.

 A ``practical reason'' for including the pressure volume term is connected with the Smarr formula,
 which now reads
\be
\frac{d-3}{d-2}M=TS+\sum_i \Omega^iJ^i  -\frac{2}{d-2}PV
+\frac{d-3}{d-2} \sum_j \Phi^j  Q^j\,,
\label{smarrBH}
\ee
generalizing the relation  \eqref{oldSmarr}  to AdS spacetimes in $d$-dimensions. 
Note the presence of the crucial $PV$ term; we shall demonstrate the need for this term in an example below.


The  relation \eqref{smarrBH} can be obtained  \cite{Caldarelli:1999xj,Kastor:2009wy} from an application of Euler's formula for homogeneous functions  $f(x,y,
\dots,z) \rightarrow f(\alpha^p x, \alpha^q y,\dots,\alpha^rz)   = \alpha^{s} f(x,y,\dots,z)$, which yields the {\it scaling relation}
\be\label{Euler}
sf(x,y,\dots,z) = p \left(\frac{\partial f}{\partial x}\right) x + q \left(\frac{\partial f}{\partial y}\right) y+\dots+
r \left(\frac{\partial f}{\partial z}\right) z\,,
\ee
upon taking the derivative with respect to $\alpha$. Taking the mass to be a homogeneous function
$M=M(A,\Lambda,Q^j,J^i)$,
and noting that the scaling dimensions of $A$ and $J^i$ are $d-2$, $M$ and $Q^j$ are both  $d-3$, and $\Lambda$ is $-2$ we have
\be\label{Euler3}
(d-3)M=(d-2)\frac{\partial M}{\partial A}A+(d-2)\sum_i \frac{\partial M}{\partial J^i}J^i+(-2)
\frac{\partial M}{\partial \Lambda}\Lambda+(d-3)\sum_j\frac{\partial M}{\partial \Phi^j}\Phi^j\,.
\ee
Employing now the first law \eqref{firstBH}, together with identification of $S$ with $A/4$ and $P$ with $-\Lambda/[8\pi]$,
we for example have $\frac{\partial M}{\partial A}A=\frac{\partial M}{\partial S}S=TS$ and so on, and so \eqref{Euler3}
yields \eqref{smarrBH}.
Note that the inclusion of the $PV$ term is {\it required} for  \eqref{smarrBH}  to hold.

Despite the fact that the preceding derivation assumes that the mass is a homogeneous function of
the other 
thermodynamic variables, the Smarr relation  \eqref{smarrBH} has been demonstrated to have very broad applicability, including $\Lambda > 0$
\cite{Dolan:2013ft}, any dimension,\footnote{This includes lower-dimensional gravity in $d = 2, 3$ \cite{Frassino:2015oca}. Let us also remark that there has recently been an investigation of a dynamical $\Lambda$ that arises as an integration constant
in two spacetime dimensions  \cite{Grumiller:2014oha}. The cosmological constant behaves as a $U(1)$ charge with a confining potential, necessitating a novel (gravitational) Born--Infeld boundary term in the action.  The relationship with the $d\to 2$ limit
of the Smarr relation \eqref{smarrBH} \cite{Frassino:2015oca} has yet to be explored.
} asymptotically Lifshitz spacetimes \cite{Brenna:2015pqa}, and more exotic black objects \cite{Caldarelli:2008pz, Altamirano:2014tva,Hennigar:2014cfa}.
 Other quoted Smarr relations \cite{Bertoldi:2009dt,Bertoldi:2009vn, Dehghani:2010kd,Liu:2014dva, Berglund:2011cp,Dehghani:2011hf,Dehghani:2013mba, Way:2012gr}
 (none incorporating a notion of volume)  have all been shown to be special cases of \eqref{smarrBH} \cite{Brenna:2015pqa}.

 Before we proceed further, let us illustrate the generalized first law and the Smarr formula for the concrete example of a charged AdS
black hole  in four dimensions. The metric and the gauge field (characterized by the gauge potential $\textsf{A}$ and
field strength $F$) read 
\ba\label{HDRN}
ds^2 &=& -f(r) dt^2 + \frac{dr^2}{f(r)} + r^{2} d\Omega^2\,,\nonumber\\
F&=&d\textsf{A}\,,\quad \textsf{A}=-\frac{Q}{r}dt\,,
\ea
 where  $f(r)$  is given by
\be\label{metfunction}
f = 1 - \frac{2M}{r} + \frac{Q^2}{r^{2}} + \frac{r^2}{l^2}\,,
\ee
and $d\Omega^2$ is the metric for the standard element on $S^2$.
 The parameter $M$ represents the ADM mass of the black hole and $Q$
its total charge. The outer (event) horizon is located at $r=r_+$, determined
from $f(r_+)=0$.  Exploiting this latter relation we have
\be
M=\frac{1}{2}\Bigl(r_+  + \frac{Q^2}{r_+  } + \frac{r_+^{3} }{l^2}\Bigr)\,,
\ee
and so we can write all thermodynamic quantities in terms of $r_+$, $l$, and $Q$, yielding
\cite{Chamblin:1999tk, Chamblin:1999hg,Wu:2000id}
\ba\label{TDchgBH}
S=\frac{A}{4}=\pi r_+^2\,,\quad \Phi=\frac{Q}{r_+}\,,\quad T&=& \frac{f'(r_+)}{4\pi}
= \frac{1}{4\pi r_{+}} \left(1 - \frac{Q^2}{r_+^{2}} + 3\frac{r_+^2}{l^2} \right)\,.
\ea
Taking the variation, we have
\be
\delta M = \frac{1}{2}  \left(1 - \frac{Q^2}{r_+^{ 2} }  + 3\frac{r_+^{ 2} }{l^2}  \right) \delta r_+
 -  \frac{r_+^{3} }{l^3} \delta l  +   \frac{Q}{r_+  }\delta Q\,, \quad
\delta S=2\pi r_+\delta r_+\,,
\ee
so that
\ba
T\delta S + \Phi \delta Q &=&  \frac{1}{2} \Bigl(1 - \frac{Q^2}{r_+^{2}} +3\frac{r_+^2}{l^2} \Bigr)  r_+ \delta r_+
+ \frac{Q}{ r_+ } \delta Q\nonumber\\
&=&\delta M+\frac{r_+^2}{l^3}\delta l=\delta M-V\delta P\,,
\ea
which is a particular case of \eqref{firstBH},
where $P$ is given by \eqref{eq:press} and
\be\label{volrp}
V  = \frac{4\pi r_+^{3}}{3}\,,
\ee
as inferred  from \eqref{Vdef}.  Checking further the Smarr relation \eqref{smarrBH}, we find
\be
\frac{1}{2}M-TS-\frac{1}{2}\Phi Q=-\frac{1}{2}\frac{r_+^3}{l^2}=-PV\,,
\ee
upon using \eqref{volrp}. It is obvious that without the $PV$ term, the Smarr formula would not hold.

\subsection{Black hole chemistry}

This  new perspective on black hole thermodynamics, with its different interpretation of black hole mass and the inclusion of   $\Lambda$ as a pressure term  \cite{Kastor:2009wy, Dolan:2011xt, Dolan:2010ha,Kubiznak:2012wp},  has   led to a different understanding of known processes and to the discovery of a broad range of new phenomena associated with black holes.  Referred to as ``Black Hole Chemistry" \cite{Kubiznak:2014zwa, Mann:2016trh, Mann:2015luq,
Frassino:2015oca, Karch:2015rpa, Caceres:2016xjz}, this approach has led  to a new understanding of  concepts such as Van der Waals fluids, reentrant phase transitions,  triple points, and polymer behavior from a gravitational viewpoint.   Both charged and rotating black holes exhibit novel  chemical-type phase behaviour that we shall discuss in the sequel.

The first observation is that the thermodynamic correspondence with black hole mechanics is completed \cite{Dolan:2010ha} to include the familiar pressure/volume terms:
\be\label{A1}
\begin{array}{|l|c|l|c|}
\hline
\multicolumn{2}{|c|}{\mbox{Thermodynamics}} & \multicolumn{2}{|c|}{\mbox{Black Hole Mechanics}} \\
\hline
\mbox{Enthalpy} &  H=E+PV & \mbox{Mass} & M\\
\hline
\mbox{Temperature} & T & \mbox{Surface Gravity} & \frac{  \kappa}{2\pi}\\
\hline
\mbox{Entropy}  &S &\mbox{Horizon Area} & \frac{A}{4} \\
\hline
\mbox{Pressure}  &P & \mbox{Cosmological Constant}  &  -\frac{\Lambda}{8\pi} \\
\hline
\mbox{First Law}  &\delta H= T \delta S +V \delta P + \ldots  & \textrm{First Law} & \delta M= \frac{\kappa}{8\pi} \delta A  +V \delta P + \ldots\\
\hline
\end{array}
\nonumber
\ee
The dots stand for the work terms. In the black hole case these are
$\sum_i \Omega^i \delta J_i + \Phi^j \delta Q^j\,$, allowing for multiply charged and spinning black hole solutions.

Black hole chemistry has a much broader range of applications than Einstein-AdS gravity.  Its concepts generalize to Lovelock gravity and to a broad range of other theories of gravity. We discuss the Lovelock case in the next subsection.

\subsection{Lovelock gravity}

One of the more fruitful applications of Black Hole Chemistry has been in Lovelock gravity \cite{Lovelock:1971yv}.  This refers to a class of gravitational theories whose actions  contain terms non-linear in the curvature.  Of course infinitely many such theories exist, but Lovelock gravity theories are  unique in that they give rise to field equations that
are generally covariant and contain at most second order derivatives of the metric.  They remain of considerable interest in the context of quantum gravity where it is expected that the Einstein--Hilbert action is only an effective gravitational action valid for small curvature or low energies, and so will be modified by higher-curvature terms.

In $d$ spacetime dimensions, the Lagrangian of a Lovelock gravity theory  is  \cite{Lovelock:1971yv}
 \begin{equation}
\mathcal{L}=\frac{1}{16\pi}\sum_{k=0}^{K}{\alpha}_k\mathcal{L}^{\left(k\right)}+{\cal L}_m\,,
\label{eq:Lagrangian}
\end{equation}
where the  ${\alpha}_k$ are the  Lovelock coupling constants. The quantities $\mathcal{L}^{\text{\ensuremath{\left(k\right)}}}$ are the $2k$-dimensional Euler densities, given by the contraction of $k$ powers of the Riemann tensor
\begin{equation}
\mathcal{L}^{\left(k\right)}=\frac{1}{2^{k}}\,\delta_{c_{1}d_{1}\ldots c_{k}d_{k}}^{a_{1}b_{1}\ldots a_{k}b_{k}}R_{a_{1}b_{1}}^{\quad c_{1}d_{1}}\ldots R_{a_{k}b_{k}}^{\quad c_{k}d_{k}}\,,
\end{equation}
where the `generalized Kronecker delta function' is totally antisymmetric in both sets of indices. The term $\mathcal{L}^{\left(0\right)}$ is defined to be unity and gives the cosmological constant term, $ \Lambda=-\alpha_0/2=-8\pi P$, $\mathcal{L}^{\left(1\right)}$ gives the Einstein--Hilbert action, $\alpha_1=1$, and $\mathcal{L}^{\left(2\right)}$ corresponds to the quadratic Gauss--Bonnet term.  The matter Lagrangian ${\cal L}_m$ in \eqref{eq:Lagrangian} describes minimal coupling to  matter.
There is an upper bound of $ K=\floor{\frac{d-1}{2}}$  on the sum, which reflects the fact that a given
 term $\mathcal{L}^{\left(k\right)}$ is a purely topological object in $d=2k$, and vanishes identically for $d< 2k$.
Only for $d>2k$ does this term contribute to the equations of motion.  General relativity, whose equations of motion are \eqref{EinEq},  is recovered upon setting ${\alpha}_k=0$ for $k\geq 2$.

Before  proceeding further, we note two special subclasses of Lovelock theories.
Introducing the rescaled coupling constants
\be
\hat \alpha_0=\frac{\alpha_0}{(d-1)(d-2)}\,,\quad
\hat \alpha_1=\alpha_1\,,\quad \hat \alpha_k=\alpha_k\prod_{n=3}^{2k}(d-n)\quad \mbox{for} \quad k\geq 2\,,
\ee
a special class of theories, called {\it Chern--Simons gravity} \cite{Zanelli:2005sa},  arises in odd dimensions for the choice
\be\label{CS-couplings}
\hat \alpha_p =\frac{\ell^{2p-2n+1}}{2n-2p-1}\left( {n-1\atop p}  \right)  \qquad p=1,2,\ldots, n-1 = \frac{d-1}{2}
\ee
of Lovelock couplings.  Here $\ell$ stands for the AdS radius; it is no longer given by the second equality in \eqref{eq:press} but instead is
a non-trivial function of  the `bare' cosmological constant $\Lambda=-\alpha_0/2$ and the higher-order Lovelock couplings.
In this particular case   the local Lorentz invariance of the Lovelock action is enhanced to a local (A)dS symmetry.
Another special case of Lovelock gravity occurs when
\be\label{SpecialCouplings}
\hat \alpha_k=\hat \alpha_K\bigl(K\hat \alpha_K\bigr)^{-\frac{K-k}{K-1}}\left({K\atop k}\right)\quad \mbox{for}\quad 2\leq k<K\,,
\ee
 where $\hat \alpha_K\neq 0$, while $\hat \alpha_1=1$ and $\hat \alpha_0$ is arbitrary. This particular choice was exploited in \cite{Dolan:2014vba} to study an {\it isolated critical point}, as we shall see in Sec.~\ref{Sec:4}.

 The equations of motion for Lovelock gravity, following from a variational principle using \eqref{eq:Lagrangian}, are
\begin{equation}
\mathcal{G}_{\, b}^{a}=\sum_{k=0}^{K}{\alpha}_k\mathcal{G}_{\,\,\quad b}^{\left(k\right)\, a}=8\pi T^a{}_b\,,
\end{equation}
where the Einstein-like tensors $\mathcal{G}_{\,\,\quad b}^{\left(k\right)\, a}$ are given by
\begin{equation}
\mathcal{G}_{\,\,\quad b}^{\left(k\right)\, a}=-\frac{1}{2^{\left(k+1\right)}}\delta_{b\, e_{1}f_{1}\ldots e_{k}f_{k}}^{a\, c_{1}d_{1}\ldots c_{k}d_{k}}R_{c_{1}d_{1}}^{\quad e_{1}f_{1}}\ldots R_{c_{k}d_{k}}^{\quad e_{k}f_{k}}\label{eq:G}\,,
\end{equation}
and each of them independently satisfies a conservation law $\nabla_{a}\mathcal{G}_{\,\,\quad b}^{\left(k\right)\, a}=0\,$.

The same arguments in App.~\ref{FirstLaw} that use  the Hamiltonian formalism can be generalized to the Lovelock case, yielding
\be\label{flawLove}
\delta M = T\delta S+V\delta P+\sum_{k=2}^K\Psi^k\delta\alpha_k+\sum_i\Omega^i\delta J^i+\sum_j\Phi^j\delta Q^j
\ee
for the first  law of black hole thermodynamics  \cite{Jacobson:1993xs,Kastor:2010gq},
and
\be\label{SmarrLovelock}
\frac{d-3}{d-2}M=TS-\frac{2}{d-2}PV+\sum_{k=2}^K\frac{2(k-1)}{d-2}\Psi^k\alpha_k+\sum_i\Omega^iJ^i+\frac{d-3}{d-2}\sum_j\Phi^jQ^j
\ee
for the Smarr formula, using the Euler scaling argument.
The Smarr formula \eqref{smarrBHLBI} for black holes can be expressed in terms of a Noether charge surface integral plus a suitable volume integral \cite{Liberati:2015xcp}.
We stress that the entropy is  no longer proportional to the horizon area but instead is given by the expression
\begin{equation}\label{S}
S= \frac{1}{4} \sum_k {\alpha}_{k} {\cal A}^{(k)}\,,\quad   {\cal A}^{(k)}  = k\int_{\mathcal{H}} \sqrt{\sigma}{\mathcal{L}}^{(k-1)}\,,
\end{equation}
where  $\sigma$ denotes the determinant of $\sigma_{ab}$,  the induced metric on the black hole horizon ${\mathcal{H}}$, and the Lovelock terms ${\mathcal{L}}^{(k-1)}$ are evaluated on that surface.  The $k=1$ case is Einstein gravity and yields the usual value of one-quarter the horizon area.
Note also that the Lovelock coupling constants ${\alpha}_k$ are regarded  as thermodynamic variables in \eqref{flawLove}.  Their  conjugate potentials (whose physical meaning has yet to be explored) were denoted by $\Psi^k$.
For spherical Lovelock black holes they can be explicitly computed \cite{Frassino:2014pha}.

A similar situation  occurs in Born--Infeld non-linear electrodynamics \cite{GunasekaranEtal:2012}.  This is a theory of
electromagnetism in which the Lagrangian  (in four dimensions) is\footnote{ Note that in the absence of magnetic fields the second electromagnetic invariant $\tilde F$ does not contribute and is often omitted in the literature. The simplified Lagrangian \eqref{BInfeld} without $\tilde F$ is then promoted to higher dimensions and called Born--Infeld theory, e.g. \cite{Zou:2013owa}. This is to be compared to the approach taken in string theory, e.g. \cite{Gibbons:2001}, where the Born--Infeld action is identified with ${\cal L}_{BI}\propto \sqrt{\det(g_{ab}+\frac{1}{b}F_{ab})}$ in all dimensions. }  \cite{Born:1934gh}
\begin{equation}\label{BInfeld}
{\cal L}_{BI} = 4{b^2}\Bigl(1-\sqrt{1+\frac{2F^2}{b^2}-\frac{\tilde F^2}{b^4}}\ \Bigr)\,,\quad F^2=\frac{1}{4}F^{ab}F_{ab}\,,\quad \tilde
F^2=\frac{1}{8}\epsilon^{abcd} F_{ab}F_{cd}\,.
\end{equation}
The parameter $b$ represents the maximal electromagnetic field strength. This quantity can be related to the string tension in the context of string theory  \cite{Gibbons:2001}, with $b=\frac{1}{2\pi\alpha'}$.
Promoting $b$ to a thermodynamic variable adds an extra term in the first law \eqref{flawLove} \cite{GunasekaranEtal:2012}
\be\label{flawLoveBI}
\delta M = T\delta S+\dots+{\cal B}\delta b\,,
\ee
where ${\cal B}=\left(\frac{\partial M}{\partial b}\right)$
is the thermodynamic conjugate to the coupling $b$.  Noting that $b$ has units of electric field and the enthalpy $M$ has units of energy, the quantity ${\cal B}$ thus has units of electric polarization  per unit volume. Consequently ${\cal B}$ has been referred to as `Born--Infeld vacuum polarization' \cite{GunasekaranEtal:2012}.
These results straightforwardly extend to higher dimensions, and there have been further investigations into Born--Infeld electrodynamics in extended phase space
\cite{Hendi:2012um, Zou:2013owa,Hendi:2014kha, Mo:2014qsa,Belhaj:2014tga, Hendi:2015soe, Zeng:2016sei}.
This yields the following generalized Smarr formula:
\ba\label{smarrBHLBI}
\frac{d-3}{d-2}M&=&TS-\frac{2}{d-2}PV+\sum_{k=2}^K\frac{2(k-1)}{d-2}\Psi^k\alpha_k+\sum_i\Omega^iJ^i\nonumber\\
&&+\frac{d-3}{d-2}\sum_j\Phi^jQ^j- \frac{1}{d-2}{\cal B}b\,,
\ea
upon incorporation of both Lovelock \cite{Kastor:2010gq} and Born--Infeld \cite{GunasekaranEtal:2012} terms.

The extended Smarr formula and the first law in the more general setting of `variable background fields'  was recently studied using the covariant formalism \cite{Wu:2016auq}; this approach offers a new perspective on variable $\Lambda$ from this more general viewpoint.

\section{What is a volume of a black hole?}\label{Sec:3}
How do we describe and characterize the geometry of horizons, and what are their general geometrical properties?
A standard answer is offered by studying a relationship between  horizon area (an intrinsic horizon property) and dynamical quantities such as the total energy or angular momentum, and results in the so called {\it Penrose  (isoperimetric) inequalities} that are closely related to cosmic censorship and Thorne's hoop conjecture \cite{Gibbons:2012ac}.  As we have seen in Sec.~\ref{Sec:2},   extended phase space thermodynamics enables one to define a new ``intrinsic'' quantity ---  {\it thermodynamic volume} --- associated with the (black hole) horizon.
It is the purpose of this section to study its  physical meaning and characteristic properties, and in particular
the associated isoperimetric inequality.

\subsection{Thermodynamic volume}\label{3a}

The black hole {\it thermodynamic volume}  is a quantity with dimensions of (length)$^{d-1}$
(in other words, a spatial volume) that characterizes a spacetime and is entirely derived from thermodynamic considerations.  For an asymptotically AdS black hole spacetime it is the quantity thermodynamically conjugate to $P$
\be\label{defV}
V\equiv \left(\frac{\partial M}{\partial P}\right)_{S,Q,J,\dots}\,,
\ee
as defined in \eqref{Vdef}.

Originally this conjugate variable was interpreted geometrically as a ``... finite, effective volume for the region outside the AdS black hole horizon" \cite{Kastor:2009wy}.
Later, it was pointed out that \eqref{Vdef} is
independent of any geometric volume \cite{Dolan:2010ha} for most black holes \cite{Cvetic:2010jb}, and should be
regarded as  a {\it thermodynamic volume}.  Furthermore, although the definition \eqref{defV}  was originally coined
for asymptotically AdS black hole spacetimes,\footnote{Note that to use definition \eqref{defV} one does not need to know the full AdS solution; a perturbative  expansion in $P$ is sufficient.
See also Sec.~\ref{Sec:6} for a similar construction in asymptotically dS spacetimes.
Other (mostly geometric) definitions of black hole volume are discussed later in this section.} it turns out \cite{Cvetic:2010jb} that a limit to asymptotically flat spacetimes, $P\to 0$, often yields a finite result for the thermodynamic volume that is `smoothly connected' to its AdS counterpart, thereby providing a way for defining a thermodynamic volume of asymptotically flat black holes.
For example, starting from the charged-AdS black hole spacetime \eqref{HDRN}
and employing the definition \eqref{defV},  the  thermodynamic volume \eqref{volrp} was obtained:
\be
V=\frac{4}{3}\pi r_+^3\,,
\ee
where $r_+$ is the black hole horizon radius. This result does not  explicitly depend on the value of $\Lambda$ (or the charge $Q$) and so can be taken to be valid for $\Lambda=0$.
Amusingly, the result is the same as if the black hole were a  ball of radius $r_+$ in Euclidean space.

A recent contrasting viewpoint \cite{Armas:2015qsv} is that thermodynamic volume should be replaced with  a more general notion of gravitational tension
that  describes the extra energy associated with the presence of gravitational fields surrounding  a black hole. Gravitational tension vanishes in the flat-space limit and is proportional $r_+^3/l^3$ for a Schwarzschild-AdS black hole.
The  relationship of this approach to the concept of thermodynamic volume we describe here remains to be explored.\footnote{
It has been further argued \cite{Armas:2015qsv}  that the volume in the asymptotically flat  limit  is either non-universal or that one must come to grips with the fact that an asymptotically flat black hole can have more than one well-defined volume.  We disagree---the definition (3.1) is unambiguous and does not imply multiple volumes in the flat space limit. Moreover, we consider the fact that the thermodynamic
volume has a smooth non-vanishing limit for  asymptotically flat black holes to be not a drawback but rather a natural feature. It is our everyday experience that volume of a body has meaning and can be non-trivial even in the absence of external pressure. Since in our setting $\Lambda$ is variable, the asymptotically flat black hole with $P = 0$ just corresponds to a ``fine-tuned'' state and there is no reason   why its volume should be trivial. Fluctuations in pressure will lead to black holes for which their volume can be calculated from the definition \eqref{defV}. Requiring a smooth limit then yields a non-trivial volume for asymptotically flat black holes.}

In the presence of rotation,  additional charges, and other thermodynamic parameters, the formula for the thermodynamic volume gets more complicated. By now a wide variety of explicit expressions for the thermodynamic volume have been found for black holes for which the exact solution with cosmological constant is known and their thermodynamics is well defined. These include higher-dimensional rotating black holes \cite{Cvetic:2010jb}, charged black holes of various supergravities \cite{Cvetic:2010jb}, superentropic black holes \cite{Hennigar:2014cfa, Hennigar:2015cja}, accelerated black holes \cite{Appels:2016uha, Liu:2016uyd}, or `ultraspinning black rings' obtained in the blackfold approximation \cite{Altamirano:2014tva, Armas:2015qsv}.  For example, for the $d$-dimensional Kerr-AdS black hole (given in App.~\ref{AppB}),  the formula \eqref{defV} yields
\be\label{VKerrAdS}
V=\frac{r_+A}{d-1}\Bigl(1+\frac{1+r_+^2/l^2}{(d-2)r_+^2}\sum_i\frac{a_i^2}{\Xi_i}\Bigr)=\frac{r_+A}{d-1}+\frac{8\pi}{(d-1)(d-2)}\sum_i a_i J_i\,,
\ee
where $a_i$ are various (up to $[(d-1)/2]$) rotation parameters, $J_i$ are the associated angular momenta, and
\be\label{AKerrAdS}
A=\frac{\omega_{d-2}}{r_+^{1-\varepsilon}}\prod\frac{a_i^2+r_+^2}{\Xi_i}\,,\quad \Xi_i=1-\frac{a_i^2}{l^2}\,,
\ee
is the horizon area. The total number of spacetime  dimensions $d=2N+1+\varepsilon$ with $\varepsilon=1$ in even and $\varepsilon=0$ in odd dimensions; in even dimensions $a_{N+1} \equiv 0$.

Equation \eqref{VKerrAdS} demonstrates that in general a simple expression for $V$ with intuitive geometrical meaning
does not hold.  It is therefore natural to ask if the quantity $V$, defined by \eqref{defV}, obeys properties that one would like to associate with the volume of black hole.
A characteristic property for the volume of a simply connected domain in Euclidean space is that it obeys an isoperimetric inequality.
We investigate this property for the thermodynamic volume in the next section.

\subsection{Reverse isoperimetric inequality}\label{3b}

In Euclidean space ${\mathbb E}^{d-1}$, the isoperimetric inequality for the volume $V$ of a connected domain
whose area is $A$ states that the ratio
\be\label{ratio}
{\cal R}= \Bigl( \frac{(d-1){V}\,}{\omega_{d-2} } \Bigr )^{\frac{1}{d-1}}\,
  \Bigl(\frac{\omega_{d-2}}{A}\Bigr)^{\frac{1}{d-2}}
  \ee
 obeys ${\cal R}\leq 1$, where
 \be\label{omeg-d}
\omega_{d}=\frac{2\pi^{\frac{d+1}{2}}}{\Gamma\Bigl(\frac{d+1}{2}\Bigr)}
\ee
is the volume of the unit $d$-sphere.
Equality holds if and only if the domain is a standard round ball.

It was {\it conjectured} in \cite{Cvetic:2010jb} that a {\it reverse isoperimetric inequality},
\be\label{ISO}
{\cal R}\geq 1\,,
\ee
holds for any asymptotically AdS black hole, upon identifying $A$ with the horizon area and $V$ with the associated thermodynamic volume,
the bound being saturated for Schwarzschild-AdS black holes. In other words, for a fixed thermodynamic volume the
entropy of the black hole is maximized for the Schwarzschild-AdS spacetime.\footnote{Note that the same inequality
can be extended to all black holes, with $\omega_d$ replaced by the corresponding unit volume of the
space transverse to the event horizon.  For example, for topological AdS black holes ($k\neq 1$ in \eqref{Schw-Ads}), $\omega_d$ would be replaced by the corresponding volume of the constant curvature space $\omega_d^{(k)}$  \cite{Mann:1997iz}.}

It is straightforward to prove the inequality \eqref{ISO} for Kerr-AdS black holes.  Following \cite{Cvetic:2010jb}, we introduce a new variable
\be
z=\frac{1+r_+^2/l^2}{r_+^2}\sum_i \frac{a_i^2}{\Xi_i}\,,
\ee
 to find that quantities \eqref{VKerrAdS} and \eqref{AKerrAdS} yield
\ba
{\cal R}^{d-1}&=&\Bigl[1+\frac{z}{d-2}\Bigr]\Bigl[\prod_i\frac{r_+^2+a_i^2}{r_+^2\Xi_i}\Bigr]^{-\frac{1}{d-2}}
\geq
\Bigl[1+\frac{z}{d-2}\Bigr]\Bigl[\frac{2}{d-2}\Bigl(\sum_i \frac{1}{\Xi_i}+\sum_i \frac{a_i^2}{r_+^2\Xi_i}\Bigr)\Bigr]^{-1/2}\nonumber\\
&=&
\Bigl[1+\frac{z}{d-2}\Bigr]\Bigl[1+\frac{2z}{d-2}\Bigr]^{-1/2}\equiv G(z)\,,
\ea
employing the arithmetic/geometric (AG) inequality $(\prod_i x_i)^{1/N}\leq (1/N)\sum_i x_i$.  Since $G(0)=1$ and $d\log G(z)/dz\geq 0$,  the reverse isoperimetric inequality \eqref{ISO} follows.

For a broad variety of (charged and/or rotating) spherical black holes \cite{Cvetic:2010jb}, as well as
for example (thin) ultraspinning black rings with toroidal horizon topology \cite{Altamirano:2014tva}, the conjecture  \eqref{ISO} has been shown to be valid. For more complicated black holes  \eqref{ISO} has been confirmed numerically.
Recently a class of exotic  black hole spacetimes was found to violate \eqref{ISO}. These are studied in the next subsection.

\subsection{Super-entropic black holes}\label{3c}

Super-entropic black holes describe an 
exotic class of rotating AdS black hole solutions with noncompact event horizons and finite horizon area, whose entropy exceeds the maximum implied from the conjectured reverse isoperimetric inequality \eqref{ISO}.
\begin{figure}
\centering
\includegraphics[width=0.7\linewidth, height=0.35\textheight]{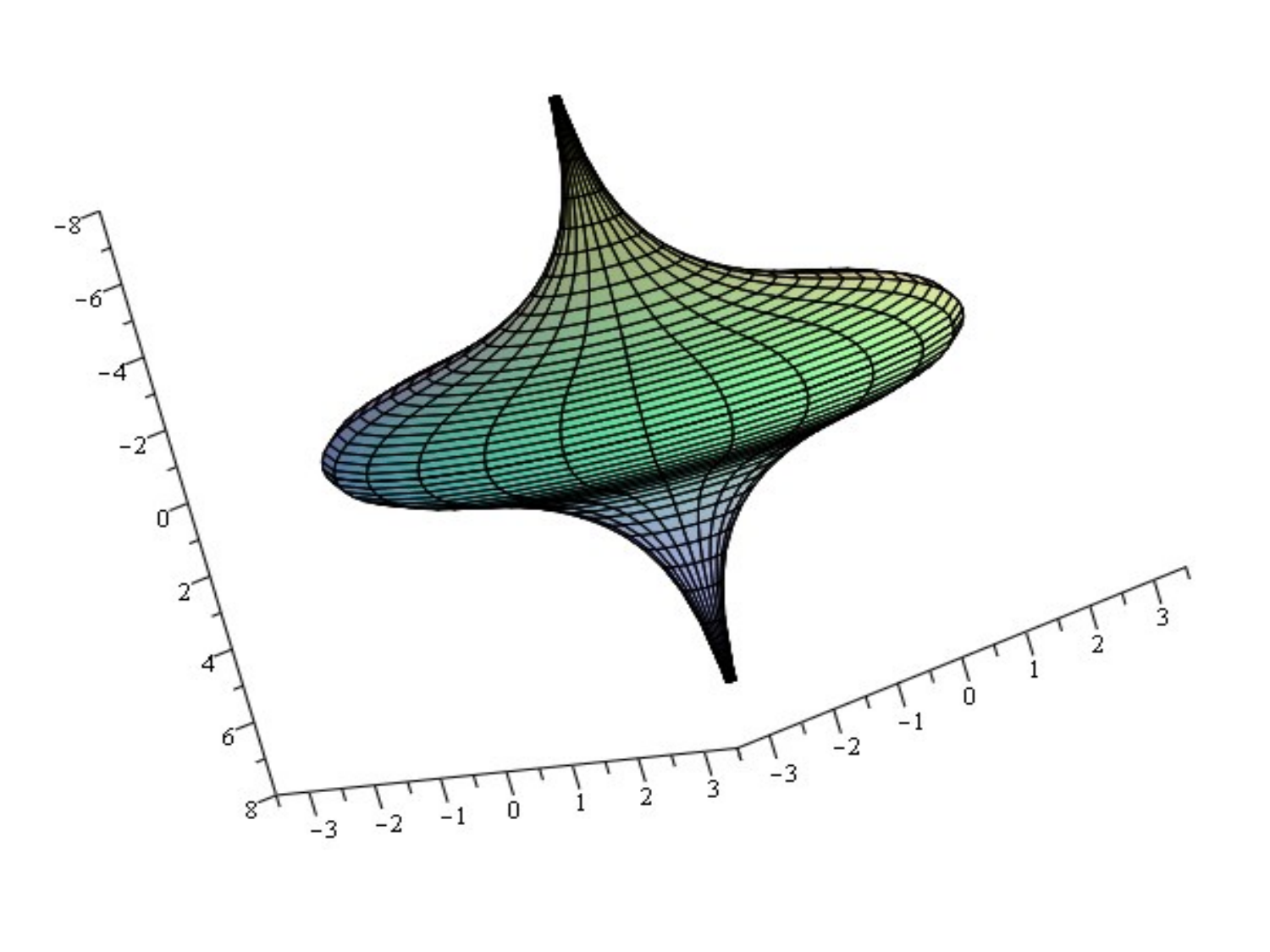}
\caption{{\bf Super-entropic black hole: horizon embedding.} The horizon geometry is embedded in $\mathbb{E}^3$ for the following choice of parameters:  $l=1$, $r_+=\sqrt{10}$ and $\mu = 2\pi$.
 }
\label{Fig:3c}
\end{figure}
First obtained by taking a particular limit of the Carter 
metric \cite{Gnecchi:2013mja}, there is now an entire class of rotating and/or charged super-entropic black holes in four and higher dimensions \cite{Klemm:2014rda, Hennigar:2014cfa, Hennigar:2015cja, Hennigar:2015gan} for which
${\cal R}\geq 1$ does not hold\footnote{Similar types of black hole spacetimes have been obtained where  the inequality remains to be checked \cite{Chen:2016rjt, Chen:2016jxv}. }.

The simplest example of such a black hole can be obtained by applying a new type of (singular) ultraspinning limit
to the $d=4$ Kerr-AdS metric (see App.~\ref{AppB}) in which the rotation parameter $a$ approaches the AdS radius $l$ \cite{Hennigar:2014cfa, Hennigar:2015cja}.
The resulting metric reads
\ba\label{KNADS2}
ds^2&=&-\frac{\Delta}{\Sigma}\left[dt-l\sin^2\!\theta d\psi\right]^2
+\frac{\Sigma}{\Delta} dr^2+\frac{\Sigma}{\sin^2\!\theta}d\theta^2
+\frac{\sin^4\!\theta}{\Sigma}\left[l dt-(r^2+l^2)d\psi\right]^2\,,\nonumber\\
\Sigma&=&r^2+l^2\cos^2\!\theta\,,\quad \Delta=\Bigl(l+\frac{r^2}{l}\Bigr)^2-2mr\,.
\ea
The thermodynamic charges are
\ba
M&=&\frac{\mu m}{2\pi}\,,\quad J=Ml\,,\quad
\Omega = \frac{l}{r_+^2+l^2}\,, \quad
T=\frac{1}{4\pi r_+}\left(3\frac{r_+^2}{l^2}-1 \right), \nonumber \\
S&=& \frac{\mu}{2}(l^2+r_+^2)=\frac{A}{4}, \quad V= \frac{r_+A}{3}=\frac{2}{3} \mu r_+ \left(r_+^2 + l^2 \right) \,,\label{eq:volume}
\label{eq:thermo_properties}
\ea
with the parameter $\mu$ denoting the periodicity of the coordinate $\psi$. The
isoperimetric ratio is straightforwardly computed
\be
\mathcal{R}=\left(\frac{r_+A}{2\mu}\right)^{1/3}\left(\frac{2\mu}{A}\right)^{1/2}
= \left(\frac{r_+^2}{r_+^2 + l^2}\right)^{1/6}< 1\,,
\ee
and obviously violates the conjecture \eqref{ISO}.

The metric \eqref{KNADS2} exhibits many exotic properties \cite{Hennigar:2014cfa, Hennigar:2015cja}:
it describes a black hole  whose horizon has the topology of
 a sphere with two punctures. Fixed $(r, t)$ sections are non-compact and near the axis of symmetry approach Lobachevsky space.
The axis itself is removed from the spacetime, and the coordinate $\psi$ becomes null as $r\to \infty$. 
The geometry of the horizon can be visualized by embedding it in Euclidean 3-space as illustrated in Fig.~\ref{Fig:3c}.

Super-entropic black holes indicate that the reverse isoperimetric inequality as stated in Sec.~\ref{3b} cannot be entirely correct, thereby  motivating the following more stringent version \cite{Hennigar:2014cfa}:\\
{\bf Conjecture} (Revised reverse isoperimetric inequality). {For an AdS black hole with thermodynamic volume $V$ and with {\it compact
horizon} of area $A$, the ratio \eqref{ratio} satisfies ${\cal R}\geq 1$.}\\
The proof of this conjecture remains an interesting open
question for further study.

\subsection{Negative volume: Taub-NUT solution}\label{3d}

So far we have limited ourselves to applications of the extended phase space thermodynamics to black hole spacetimes.
However if  taken seriously it should apply to all geometries and spacetimes, even those without horizons \cite{Johnson:2014xza}. In this subsection we look into its applications to the Taub-NUT-AdS class of solutions.

The Euclidean Taub-NUT-AdS metric\footnote{Lorentzian Taub-NUT solutions have many peculiar properties and are usually discarded as physically irrelevant; however this viewpoint has recently been challenged \cite{Clement:2015cxa}.} is \cite{Stephani:2003tm}
\ba
ds^2&=&f(d\tau+2n\cos\theta d\phi)^2+\frac{dr^2}{f}+
(r^2-n^2)(d\theta^2+\sin^2\!\theta d\phi^2)\,,\nonumber\\
f&=&\frac{(r^2+n^2)-2mr+(r^4-6n^2r^2-3n^4)/l^2}{r^2-n^2}\,,
\ea
and represents a `gravitational analogue' of magnetic monopole, with the NUT charge $n$ playing the role of the dyonic charge to gravitational mass $m$. In order to ensure the invisibility of Misner strings, the Euclidean time $\tau$ has to be identified with periodicity $\beta=8\pi n$. Asymptotically, the solution  approaches a squashed 3-sphere, written as an $S^1$ fibration over $S^2$ with first Chern class $n$.

Concentrating on thermodynamics, the most peculiar feature of the solution is that
the entropy
\be
S=4\pi n\Bigl[\frac{r_+^2+n^2}{2r_+}+\frac{4\pi P}{3}\Bigl(3r_+^3-12n^2r_+-\frac{3n^4}{r_+}\Bigr)\Bigr]\,,
\ee
which can be calculated from the corresponding action  \cite{Mann:1999,Emparan:1999pm} (and can be understood
 as a Noether charge \cite{Garfinkle:2000ms}) no longer obeys the Bekenstein--Hawking law \eqref{ent}.
The temperature reads
\be\label{Tn}
T=\frac{|f'(r_+)|}{4\pi} = \frac{1}{8\pi n}\,,
\ee
where the latter equality follows from regularity requirements and  imposes a restriction\footnote{Although natural in the case of spherical Taub-NUT solutions \cite{Hawking:1998ct}, a similar restriction is often imposed for the planar and hyperbolic counterparts as well \cite{Mann:2004mi, Lee:2015wua}; recently this has been questioned
 \cite{Hennigar:2015gan}.}  $n=n(r_+)$.
Dependent on the nature of the fixed point of the vector $\partial_\tau$ for which $f(r_+)$ vanishes, this equation has two solutions:
i) the {\it Taub-NUT} case $n=r_+$, for which the fixed point set is zero dimensional and ii) the {\it Taub-Bolt} case where it remains two dimensional.

Together with the expression for the gravitational enthalpy $H=M$, equation \eqref{Tn} yields
the extended first law provided we identify the following thermodynamic volume \cite{Johnson:2014xza}:
\be
V=\frac{4\pi r_+^3}{3}\Bigl(1-\frac{3n^2}{r_+^2}\Bigr)\,.
\ee
Of particular interest  is the  Taub-NUT case $r_+=n$, which yields
\be
V_{\mbox{\tiny NUT}}=-\frac{8\pi n^3}{3}\,,
\ee
or in other words a  {\it negative} thermodynamic volume!
First noted in \cite{Johnson:2014xza}, this peculiar feature has been interpreted as the fact that in the Taub-NUT case, it is the
environment that has to do work on the system to create the solution while the universe has to increase its volume. This is
in contrast to the black hole case where part of the universe had to be removed to `make a place' for the black hole.

Picking up the threads of \cite{Johnson:2014xza}, the thermodynamic properties and possible phase transitions of Taub-NUT-AdS solutions and their generalizations have been further studied \cite{Lee:2014tma,Lee:2015wua, Hennigar:2015gan} and extended to include rotation
 \cite{MacDonald:2014zaa} and deformations to dyonic black holes \cite{Johnson:2014pwa}.  It is somewhat remarkable that extended phase space thermodynamics provides a sensible framework for the study of these unusual solutions and that a plausible explanation may exist for objects characterized by negative volume.

\subsection{Black hole compressibility}\label{3compr}

Having defined the concept of black hole volume, one can start studying its associated physical quantities. One of them, the black hole {\it  adiabatic compressibility}
\cite{Dolan:2011jm} has attracted attention in connection with black hole stability \cite{Dolan:2012jh, Dolan:2013dga, Dolan:2014woa} (see also \cite{Dolan:2013yca, Dolan:2014lea}).

Adiabatic compressibility is defined as
\be\label{kappaS}
\kappa_S=-\frac{1}{V}\left(\frac{\partial V}{\partial P}\right)_{S,J,Q}\,,
\ee
and for Kerr-AdS black holes is manifestly positive and regular; non-rotating black holes are adiabatically incompressible.  In four dimensions $\kappa_S$ reaches its maximum in the extremal black hole case, while in higher dimensions the `softness' of the equation of state can be used as an indicator of the ultraspinning instability \cite{Emparan:2003sy}.

Associated with the adiabatic compressibility is the {\it ``speed of sound''} $v_s\in[0,1]$ \cite{Dolan:2011jm}.
 Defining an average density, $\rho=M/V$, this reads
\be
v_S^{-2}=\left(\frac{\partial \rho}{\partial P}\right)_{S,J,Q}=1+\rho \kappa_S=
1+\frac{9(2\pi J)^4}{[6S^2+16PS^3+3(2\pi J)^2]^2}\,,
\ee
where the last equality applies to the 4-dimensional Kerr-AdS black hole. One can think of $v_s$ as a velocity of a ``breathing mode'' due to changing volume at constant $S$.
It has been speculated  that a collection of primordial black holes might affect the speed of sound through the medium in the early universe  in a manner similar to how a suspension of compressible spheres affects the speed of sound in a fluid \cite{Dolan:2011jm}.

\subsection{Killing co-potential volume}\label{3aa}

The concept of thermodynamic volume was given a {\it geometric footing} when the first law of black hole mechanics was extended to include a cosmological constant  \cite{Kastor:2009wy}. It stems from the following simple idea.

Consider a Killing horizon $H$ generated by the corresponding Killing vector $\xi$. Due to the Killing equation such a vector is divergence-less, $\nabla\cdot \xi=0$, and hence (at least locally) there must exist a {\it Killing co-potential} $\omega$ (defined up to a co-closed 2-form), such that $\nabla_a \omega^{ab} =\xi^b$.    The arguments in App.~\ref{FirstLaw} then yield\footnote{Expressions similar to the quantity \eqref{vdef2} can be also written for Lovelock thermodynamic potentials $\Psi^k$ \cite{Kastor:2010gq} that appear in the thermodynamic first law of Lovelock black holes, \eqref{flawLove}.}
\begin{equation}\label{vdef2}
V=  \int _{\infty} d\mathcal{S} r_c n_b \left(\omega ^{cb} - \omega_{AdS}^{cb}\right) -\int _{H} d\mathcal{S} r_c n_b \omega ^{cb}\,,
\end{equation}
as a candidate definition for the volume of a black hole.  There are likewise expressions for the variations  of the conserved mass \eqref{flatdm} and angular momenta \eqref{flatdj} respectively. One
would like to integrate these relations to obtain expressions for the total energy and  angular momenta of the black hole.

Unfortunately the definition \eqref{vdef2} is not unique, since $ \tilde \omega_{ab}=  \omega_{ab} + \zeta_{ab}$ (where
$ \nabla ^a \zeta_{ab} =0$) also satisfies $ \nabla ^a \tilde \omega_{ab} = \xi_b$ and so is an equally valid co-potential.
This renders ambiguous the definition of energy based on \eqref{flatdm}.    The best that can be
done is to make a gauge choice for $\omega$ such that \cite{Cvetic:2010jb}
\be
M = -\frac{(d-2)}{16\pi (d-3)}\,
   \int_{\infty}d\mathcal{S} r_c n_b (\nabla^{[c}k^{b]} + 2\Lambda  \omega ^{cb}_{AdS}) \label{komaradsmass}
\ee
is the total mass $M$ (with $k=\partial_t$ the timelike Killing vector), which itself must be determined
by other means.  This in turn implies from \eqref{vdef2} that
\begin{equation}\label{vdef3}
V=  { -\int_{H} d\mathcal{S} r_c n_b \omega ^{cb}  }
\end{equation}
is the geometric definition of volume.

The conformal approach to calculating conserved charges \cite{Ashtekar:1984zz, Ashtekar:1999jx, Das:2000cu} provides the most straightforward means for computing mass and angular momenta.  It has a great advantage over other methods (such as that of Abbott and Deser \cite{Abbott:1981ff}) insofar as   it involves an integration at infinity of a finite quantity
computed from the Weyl tensor.  No infinite subtraction of a pure AdS background is required.


\subsection{Other definitions of black hole volume}\label{3e}
We conclude this section with an overview of
existing alternate approaches towards defining   black hole volume:
\begin{itemize}
\item
{\it Geometric volume}, due to Parikh \cite{Parikh:2005qs}, is probably the first ever notion of black hole volume.
To find geometric volume, one essentially integrates the full $d$-dimensional volume element over a $t=const.$ slice, yielding
\be\label{Vgeo}
V_{\mbox{\tiny geom}}=\int_{r_0}^{r_+} \!\!dr \int d\Omega_{d-2}\sqrt{-g_{(d)}}\,,
\ee
which is independent of the choice of `stationary time-slicing' \cite{Parikh:2005qs}.
The  lower bound $r_0$ of integration is identified with the `position of the singularity' and becomes problematic to define in the case of rotating black holes.  Geometric volume was further studied in \cite{Grumiller:2005zk, Ballik:2010rx} and has been compared to  thermodynamic volume in \cite{Cvetic:2010jb}, and to  vector volume \cite{Ballik:2013uia} and Hayward's volume \cite{Hayward:1997jp} in \cite{Ballik:2013uia}.
 The volume \eqref{Vgeo} seems to satisfy the standard isoperimetric inequality, ${\cal R}\leq 1$, \cite{Cvetic:2010jb}. It was also implicitly used for the study of horizon thermodynamics \cite{Padmanabhan:2002sha, Paranjape:2006ca, Hansen:2016ayo}, the subject of Sec.~\ref{Sec:6}.

\item
{\it Vector volume},  due to Ballik and Lake \cite{Ballik:2013uia},  is a more mathematically rigorous formulation of geometric volume.

\item
{\it Dynamical volume} is due to Christodoulou and Rovelli \cite{Christodoulou:2014yia} and is based on the following simple observation in Euclidean space: the volume inside a two-sphere $S$ is the volume of the largest spacelike spherically symmetric 3d surface $\Sigma$ bounded by $S$.
Generalizing to curved space,
the horizon of a spherically symmetric black hole is foliated by (spacelike) spheres $S_v$, labeled by the null coordinate $v$ (setting $v=0$ at collapse time).
{At a given `time' $v$, the spacelike slice $\Sigma_v$ bounded by $S_v$ of maximal volume is taken to correspond to the  volume of the black hole.}  Interestingly, the dynamical volume grows with $v$ and quickly approaches `large' asymptotic values
\be
V_{\mbox{\tiny dyn}}\propto m^2v \quad \mbox{as} \quad v\to \infty\,,
\ee
providing  `plenty of room' to store information \cite{Christodoulou:2014yia}.
Obviously the definition leads to a complicated maximization problem, the resulting volume being very different from the thermodynamic volume and the other three volumes mentioned in this subsection. For this reason we will not discuss this notion any further in this paper and refer interested readers to the original paper \cite{Christodoulou:2014yia} and to recent studies    \cite{Bengtsson:2015zda, Ong:2015tua, Zhang:2015gda}.
We just mention
 that  dynamical volume is closely related (with the main difference corresponding to a different choice of boundary conditions) to the time-dependent volume of an Einstein--Rosen bridge that is conjectured to describe the computational complexity of the dual quantum state \cite{Susskind:2014rva, Stanford:2014jda}.
  \end{itemize}

With a variety of definitions for the black hole volume one may ask if there is any consensus and connection among them. Interestingly, it turns out that all the volumes (apart  from the dynamical one) coincide for the simple spherically symmetric case, but produce different results for more complicated spacetimes, for example in the presence of rotation. Concretely, for the Kerr-AdS black hole the geometric volume, or the vector volume \cite{Ballik:2013uia}, both read
\be
V_{\mbox{\tiny geom}}=\frac{r_+A}{d-1}\,.
\ee
This is to be compared with the expression \eqref{VKerrAdS} for the thermodynamic volume. We observe that the two differ
by the presence of a ``rotational'' part and coincide in the limit $a_i\to 0$.

For  black holes in the ultraspinning regime $a \to l$, the rotational part completely dominates and the thermodynamic volume becomes very different from its geometric counterpart  \cite{Altamirano:2014tva}.
The two volumes are also completely different for the Taub-NUT-AdS geometries studied in the previous subsection  \cite{Johnson:2014xza}.

\section{Black hole chemistry}\label{Sec:4}
With thermodynamic pressure and volume defined, we can {\it extend the thermodynamic phase space} and study the thermodynamics of black holes in a new framework, sometimes referred to as {\it Black Hole Chemistry} \cite{Kubiznak:2014zwa}. This change of perspective has been shown to have a number of remarkable consequences---black holes now seem to behave in ways that are analogous to a variety of ``everyday" chemical phenomena, such as Van der Waals behavior, solid/liquid phase transitions, triple points, reentrant phase transitions, and heat engines. These will be described in this section, employing the
following thermodynamic machinery:
\begin{itemize}
\item
We shall study the thermodynamics of charged and/or rotating AdS black holes in a canonical (fixed $Q$ or $J$) ensemble. This will then be related to {\it fluid thermodynamics}, by comparing the ``same physical quantities'':
cosmological pressure is identified with the pressure of the fluid, thermodynamic volume of a black hole with the volume of the fluid, temperature of the black hole with the temperature of the fluid, and so on. Although quite natural, note that this ``{\it dictionary}'' is quite different from the extended AdS/CFT dictionary discussed in the next section.

\item
The thermodynamic potential of interest is the {\it Gibbs free energy}
\be
G=M-TS=G(P,T,J_1,\dots,J_N,Q_1,\dots, Q_n)\,.
\ee
The equilibrium state corresponds to the global minimum of $G$.

\item
{\it Local thermodynamic stability} corresponds to positivity of the specific heat
\be
C_P\equiv C_{P,J_1,\dots, J_N,Q_1,\dots,Q_n}=T\Bigl(\frac{\partial S}{\partial T}\Bigr)_{P,J_1,\dots,J_N,Q_1,\dots, Q_n}\,.
\ee
\end{itemize}

The aim of this program is to construct $P-T$ {\it phase diagrams}, find {\it critical points}, study their {\it critical exponents}, and determine whatever other interesting transitional behavior might arise.

We shall start from the simple example of a Schwarzschild-AdS black hole and its associated Hawking--Page transition, and then proceed to more complicated spacetimes that demonstrate more elaborate phase phenomena.

\subsection{A new look at the Hawking--Page transition} \label{Sec:41}

The spherically symmetric ansatz  \eqref{Schw-Ads} in $d=4$ has the metric function
\be
f=k-\frac{2M}{r}+\frac{r^2}{l^2}\,,
\ee
valid for spherical ($k=1$), planar $(k=0$), or hyperbolic $(k=-1$) horizon geometries.
The thermodynamic quantities are similar to the $Q=0$ versions of those in \eqref{TDchgBH}  and read
\be\label{S}
M = \frac{r_+ A_k}{8}\Bigl(k + \frac{r^2_+}{l^2}\Bigr)\,, \quad S=\frac{\pi A_k}{4} r^2_+\,,\quad
T = \frac{k l^2 + 3r^2_+}{4\pi l^2 r_+}\,,\quad V = \frac{\pi A_k}{3} r^3_+\,,
\ee
where  $\pi A_k$ is the area of the constant-curvature space.\footnote{For a sphere, $A_{k=1} = 4$;
for a torus, $A_{k=0} = A B$, where $A$ and $B$ are the sides of the torus. There is no simple formula for $A_{k=-1}$.}

\begin{figure*}
\centering
\begin{tabular}{cc}
\includegraphics[width=0.49\textwidth,height=0.3\textheight]{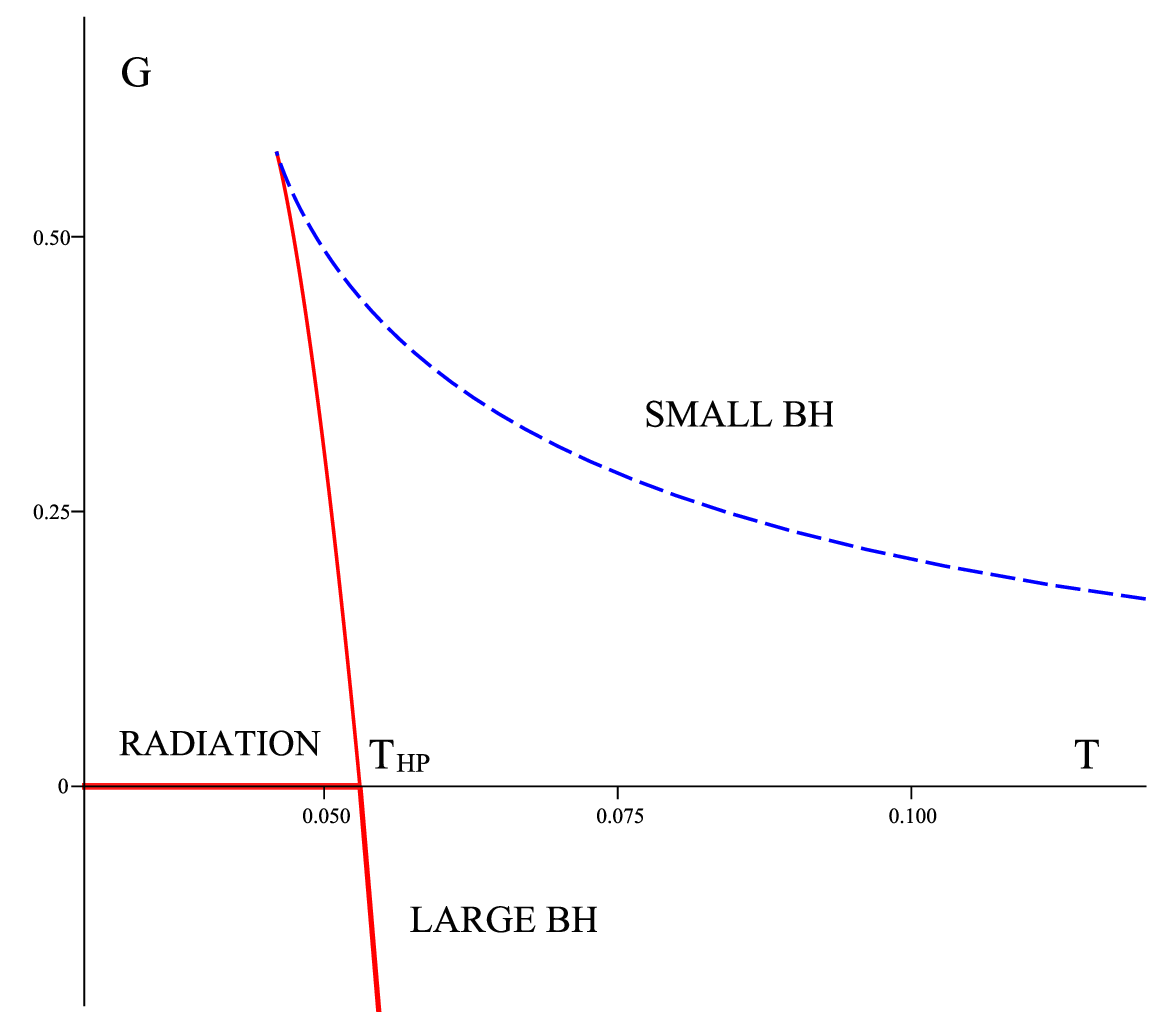}
&
\includegraphics[width=0.49\textwidth,height=0.3\textheight]{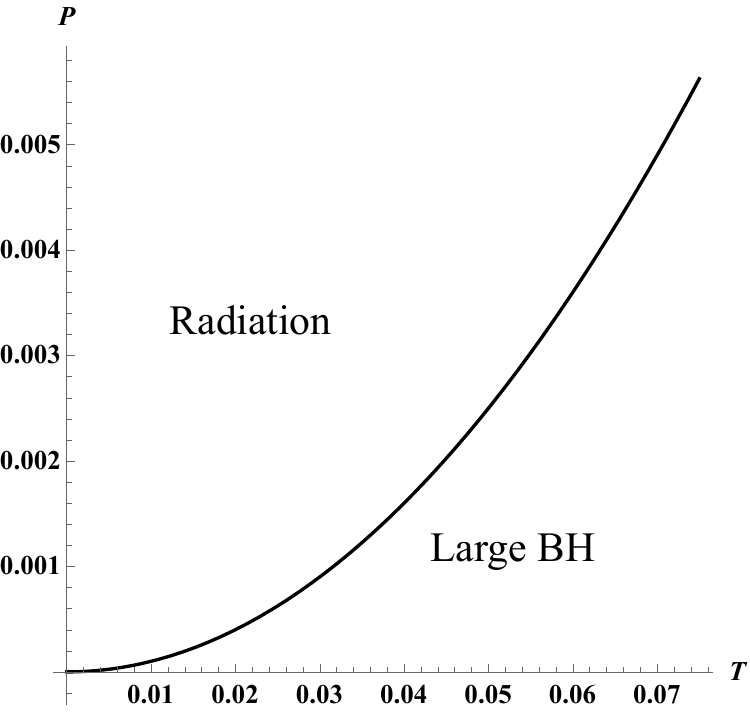}
\end{tabular}
\caption{{\bf Hawking--Page transition.} {\it Left.} The Gibbs free energy of a Schwarzschild-AdS black hole is displayed as a function of temperature for fixed pressure $P=1/(96\pi)$.
The upper branch of small black holes has negative specific heat and is thermodynamically unstable. For $T>T_{\mbox{\tiny  HP}}$ the lower branch of large black holes (with positive specific heat) has negative Gibbs free energy and corresponds
to the globally thermodynamically preferred state. At $T_{\mbox{\tiny  HP}}$ we observe a discontinuity in the first derivative of the radiation/black hole Gibbs free energy characteristic of the first order phase transition. {\it Right.} The $P-T$ phase diagram has a coexistence line of infinite length and is reminiscent of the solid/liquid phase portrait.
}
\label{Fig:HP}
\end{figure*}
Concentrating on the $k=1$ spherical case, we display the Gibbs free energy $G=M-TS$  in the left part of Fig.~\ref{Fig:HP}. We observe two branches of black holes that meet at a cusp. The upper branch displays `small' ($r_+<l/\sqrt{3}$) thermodynamically unstable black holes with negative specific heat, the lower branch corresponds to `large' black holes with positive specific heat. Large black holes with $r_+ > r_{\mbox{\tiny  HP}}=l$ have negative Gibbs free energy (which is lower than that of an AdS space filled with hot radiation) and represent the globally preferred state.
This means that at $T=T_{\mbox{\tiny  HP}}=1/(\pi l)^{-1}$ there is a first order {\it Hawking--Page} \cite{Hawking:1982dh} phase transition between thermal radiation and large black holes. As noted in Sec.~\ref{Sec:2}, this
can be interpreted as a confinement/deconfinement phase transition in the dual quark
gluon plasma \cite{Witten:1998zw}.

Considering the fluid interpretation in an extended phase space, the {\it coexistence line} of thermal radiation/large black hole phases, determined from $G=0$,
reads
\be\label{HPcoexistence}
P|_{\mbox{\tiny  coexistence}}=\frac{3\pi}{8} T^2\,.
\ee
One can easily verify (taking $S_r\approx 0$ and $V_r\approx 0$) that its slope satisfies the {\it Clausius--Clapeyron} equation
\be
\frac{dP}{dT}\Bigr|_{\mbox{\tiny  coexistence}}=\frac{\Delta S}{\Delta V}=\frac{S_{bh}-S_r}{V_{bh}-V_{r}}=\frac{S_{bh}}{V_{bh}}\,,
\ee
a result not previously noted in the literature. From the right side of Fig.~\ref{Fig:HP} we see that
the corresponding $P-T$ phase diagram for this black hole has no terminal point,
indicating that this phase transition is present for all pressures. It is reminiscent of a solid/liquid phase transition,  with the radiation phase playing role of a solid \cite{Kubiznak:2014zwa}.

By rewriting the temperature equation \eqref{S} whilst using the definition of pressure \eqref{eq:press}, we get a corresponding ``{\it fluid equation of state}''
\be\label{HPstate}
P=\frac{T}{v}-\frac{k}{2\pi v^2}\,,\quad v=2r_+l_P^2=2\Bigl(\frac{3V}{4\pi}\Bigr)^{1/3}=6\frac{V}{N}\,,
\ee
where we have, in this section only, explicitly restored the Planck length $l_P = \sqrt{\hbar G/c^3}$. The quantity $v$ plays the role of a
`{\it specific volume}' \cite{Kubiznak:2012wp, Altamirano:2014tva}, given by the thermodynamic volume $V$ divided by the `number of states' associated with the horizon, $N=A/l_P^2$. Note that for $k=0$ planar black holes we obtain the {\it ideal gas law}, $T=Pv$.

\subsection{Charged AdS black holes and Van der Waals fluids} \label{Sec:42}

Can we go beyond the ideal gas law and obtain a more realistic equation of state?  Consider adding charge to the black hole, which implies the metric function becomes
\be
f=1-\frac{2M}{r}+\frac{Q^2}{r}+\frac{r^2}{l^2}
\ee
for $k=1$. This is an exact solution to the Einstein--Maxwell-AdS equations, corresponding to a  charged-AdS black hole \eqref{HDRN}.
\begin{figure*}
\centering
\begin{tabular}{cc}
\includegraphics[width=0.49\textwidth,height=0.3\textheight]{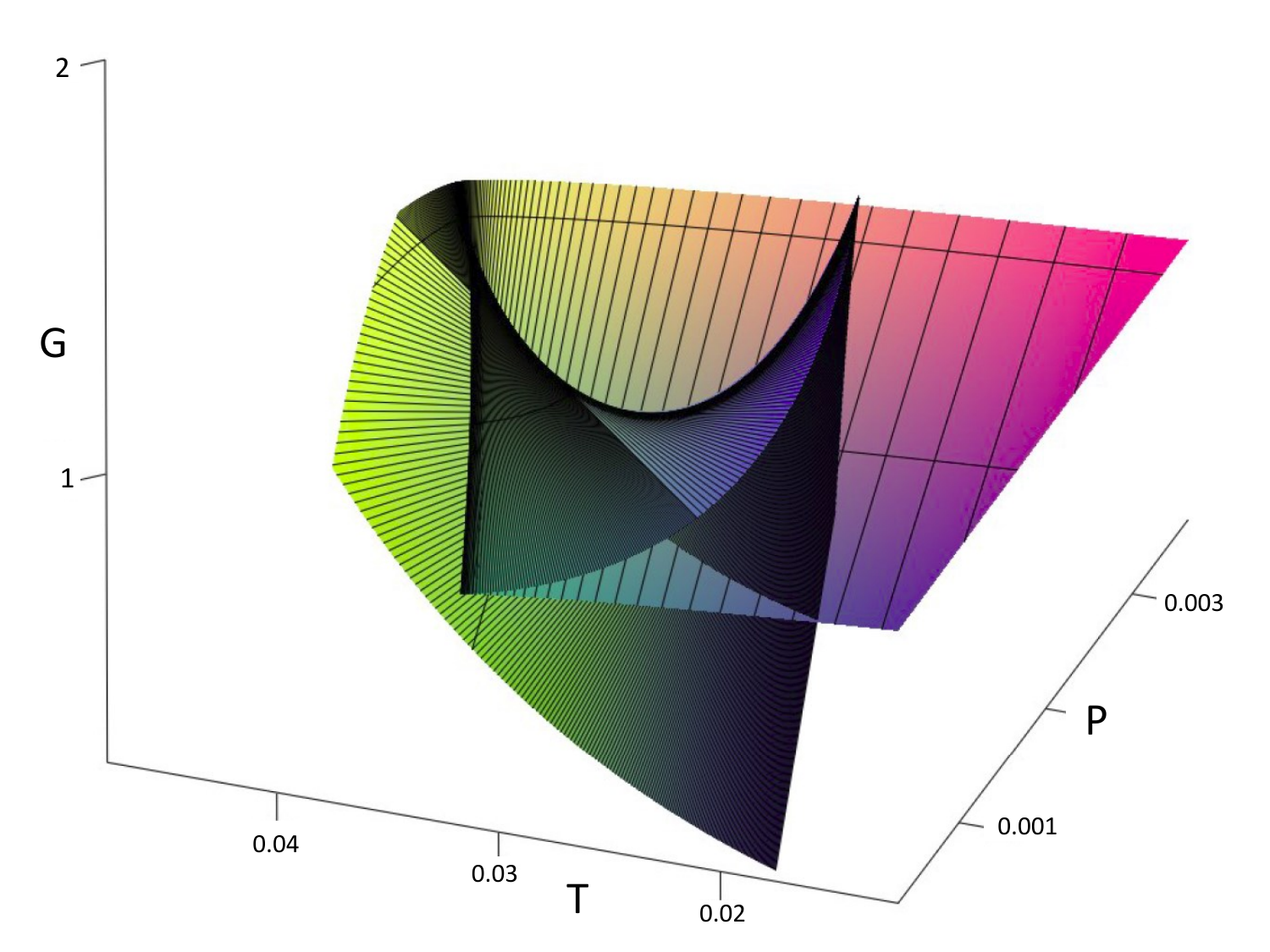} &
\includegraphics[width=0.49\textwidth,height=0.3\textheight]{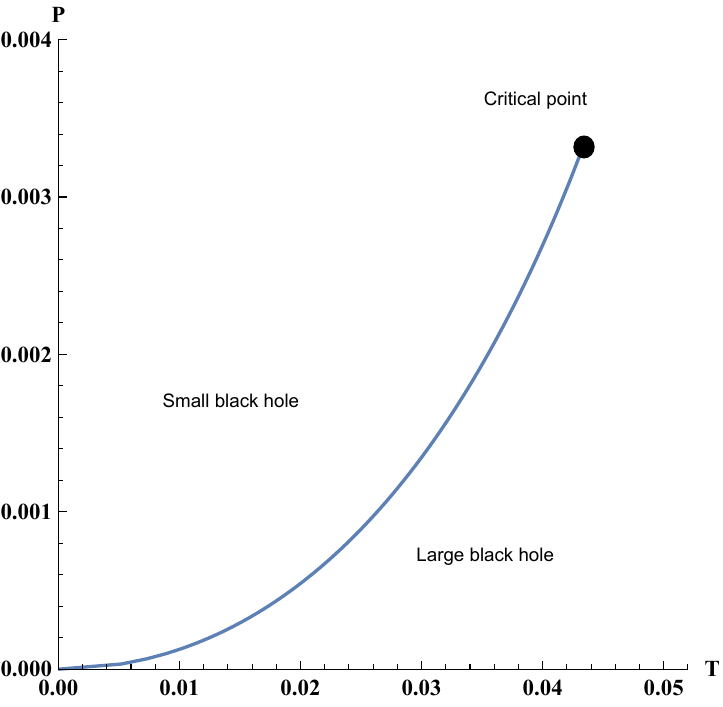}
\end{tabular}
\caption{{\bf Analogue of Van der Waals behavior.} {\it Left.} A characteristic swallowtail behavior of the Gibbs free energy of a charged-AdS black hole is displayed for fixed $Q=1$.
{\it Right.} The $P-T$ phase diagram shows SBH/LBH phase transition reminiscent of the liquid/gas phase transition. The coexistence line terminates at a critical point where the phase transition is of the second order.
}
\label{Fig:Swallow}
\end{figure*}

Charged AdS black holes allow for a first order {\it small-black-hole/large-black-hole} (SBH/LBH) phase transition,
in a canonical (fixed charge) ensemble \cite{Chamblin:1999tk,Chamblin:1999hg,Wu:2000id} (see also \cite{Cvetic:1999ne, Cvetic:1999rb}),  which is in many ways reminiscent of the liquid/gas transition of a non-ideal fluid described by the Van der Waals equation \eqref{VdWstate}.  In extended phase space this analogy
becomes even more complete since it allows proper identification between intensive and extensive variables \cite{Dolan:2011xt, Kubiznak:2012wp}.
From \eqref{TDchgBH} the thermodynamic quantities then read
\be
T=\frac{l^2(r_+^2-Q^2)+3r_+^4}{4\pi r_+^3 l^2}\,,\quad S=\pi r_+^2\,,\quad
V=\frac{4}{3}\pi r_+^3\,,\quad \Phi=\frac{Q}{r_+}\,,
\ee
giving rise to the following equation of state:
\be\label{RNstate}
P=\frac{T}{v}-\frac{1}{2\pi v^2}+\frac{2Q^2}{\pi v^4}\,,\quad v=2r_+=2\Bigl(\frac{3V}{4\pi}\Bigr)^{1/3}\,,
\ee
which qualitatively mimics the behaviour of the Van der Waals equation
\be\label{VdWstate}
\Bigl(P+\frac{a}{v^2}\Bigr)(v-b)=T\,,
\ee
where the parameter $a>0$ measures the attraction between particles and the parameter $b$ corresponds to the ``volume of fluid particles''.
The  corresponding black hole Gibbs free energy
\be
G=M-TS=\frac{l^2r_+^2-r_+^4+3Q^2l^2}{4l^2r_+}\,
\ee
demonstrates  swallowtail behavior, shown in the left part of Fig.~\ref{Fig:Swallow},  characteristic of first-order phase transitions.  The swallowtail  appears for pressures less than the critical value $P_c$, and terminates at a {\it critical point} at
\be
P_c=\frac{1}{96\pi Q^2}\,, \quad v_c=2\sqrt{6}Q\,, \quad T_c=\frac{\sqrt{6}}{18\pi Q}\,,
\ee
at which point the phase transition becomes  {\it second-order}.  The $P-T$ phase diagram displayed in the right part of Fig.~\ref{Fig:Swallow} illustrates  the {\it coexistence line} for such a first order phase transition and its terminal critical point.

Amusingly, the critical ratio, $P_c v_c/T_c = 3/8$ is exactly the same as for the Van der Waals fluid\footnote{The fact that
this ratio has no dependence on black hole charge $Q$ is unsurprising by
dimensional arguments. However, there is currently no obvious explanation as to
why both ratios are the same.  This is true, however, only in four dimensions---for charged-AdS black holes in higher dimensions the ratio becomes $P_c v_c/T_c = (2d-5)/(4d-8)$ \cite{Gunasekaran:2012dq}.}
and the critical point is characterized by standard
mean field theory exponents\footnote{The critical exponents characterize the behavior of various physical quantities in the vicinity of a critical point. Specifically,
denoting by $t=T/T_c-1$, the critical exponents for the black hole are defined as follows \cite{Kubiznak:2012wp}:
\begin{itemize}
\item
Exponent $\alpha$ governs the behaviour of the specific heat at constant volume,
\be
C_V=T \frac{\partial S}{\partial T}\Big|_{V}\propto |t|^{-\alpha}\,.
\ee
\item Exponent $\beta$ describes the behaviour of the {\it order parameter} $\mathfrak{M}=V_l-V_s$, a  difference between the volume of a large black hole $V_l$ and the volume of a small black hole $V_s$  on the given isotherm
\be
\mathfrak{M} =V_l-V_s\propto |t|^\beta\,.
\ee
[Alternatively, one could define the order parameter as the difference between the specific volumes, $\mathfrak{m}=v_l-v_s$.]
\item
Exponent $\gamma$ determines the behaviour of the {\it isothermal compressibility} $\kappa_T$
\be
\kappa_T=-\frac{1}{V}\frac{\partial V}{\partial P}\Big|_T\propto |t|^{-\gamma}\,,
\ee
 c.f. adiabatic compressibility $\kappa_S$, \eqref{kappaS}, discussed in Sec.~\ref{Sec:3}.
\item Exponent $\delta$ governs the following behaviour on the critical isotherm $T=T_c$:
\be
|P-P_c|\propto |V-V_c|^\delta\,.
\ee
\end{itemize}
} \cite{Kubiznak:2012wp}:
\begin{equation}\label{eqn:meanfield}
\alpha = 0\,, \quad \beta = \frac{1}{2}\,, \quad \gamma=1\,, \quad \delta=3\,.
\end{equation}
An example of black holes with a critical point characterized by different critical exponents is given in Sec.~\ref{Sec:45}.

\begin{figure*}
\centering
\begin{tabular}{cc}
\includegraphics[width=0.49\textwidth,height=0.3\textheight]{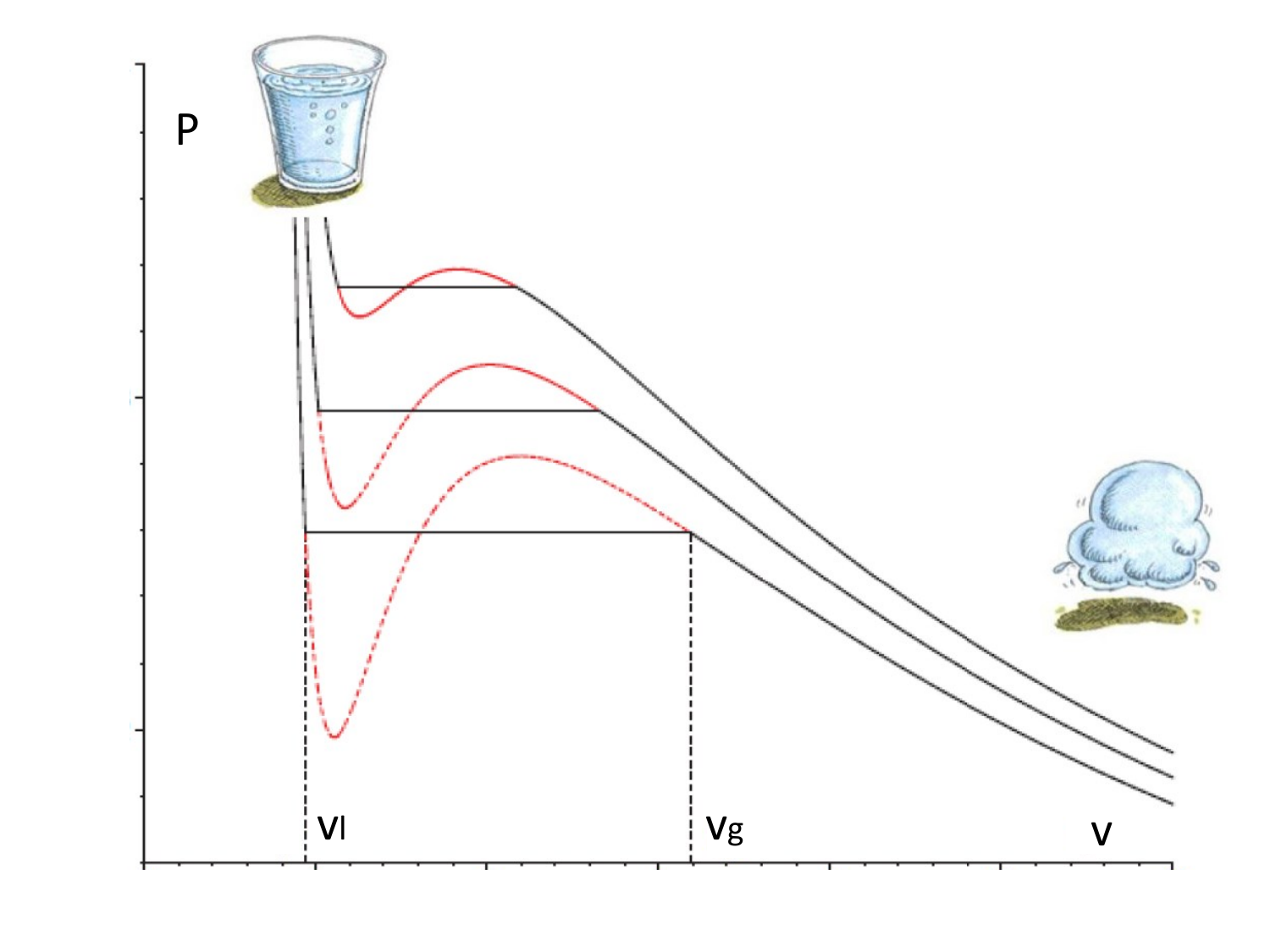} &
\includegraphics[width=0.49\textwidth,height=0.3\textheight]{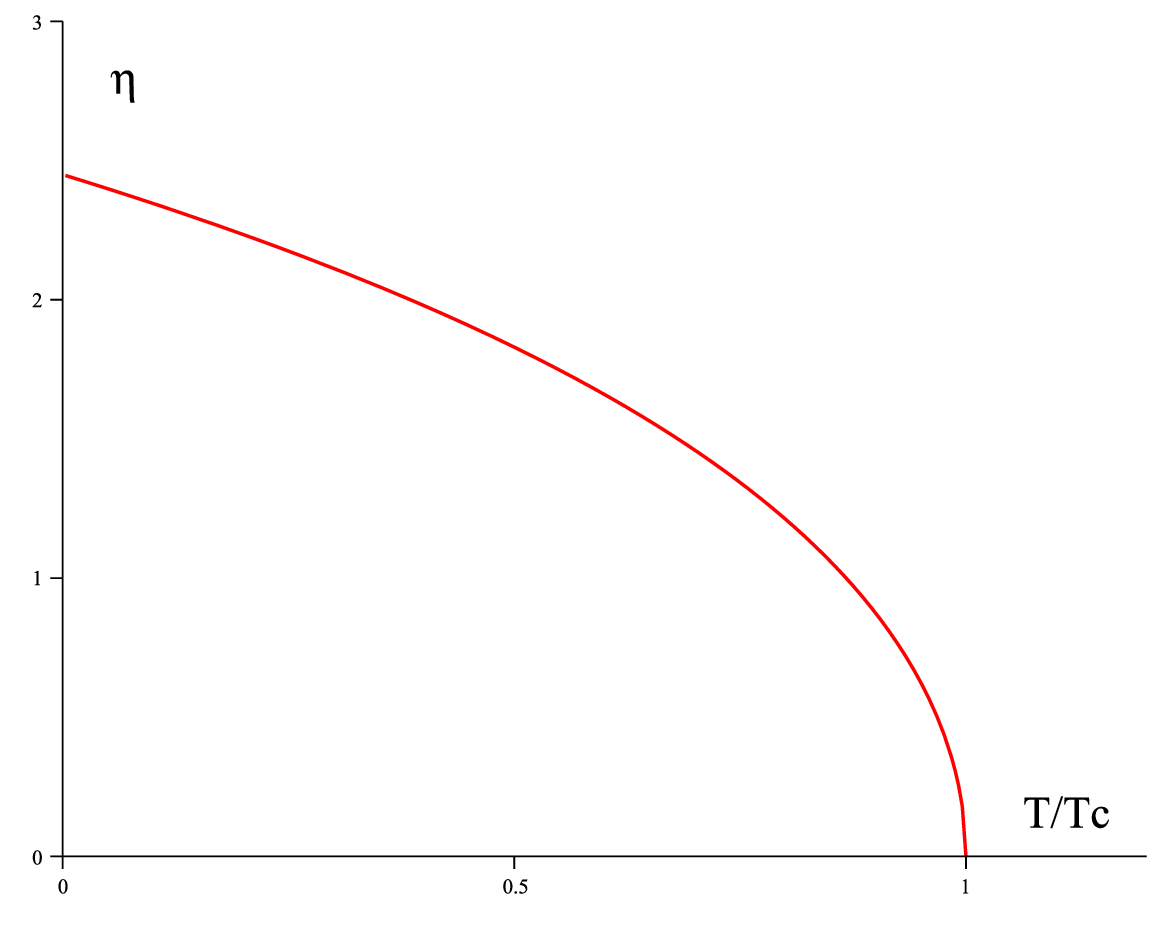}
\end{tabular}
\caption{{\bf Further analogies with Van der Waals fluid.} {\it Left.} The figure
schematically displays Maxwell's equal area law describing the liquid/gas phase transition of
the Van der Waals fluid: the `oscillating' (dashed) part of the isotherm $T<T_c$ is replaced by an isobar, such that the areas above and below the isobar are equal one another. Similar law holds for the SBH/LBH phase transition of the charged AdS black hole in the $P-V$ diagram with the specific heat of the fluid $v$ replaced by the thermodynamic volume $V$ of the black hole.
{\it Right.} The ratio of micromolecular densities, $\eta=(n_s-n_l)/n_c$, is displayed as a function of $T/T_c$.
}
\label{Fig:n}
\end{figure*}

The coexistence line is governed by the behavior of the Gibbs free energy (the bottom line of the swallowtail). Alternatively, it can be obtained \cite{Spallucci:2013osa} by imposing   {\it Maxwell's equal area law}, which states that the two phases coexist when the areas above and below a line of constant pressure drawn through a $P-V$ curve are equal\footnote{As correctly pointed out in \cite{Lan:2015bia, Xu:2015hba} Maxwell's equal area law is only qualitatively but not quantitatively right when imposed in the $P-v$ plane. This is  due to the fact that $v\propto V/N$ where $N$ is no longer a constant but $N=N(r_+)$.
}, as shown in the left part of Fig.~\ref{Fig:n}.\footnote{ The equal area law for the holographic entanglement entropy was studied in
\cite{Nguyen:2015wfa}.}

For various black holes the approximate coexistence line has been constructed numerically \cite{Wei:2014qwa, Wei:2015ana, Cheng:2016bpx, Wei:2015iwa}. For the particular case of the 4d charged-AdS black hole that we are considering an analytic formula exists \cite{Mo:2016sel}. As we cross the coexistence line, for a given $Q$ and $3/[8\pi l^2]=P\in(0,P_c)$, the size of the black hole `jumps' from a small radius, $r_s$, to a large one, $r_l$, given by
\be
r_s=\frac{2Ql}{\sqrt{l^2-2Ql}+\sqrt{l^2-6Ql}}\,,\quad r_l=\frac{1}{2}\Bigl(\sqrt{l^2+2Ql}+\sqrt{l^2-6Ql}\Bigr)\,,
\ee
both determining the same temperature $T(r_s)=T(r_l)$ and the same Gibbs free energy $G(r_s)=G(r_l)$. So we arrive at the following analytic formula for the coexistence line:
\be
T|_{\mbox{\tiny  coexistence}}=\frac{r_s^2-Q^2}{4\pi r_s^3}+\frac{3r_s}{4\pi l^2}\in(0,T_c)\,,
\ee
correcting the result obtained previously \cite{Mo:2016sel}.
Equipped with this expression, one can easily verify the {\it Clausius--Clapeyron} equation,
\be\label{CCeqn}
\frac{dP}{dT}\Bigr|_{\mbox{\tiny  coexistence}}=\frac{\Delta S}{\Delta V}=\frac{S_l-S_s}{V_l-V_s}\,,
\ee
governing the slope of the coexistence curve. Equation \eqref{CCeqn} has been previously verified using an approximate coexistence formula \cite{Wei:2014qwa}.

Of course, the SBH/LBH first order phase transition requires non-trivial latent heat, given by
$\Delta Q=T\Delta S$, which vanishes at the critical point where the phase transition is  second order.
Similar to the Clausius--Clapeyron equation \cite{Zhao:2014fea}, one can verify the validity of {\it Ehrenfest's equations} at the critical point
 \cite{Mo:2013ela, Mo:2014wca, Mo:2014mba}.

 To further push the analogy with fluids and to investigate the possible ``microscopic structure'' of a charged AdS black hole, a toy model introducing the concept of ``{\it black hole molecular density}''
\be
n=\frac{1}{v}=\frac{1}{2l_p^2 r_+}
\ee
was developed  \cite{Wei:2015iwa}.
 As we cross the coexistence line, the number densities of large and small black holes jump.  This is illustrated in Fig.~\ref{Fig:n}, where the molecular density ratio $\eta=(n_s-n_l)/n_c$ is plotted as a function of $T/T_c$. This picture is reminiscent of the magnetization/temperature diagram of an Ising ferromagnet. However in Fig.~~\ref{Fig:n}  different points on the curve correspond to different pressures, determined for a given temperature from the coexistence curve.
We refer the reader to \cite{Wei:2015iwa, Wei:2016hkm, Dutta:2016urd, Zangeneh:2016snh, Mo:2016sel} for further developments on this model.

Another approach involves consideration of the quasinormal modes of a scalar field around a charged AdS black hole
\cite{Chan:1996yk,Chan:1999sc}. Different phases of the black hole can be identified from the behavior of the quasinormal modes \cite{Liu:2014gvf, Chabab:2016cem}.

We stress that although qualitatively similar, the black hole equation of state \eqref{RNstate} is not exactly that of   Van der Waals
\eqref{VdWstate}. A similar situation occurs for black hole solutions in the presence of rotation, higher dimensions, or higher curvature corrections. Asymptotically AdS black holes whose thermodynamics {\it match exactly} that of the Van der Waals fluid were constructed in \cite{Rajagopal:2014ewa, Delsate:2014zma} in the context of Einstein gravity. Surprisingly, the solution is supported by `exotic matter' that does not obey any of the energy conditions everywhere outside of the horizon.  Extensions to a polytropic
 black hole found that the energy-momentum tensor obeyed the three energy conditions \cite{Setare:2015xaa}.

Another interesting result was found for certain cases of (3+1)-dimensional STU black holes \cite{Caceres:2015vsa}. The STU black hole solution has  up to four $U(1)$ charges. An interesting  phase structure in the fixed charge ensemble interpolating between the Hawking--Page transition of the Schwarzschild-AdS solution and the Van der Waals transition of the charged-AdS case was discovered.  The latter behaviour occurs when three or four of these charges are nonzero, with the usual charged AdS black hole recovered when all four charges are equal.   With only one charge switched on we observe a Hawking--Page-like transition.
Two nonzero charges yield a situation intermediate between these two \cite{Caceres:2015vsa}.

More interesting phase behavior can take place in more complicated black hole spacetimes that generally require   dimensions $d>4$.  We consider these cases next.

\subsection{Reentrant phase transitions} \label{Sec:43}

A system undergoes a {\it reentrant phase transition (RPT)} if a monotonic variation of any thermodynamic quantity results in two (or more) phase transitions such that the final state is macroscopically similar to the initial state.
\begin{figure}
\begin{center}
\rotatebox{0}{
\includegraphics[width=0.55\textwidth,height=0.35\textheight]{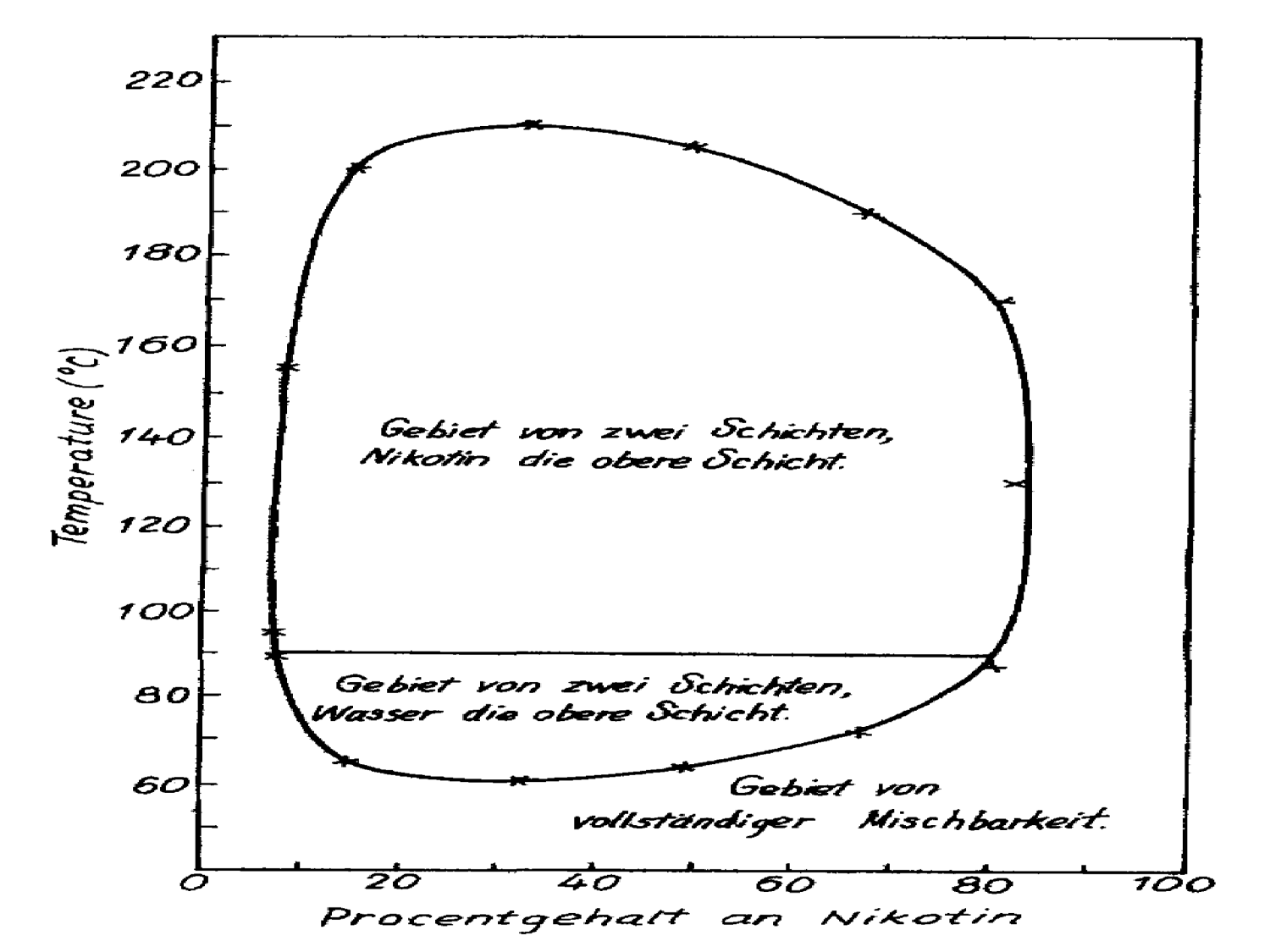}
}
\caption{{\bf Reentrant phase transition in nicotine/water mixture.}
The diagram displays possible phases of the mixture dependent on the temperature and  percentage of the nicotine. 
`Outside of the bubble' the mixture is in a homogeneous state. Inside, two layers of nicotine and water exist separately: in the upper half the nicotine layer is above the water layer while the layers swap in the bottom half of the bubble. Fixing the percentage of nicotine at, for example, $40\%$  and increasing the temperature from low to high, we observe the following phases: homogeneous mixture (low temperatures), water above nicotine (bottom half of the bubble) nicotine above water (upper half of the bubble), homogeneous mixture (high temperatures). Since the initial and final states are macroscopically similar, this is an example of an RPT.  Reproduced  from ref. \cite{Hudson:1904} (
C. Hudson, {\it Die gegenseitige lslichkeit von nikotin in wasser} Z. Phys. Chem. {\bf 47} (1904) 113)
with permission from De Gruyter.
}
\label{Nicotine}
\end{center}
\end{figure}
An RPT was first observed by Hudson in 1904 in a nicotine/water mixture \cite{Hudson:1904}. As the temperature of the mixture for a sufficient fixed percentage of nicotine increases,  the homogeneous mixed state separates into distinct
nicotine/water phases, as illustrated in the phase diagram in Fig.~\ref{Nicotine}. For sufficiently high temperatures the homogeneous state reappears.  Since their discovery, reentrant phase transitions have been commonly observed in multicomponent fluid systems, gels, ferroelectrics, liquid crystals, and binary gases, where the reentrant behavior often emerges as a consequence of two (or more) `competing driving mechanisms'. It can also take place in non-commutative spacetimes \cite{Panella:2016nkr}. We refer the interested reader to a topical review \cite{narayanan1994reentrant} for more details.
 \begin{figure*}
\centering
\begin{tabular}{cc}
\includegraphics[width=0.49\textwidth,height=0.3\textheight]{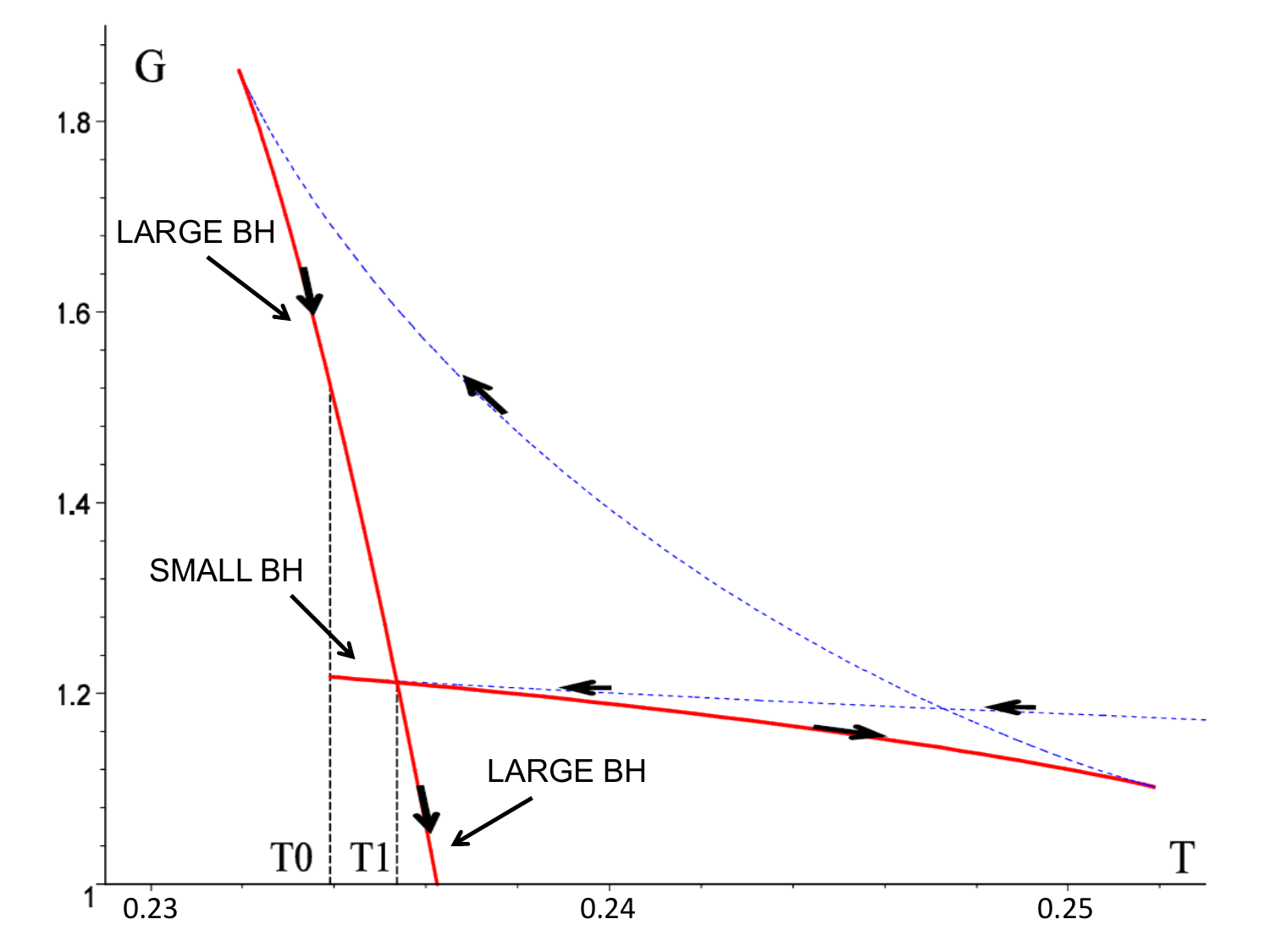} &
\includegraphics[width=0.49\textwidth,height=0.3\textheight]{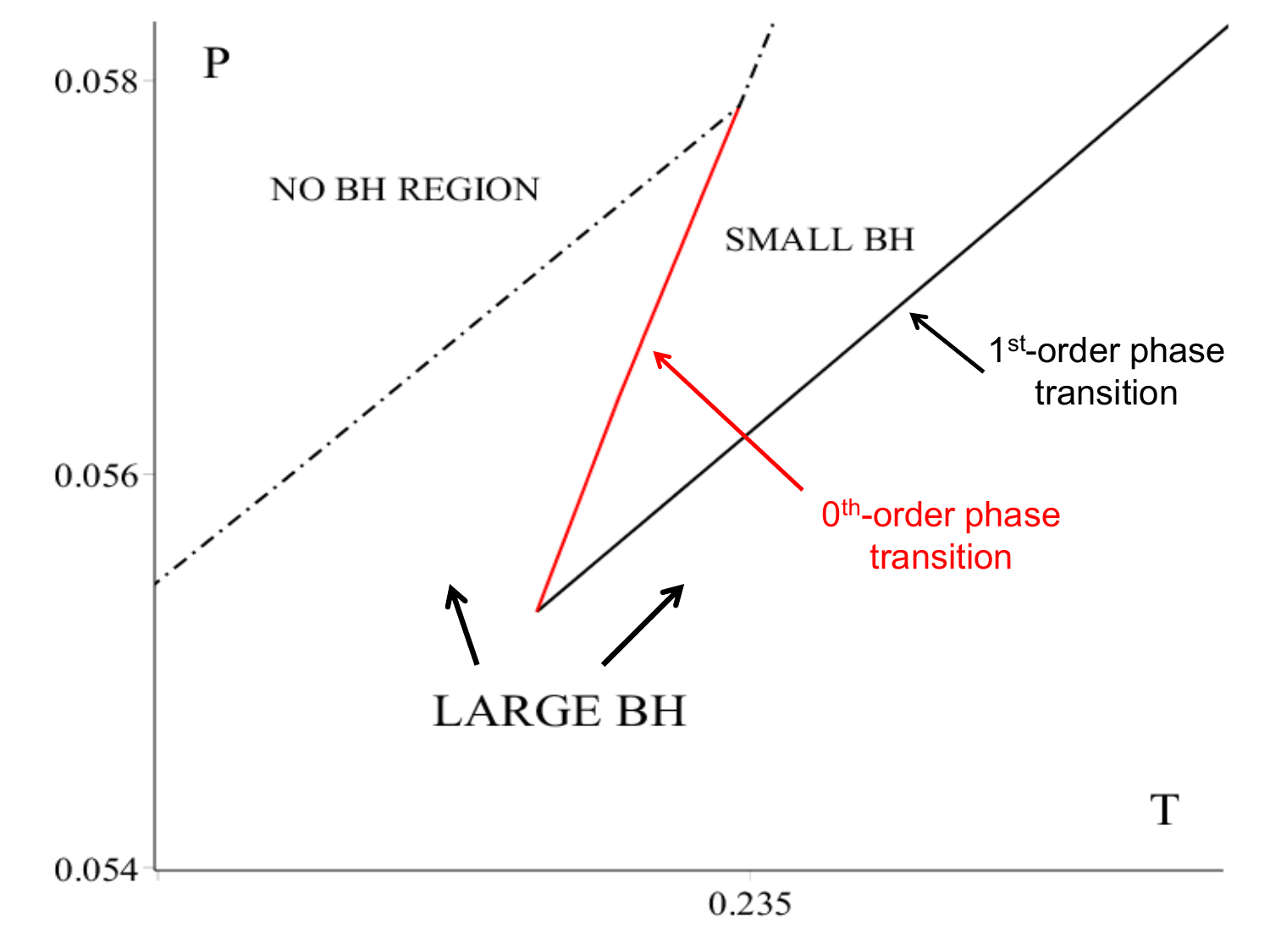}
\end{tabular}
\caption{{\bf Reentrant phase transition: singly spinning Kerr-AdS black hole in $d=6$.} {\it Left.}  The figure illustrates the typical behaviour of $G$ when the reentrant phase transition is present, $P\in(0.0553,0.0579)$.
Black arrows indicate increasing $r_+$. If we start decreasing the temperature from, say $T=0.24$, the system follows the lower vertical solid
red curve of large stable black holes until it joins the upper horizontal solid red curve of small stable black holes---this corresponds to a first order LBH/SBH phase transition at $T=T_1$. As $T$ continues to decrease the system follows this upper curve until
$T=T_0$, where $G$ has a discontinuity at its global minimum. Further decreasing $T$, the system jumps to the
uppermost vertical red line of large stable black hole---this corresponds to the zeroth order phase transition between small and large black holes.
In other words, as $T$ continuously decreases, we observe LBH/SBH/LBH  reentrant phase transition.
{\it Right.} The corresponding $P-T$ diagram clearly illustrates 3 possible phases: a region where there are no black holes, an LBH region and an SBH region, the last two being separated by the coexistence lines of 1st-order (black curve) and 0th-order (red curve) phase transitions. The 1st-order coexistence line eventually terminates at a critical point (not displayed).}
\label{Fig:RPT}
\end{figure*}

The first example of a reentrant phase transition for black holes was discovered in \cite{Gunasekaran:2012dq} in the context of four-dimensional black holes in Born--Infeld theory\footnote{ Interestingly, no such phenomena seem to exist for Born--Infeld-AdS black holes described by \eqref{BInfeld} (with $\tilde F=0$) in higher dimensions \cite{Zou:2013owa};  however see the corresponding footnote in Sec.~\ref{Sec:2}.} and later  in the simpler setting of vacuum black holes in higher dimensions \cite{Altamirano:2013ane}, further studied in \cite{Altamirano:2013uqa, Altamirano:2014tva, Wei:2015ana}.  This phenomenon has since been seen in higher curvature settings
 \cite{Frassino:2014pha,Wei:2014hba,Hennigar:2015esa, Sherkatghanad:2014hda} and for higher-curvature hairy black holes\footnote{These latter objects exhibit a number of interesting features -- for example their entropy becomes negative for certain choices of the parameters  \cite{Hennigar:2015wxa}.  }
 \cite{Hennigar:2015wxa}.

Specifically, for a 6-dimensional singly-spinning Kerr-AdS black hole (see App.~\ref{AppB} for details),
for a certain range of pressures (and a given angular momentum)
a monotonic lowering of the temperature yields a LBH/SBH/LBH reentrant phase transition, illustrated in Fig.~\ref{Fig:RPT}
 \cite{Altamirano:2013ane}.
This effect is accompanied by a discontinuity in the global minimum of the Gibbs free energy, referred
to as a zeroth-order phase transition \cite{Gunasekaran:2012dq}, a phenomenon seen for example in superfluidity and superconductivity \cite{maslov2004zeroth}.
We are thus led to the following analogy:
\be
\begin{array}{|c|c|c|}
\hline
{\mbox{Low $T$}} & \mbox{Medium $T$} & \mbox{High $T$} \\
\hline
{\mbox{mixed}} & \mbox{water/nicotine} & \mbox{mixed} \\
\hline
{\mbox{large BH}} & \mbox{small BH} & \mbox{large BH} \\
\hline
\end{array}
\ee

\begin{figure}
\begin{center}
\rotatebox{0}{
\includegraphics[width=0.6\textwidth,height=0.3\textheight]{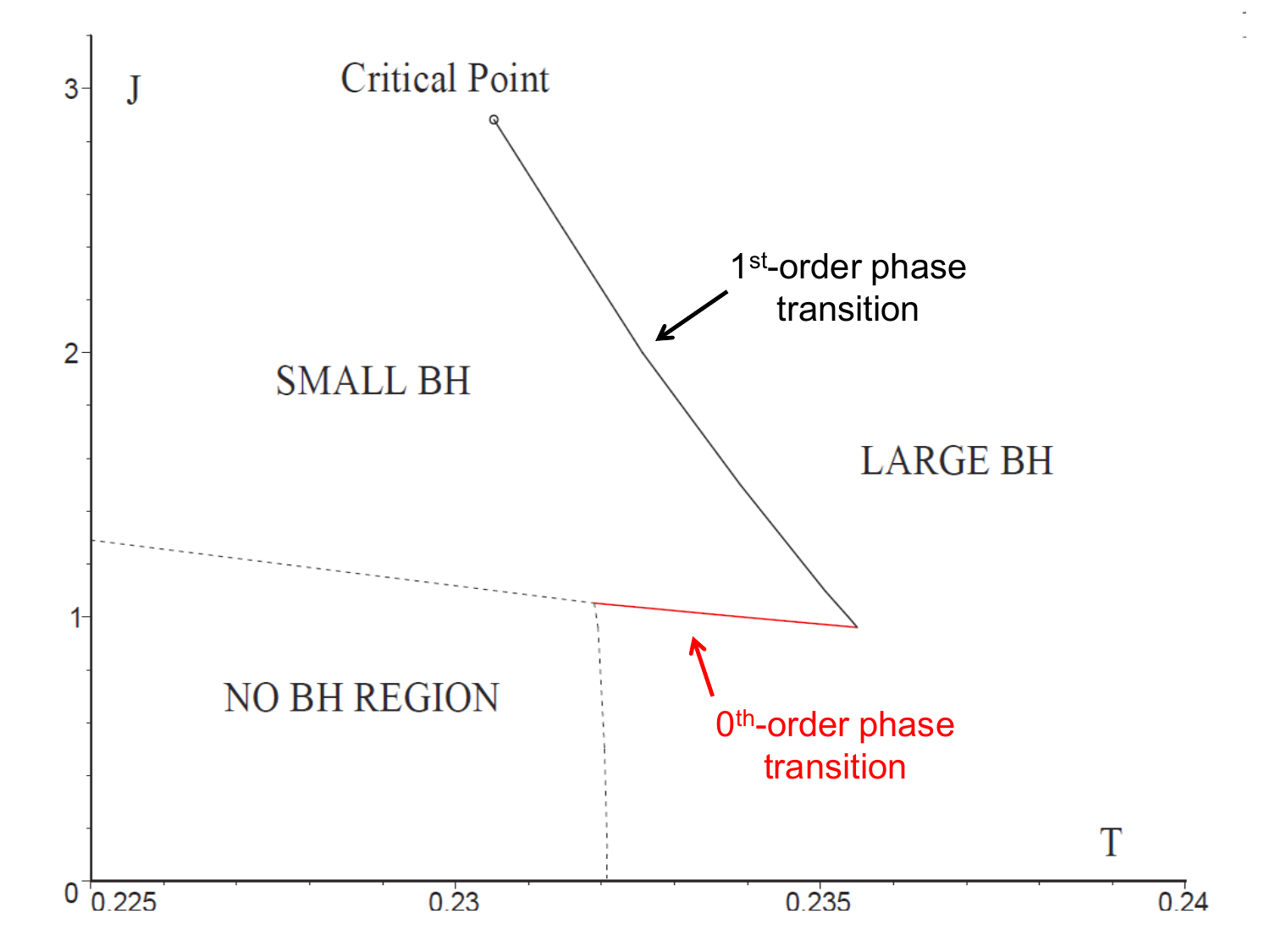}
}
\caption{{\bf Reentrant phase transition in $J-T$ plane.}
The diagram is displayed for a single spinning Kerr-AdS black hole in 6 dimensions for a fixed pressure $l=2.656$.
}
\label{Fig:RPT2}
\end{center}
\end{figure}

We conclude this subsection with three remarks:
i) An RPT does not require variable $\Lambda$. The phenomenon will be present for (properly chosen) fixed cosmological constant and can be, for example, studied in the $J-T$ plane, as shown in Fig.~\ref{Fig:RPT2}.
ii) For RPT phase transitions one often needs at least two competing phenomena: one driving the phase change and the other returning the system back to its original state. In the case of rotating black holes in 6 dimensions it is plausible \cite{Altamirano:2014tva} that behind the observed RPT there is a competition between the blackbrane behavior of small almost ultraspinning black holes \cite{Emparan:2003sy} and the completely different behavior of slowly rotating large Schwarzschild-like black holes. If this intuition is correct, it also explains why such phenomena have not been observed in rotating black hole spacetimes of dimensions   $d< 6$ where   ultraspinning black holes are absent.
iii) In higher curvature gravities it is possible to observe {\it multiple RPTs}, and/or RPTs where the zeroth-order phase transition no longer plays a central role (the  RPT is achieved by a succession of two first order phase transitions) \cite{Frassino:2014pha, Hennigar:2015esa, Sherkatghanad:2014hda}.

\subsection{Analogue of solid/liquid/gas phase transition: triple points} \label{Sec:44}
\begin{figure*}
\centering
\begin{tabular}{cc}
\includegraphics[width=0.49\textwidth,height=0.3\textheight]{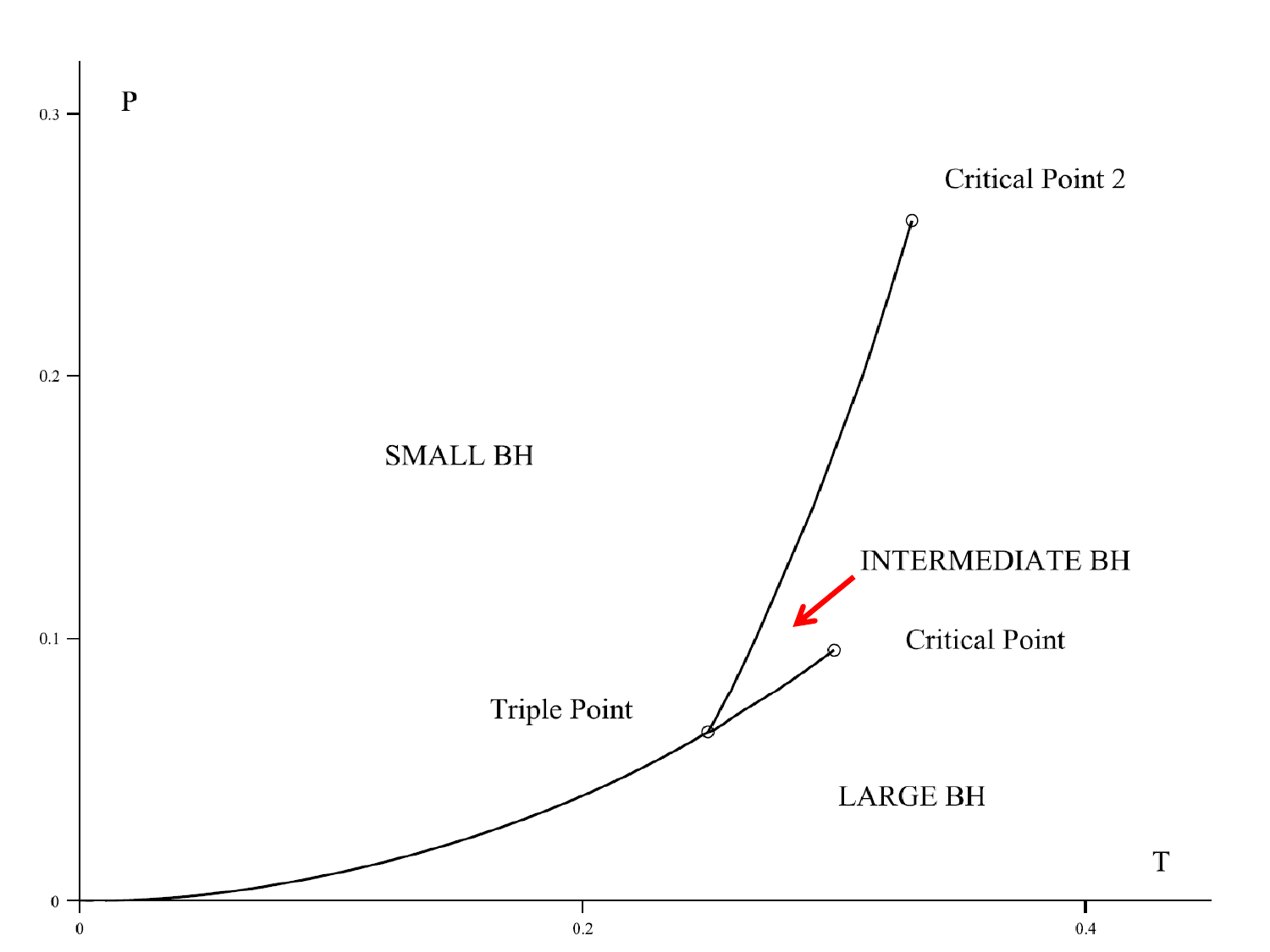} &
\includegraphics[width=0.49\textwidth,height=0.3\textheight]{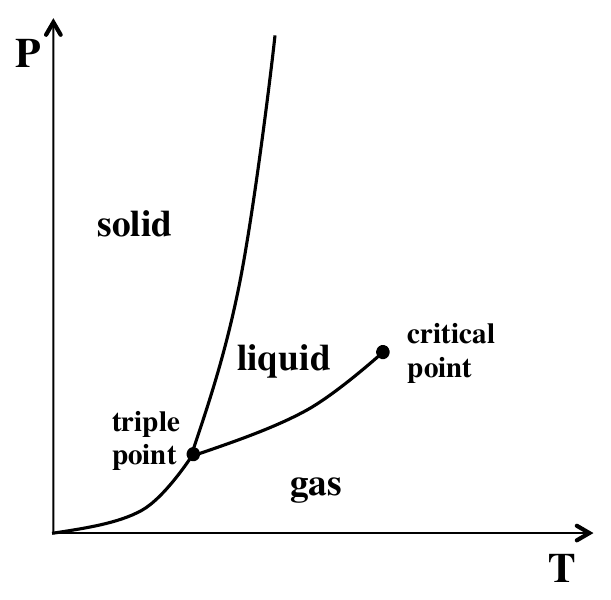}
\end{tabular}
\caption{{\bf Triple point.} The {\it left} figure displays a $P-T$ phase diagram for a doubly-spinning Kerr-AdS black hole in $d=6$ and fixed angular momenta ratio $J_2/J_1=0.05$.  The portrait is in many ways analogous to the solid/liquid/gas phase diagram, displayed in the {\it right} figure, including the existence of a triple point where three coexistence lines merge together. Note however, that in the black hole case there is an additional critical point: the small black hole/intermediate black hole coexistence line is no longer semi-infinite (as in the solid/liquid case) and terminates, similar to the ``liquid/gas'' coexistence line, in a critical point (denoted here as the `Critical point 2').
}
\label{Fig:Triple}
\end{figure*}
One can also obtain a gravitational analogue of a solid/liquid/gas phase transition and that of a {\it triple} (tricritical) point
\cite{Altamirano:2013uqa}. For example, a doubly spinning Kerr-AdS black hole in $d=6$ with a properly chosen ratio of angular momenta, exhibits this
phenomenon, with the associated   phase diagram displayed in left Fig.~\ref{Fig:Triple}. Three `phases' of black holes are evident: small, intermediate, and large, all meeting at a triple point. The main difference from the solid/liquid/gas phase transition is the absence of a semi-infinite coexistence line between the solid and liquid phases, which is now replaced by a finite coexistence line that separates small and intermediate black holes and terminates at a second critical point\footnote{Of course, the existence of an additional critical point makes the two diagrams displayed in Fig.~\ref{Fig:Triple} fundamentally different. A semi-infinite coexistence line clearly separates two phases: to obtain liquid from solid we actually have to melt the crystal. There is no possibility of ``going around a critical point'' as in the liquid/gas case where one can by choosing a `clever path' avoid undergoing a first order phase transition.}.
Similar behavior has also recently been observed for black holes in higher curvature theories of gravity \cite{Wei:2014hba, Frassino:2014pha, Hennigar:2015esa}.

\subsection{Beyond mean field theory} \label{Sec:45}

\begin{figure*}
\centering
\begin{tabular}{cc}
\includegraphics[width=0.49\textwidth,height=0.3\textheight]{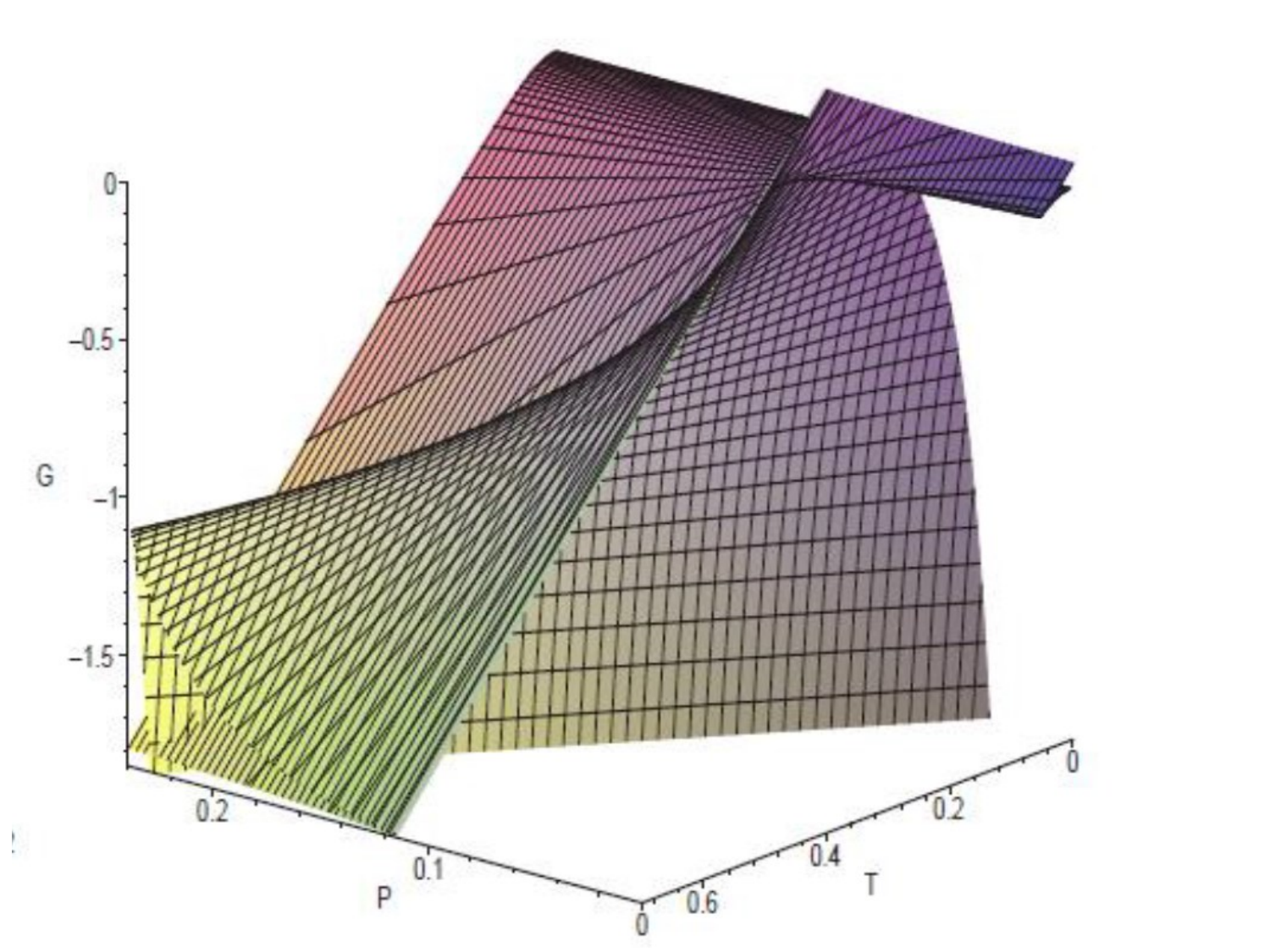} &
\includegraphics[width=0.49\textwidth,height=0.3\textheight]{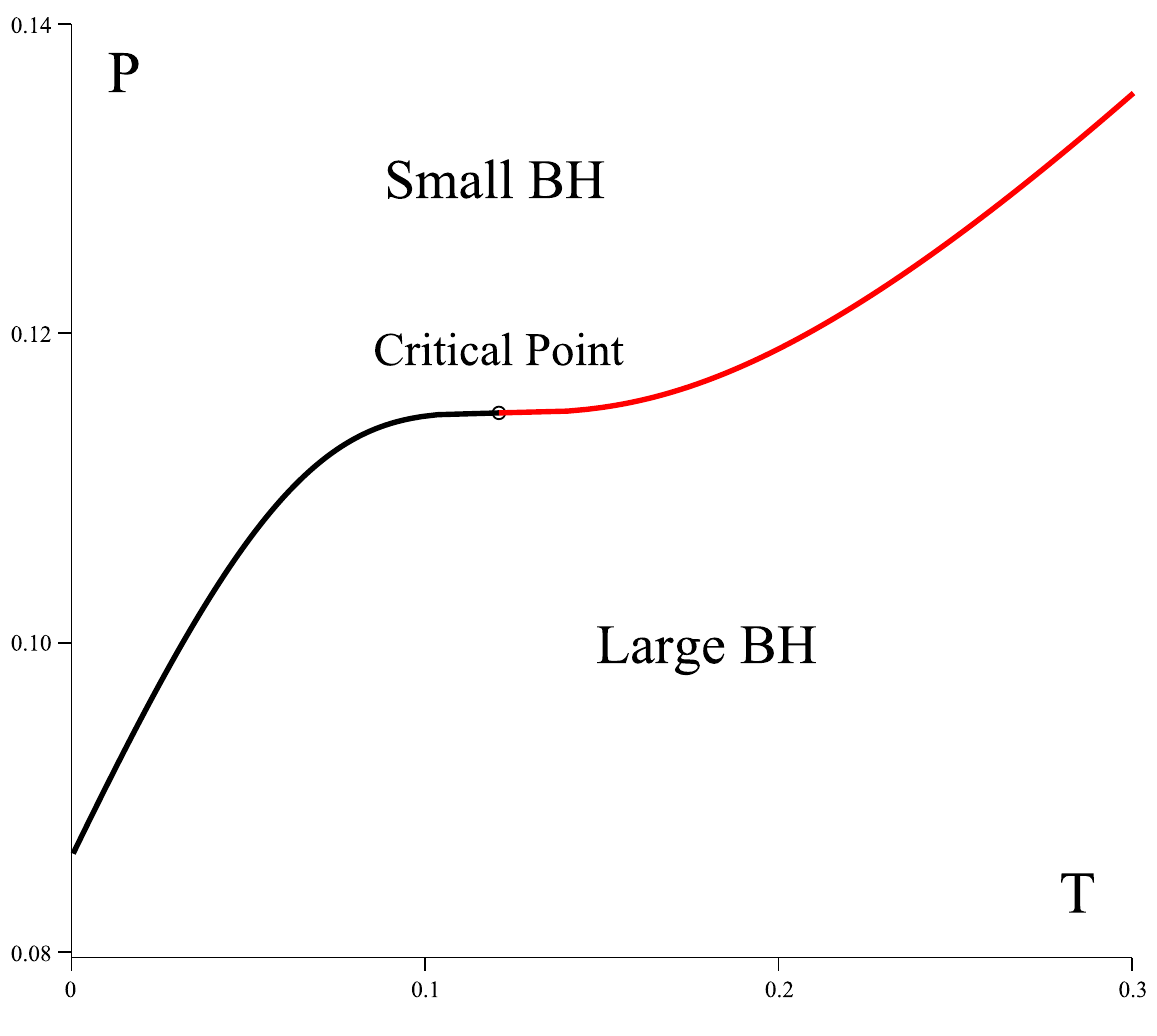}
\end{tabular}
\caption{{\bf Isolated critical point.} {\it Left.} The Gibbs free energy
exercises two swallowtails both emanating from the same isolated critical point.
{\it Right.} The corresponding $P-T$ diagram displays two phases of black holes: large and small, separated by two
first order phase transitions (with coexistence lines denoted by black and red curves) that both emerge from a single
isolated critical point where the phase transition becomes second order and is characterized by non-standard critical exponents.
In both figures we have set $K = 3$ and  $d = 7$; higher $d$ and higher (odd) $K$ have similar qualitative behavior.
}
\label{Fig:Isolated}
\end{figure*}
We have seen that in the context of black hole chemistry, phase diagrams of various black hole spacetimes admit critical points. These often terminate a coexistence line describing a first order phase transition between black holes of various sizes and are characterized by standard mean field theory critical exponents.

It is natural  to ask if this is generic. Does every black hole that can be obtained by a geometric theory of gravity have all its critical points characterized by the same critical exponents? The answer to this question seemed to be `yes' until a new kind of {\it isolated critical point} was discovered \cite{Frassino:2014pha,Dolan:2014vba}.

Lovelock gravity was already known to  have multiple critical points  \cite{Xu:2014tja}: the Gibbs free energy could have
more than one swallowtail.  Unusual behaviour  for AdS topological black holes ($k=-1$ in \eqref{Schw-Ads}) was
subsequently observed: such black holes have small/large coexistence phases both below {\it and above} the critical point
\cite{Mo:2014qsa}, though this  behavior is contingent on the ensemble chosen \cite{Zou:2014mha}.

Isolated critical points were first seen in a very special setting of   higher-curvature Lovelock gravity (see Sec.~\ref{Sec:2}) and satisfied several conditions.  i) The Lovelock couplings must  be fine-tuned,  given by Eq. \eqref{SpecialCouplings}. ii) The order $K\geq 3$ of the Lovelock gravity has to be odd.  iii) The geometry of the horizon must be  hyperbolic $(k=-1$).

 Under these conditions, the Gibbs free energy develops two swallowtails whose tips coincide. This results in an interesting phase diagram  in which two coexistence lines of first order phase transitions  meet in a single point where the phase transition is of
second-order, as depicted in Fig.~\ref{Fig:Isolated}.   The associated critical exponents now read \cite{Dolan:2014vba}
 \begin{equation}
\alpha = 0\,, \quad \beta = 1\,, \quad \gamma=K-1\,, \quad \delta=K\,,
\end{equation}
and are obviously different from the standard exponents \eqref{eqn:meanfield}. Interestingly, these exponents still satisfy the Widom scaling relation and the Rushbrooke inequality
\be
\gamma=\beta(\delta-1)\,,\quad \alpha+2\beta+\gamma\geq 2\,,
\ee
(both derivable from general thermodynamic considerations) but no longer, as per usual, saturate the latter inequality.
Black hole solutions in  quasi-topological gravity \cite{Oliva:2010eb,Myers:2010ru}
can also exhibit isolated critical points \cite{Hennigar:2015esa}, with the conditions and results holding for this class of black holes as well.

The Prigogine--Defay ratio \cite{prigogine1974chem}  describes the ratio of the jumps of the isobaric heat capacity $\Delta C_P$, isothermal compressibility $\Delta \kappa_T$, and  isobaric thermal expansion coefficient $\Delta \alpha_P$,  and reads:
\be\label{PiK}
\Pi=\frac{1}{VT}\Bigl(\frac{\Delta C_P\Delta \kappa_T}{(\Delta \alpha_P)^2}\Bigr)_{T}=\frac{1}{K}\,,
\ee
indicating that the phase transition has more than one order parameter and is perhaps
a glass phase transition \cite{gupta1976prigogine,gundermann2011predicting}.

One curiosity about this class of black holes is that they have zero mass.  Asymptotically AdS zero-mass black holes have been known to exist for some time \cite{Mann:1997jb}, and it is quite intriguing that they can exhibit such unusual phase behaviour under the right circumstances.
The origin and physical meaning of this peculiar critical point remains to be understood.

\subsection{Holographic heat engines} \label{Sec:46}

With black hole chemistry where $P$ and $V$ are thermodynamic variables, one can also explore how a black hole heat engine cycle (closed loop on the $P-V$ plane) can be precisely realized \cite{Johnson:2014yja}. For example, we can consider black holes undergoing the {\it Carnot cycle} and calculate  the corresponding
efficiency.

The efficiency of the Carnot cycle does not depend on the equation of state and hence should be the same for all black hole systems. It is given by
\begin{equation}
\eta_{\text{\tiny Carnot}} = 1-\frac{Q_{C}}{Q_{H}} = 1-\frac{T_{C}}{T_{H}}\,,
\end{equation}
and is the maximum possible for any given heat engine.
\begin{figure*}
\centering
\begin{tabular}{cc}
\includegraphics[width=0.49\textwidth,height=0.3\textheight]{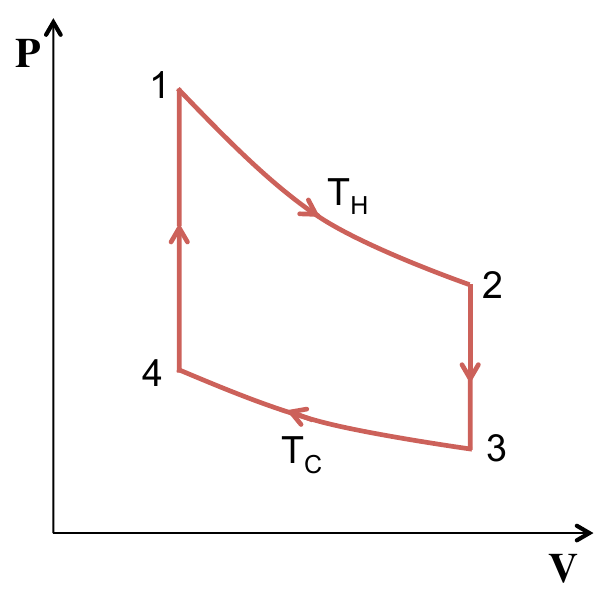} &
\includegraphics[width=0.49\textwidth,height=0.3\textheight]{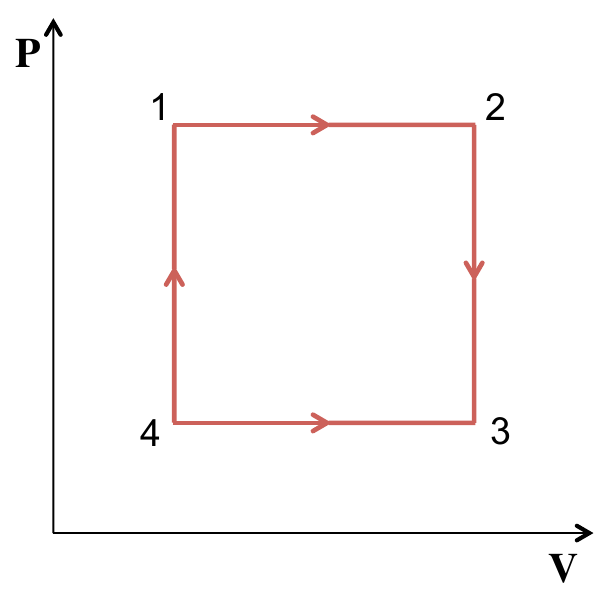}
\end{tabular}
\caption{{\bf $P-V$ diagram of thermodynamic cycles.} {\it Left.} Carnot cycle. Paths 12 and 34 are isothermal. Paths 23 and 41 are adiabatic. For a static black hole, they also correspond to isochoric (constant volume) paths.
{\it Right.} Rectangular path.}
\label{Fig:cycles}
\end{figure*}
For non-rotating (i.e. static) black holes, a path of fixed entropy (i.e. an adiabat) is a path of fixed volume (i.e. an isochore), since both of these depend only on $r_{+}$. Thus, the Carnot engine coincides with the {\it Stirling engine}, and is given in the left of Fig.~\ref{Fig:cycles}. The Carnot cycle has been studied for static black holes in Einstein gravity \cite{Johnson:2014yja, Belhaj:2015hha, Setare:2015yra}, in Gauss--Bonnet gravity \cite{Johnson:2015ekr}, with Born--Infeld sectors \cite{Johnson:2015fva}, in the presence of a dilaton \cite{Bhamidipati:2016gel}, and in Horava--Liftshitz gravity \cite{Sadeghi:2016xal}. It was also studied for the Kerr black hole in the limit of slow rotation and high pressure \cite{Sadeghi:2015ksa}.

In addition, a simple formula for the efficiency has been derived for rectangular paths in the $P-V$ plane (see right of Fig.~\ref{Fig:cycles}) \cite{Johnson:2016pfa}. Because heat flow only occurs for the horizontal paths which are at constant pressure, using the first law
\begin{equation}
\delta M = \delta H = T\delta S + V \delta P
\end{equation}
with $\delta P = 0$, we get the efficiency
\begin{equation}
\eta = 1- \frac{M_{3}-M_{4}}{M_{2}-M_{1}}\,.
\end{equation}
This only depends on the mass of the black hole at the four corners of the $P-V$ diagram. The $P-V$ rectangles can be made arbitrarily small and tiled to give the efficiency of arbitrary cycles.
\begin{figure*}
\centering
\includegraphics[width=0.49\textwidth,height=0.3\textheight]{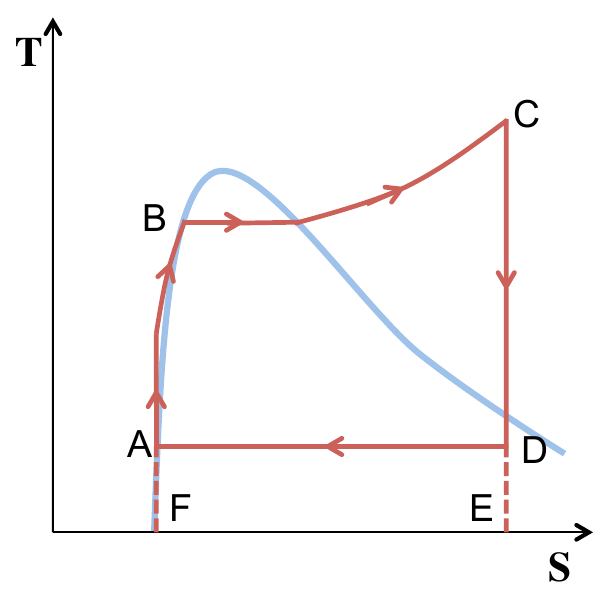}
\caption{{\bf $T-S$ diagram of an ideal Rankine cycle.} The Rankine cycle is displayed by the red curve, whereas the blue curve corresponds to the coexistence line.}
\label{Fig:rankine}
\end{figure*}

Another cycle that has been studied \cite{Wei:2016hkm} in the black hole context is the {\it Rankine cycle} (illustrated in Fig.~\ref{Fig:rankine}), which is used in for example, steam turbines, and makes use of the liquid/gas phase transition. The efficiency for this cycle is given by
\begin{equation}
\eta = 1-\frac{\text{Area}(ADEFA)}{\text{Area}(FABCDEF)} \,.
\end{equation}
The study of black hole heat engines in extended thermodynamics is a novel direction, which will be exciting in the context of holography (the topic of the next section).

\subsection{Superfluid Black Holes}

A new development in black hole chemistry occurred very recently with the discovery of a {\it $\lambda$-line} \cite{Hennigar:2016xwd}.   This is a line of second order (continuous) black hole phase transitions that strongly resemble those occurring in condensed matter systems such as  the onset of superfluidity in liquid helium~\cite{RevModPhys.71.S318}.   This phenomenon was observed to take place in
a broad class of  asymptotically AdS  black holes with scalar hair.  These black holes are exact solutions  \cite{Giribet:2014bva} to a class of theories in which a scalar field is conformally coupled to the higher-curvature terms in  Lovelock gravity \cite{Oliva:2011np}, and evade no-go results that had been previously reported ~\cite{nogo_hairy}.

\begin{figure*}
\centering
\begin{tabular}{cc}
\includegraphics[width=0.49\textwidth,height=0.3\textheight]{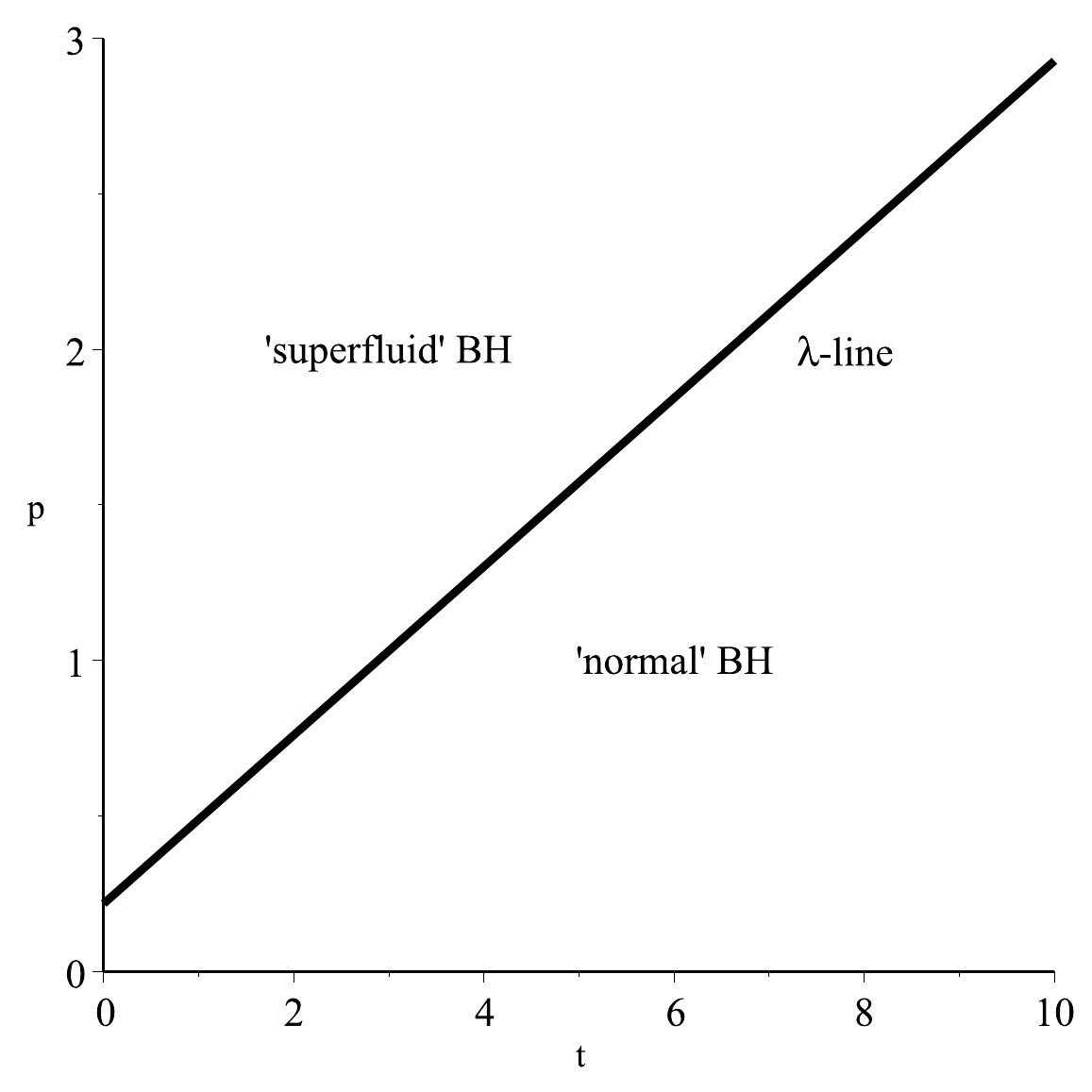} &
\includegraphics[width=0.49\textwidth,height=0.3\textheight]{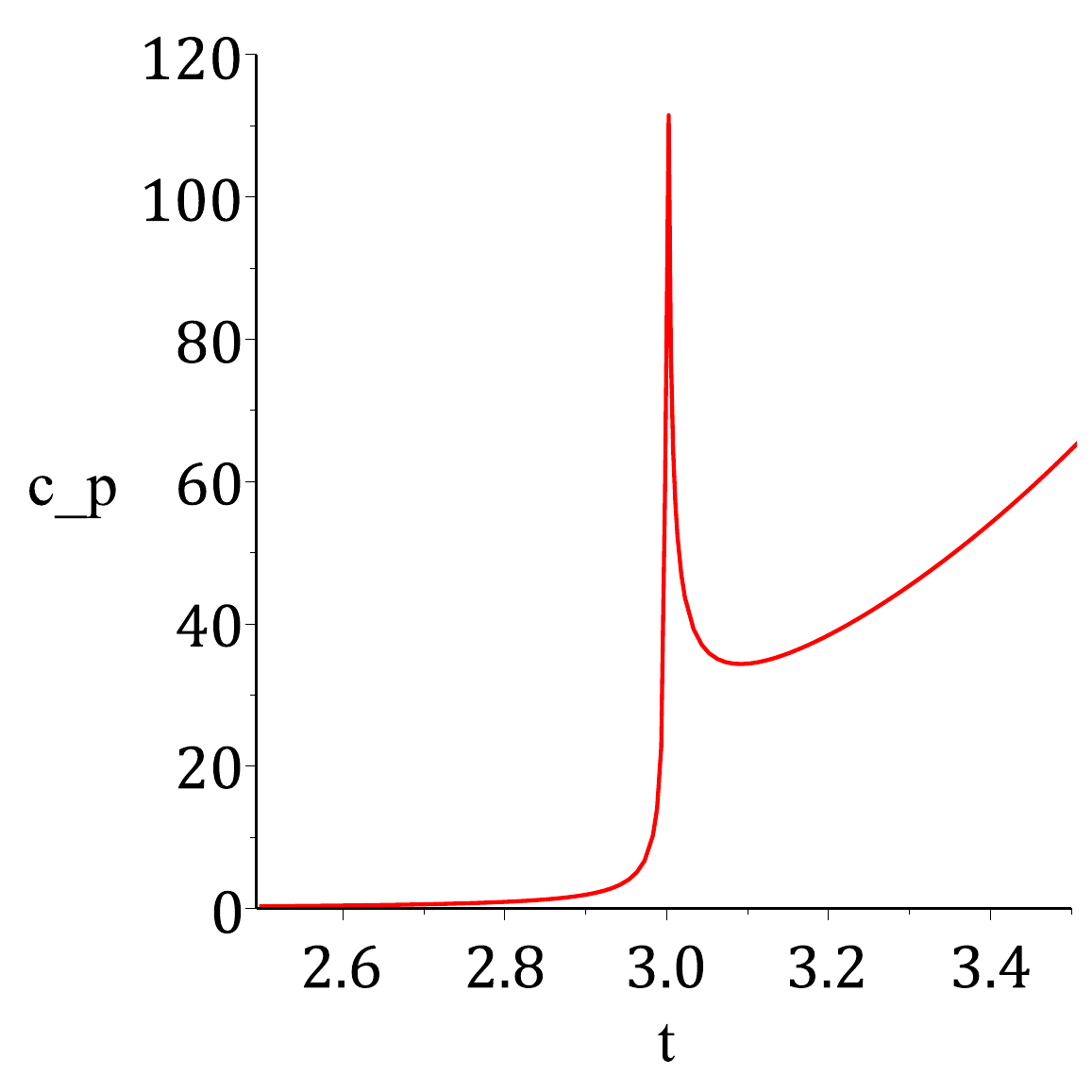}
\end{tabular}
\caption{ {\bf The $\lambda$-line of superfluid black holes.} $p-t$ diagram (left) illustrates an infinite `coexistence line' of second-order phase transitions separating the `superfluid black hole phase' from the normal black hole phase.
For every point on this line, the specific heat as a function of temperature has the characteristic $\lambda$-shape, illustrated in the right figure for $d=7$ and $t_c=3$. }
\label{Fig:weird}
\end{figure*}

To take a specific example, in the cubic-curvature case in  $d$-dimensions
the  equation of state for a static spherically symmetric hairy charged black hole
reads \cite{Hennigar:2016xwd}
\ba\label{eos}
p &=& \frac{t}{v}-\frac{k(d-3)(d-2)}{4\pi v^2}+\frac{2\alpha k t}{v^3}-\frac{\alpha(d-2)(d-5)}{4\pi v^4}+ \frac{3t}{v^5}\nonumber\\
&&-\frac{\sigma(d-7)(d-2)}{4\pi v^6}+\frac{q^2}{v^{2(d-2)}}-\frac{h}{v^d}\,,
\ea
where all quantities have been rescaled in terms of the cubic coupling constant $\alpha_3$, with
$\alpha=\frac{\alpha_2}{\sqrt{\alpha_3}}$ the quadratic curvature coupling,
$p$ the pressure, $v$ the volume, $t$, the temperature, $q$ the charge, $h$ the hair parameter, and
$k = -1,0,1$ the curvature parameter of the transverse constant curvature space.
 It is straightforward to show that the conditions for a critical point
\be\label{eqn:cp_condition}
\frac{\partial p}{\partial v}=\frac{\partial^2 p}{\partial v^2}=0
\ee
have for $k=-1$ the solution
\be
p_c  = \left[ \frac{8}{225} v^3_c   \right]t_c + \frac{v^2_c  (11d-40)(d-1)(d-2)}{900 \pi d}
\ee
and critical volume $v_c = 15^{1/4}$
\textit{for all temperatures} $t_c$, provided $\alpha = \sqrt{5/3}$, $h=\frac{4(2d-5)(d-2)^2v_c^{d-6}}{\pi d (d-4)}$, and $q^2= \frac{2(d-1)(d-2) v_c^{2d-10}}{ \pi (d-4)}$.
 In other words, this black hole exhibits \textit{infinitely many critical points}! In the $p-v$ plane, every isotherm has an inflection point at $v= 15^{1/4}$, and
 there is no first order phase transition (in the variables $(t, p)$) but rather a line of second order phase transitions, see left Fig.~\ref{Fig:weird}, characterized by a diverging specific heat at the critical values, as shown in right Fig.~\ref{Fig:weird}, with the characteristic $\lambda$  shape  clearly visible.
  Further investigation \cite{Hennigar:2016xwd}
 indicates that these black holes have no pathological properties except for the generic singularity inside the event horizon common to all black holes.

A plot of the specific heat as a function of temperature reveals a striking
resemblance to the fluid/superfluid $\lambda$-line transition of $^4$He \cite{Hennigar:2016xwd}.   The phase diagram for helium is actually more complicated (as expected), including solid and gaseous states. Yet with only 4 parameters the essence of the $\lambda$-line can be captured in black hole physics.   The  interesting properties of a superfluid are either dynamical or require a full quantum description to understand \cite{RevModPhys.71.S318}, and so one might expect some underlying
quantum theory of gravity will allow exploration of  the black hole analogues of these properties at a deeper level.

The conditions for a critical point yielding a  $\lambda$-line are quite general, holding for (at least)
all Lovelock theories cubic and higher.    For an equation of state of the form,
\be\label{lambda-cond}
P = a_1(V, \varphi_i) \,  T + a_2(V, \varphi_i)
\ee
where $V$ is thermodynamic volume and the $\varphi_i$ represent additional constants in the equation of state, equation \eqref{lambda-cond} will exhibit a $\lambda$-line
provided the equations
\begin{align}\label{supercriteqs}
\frac{\partial a_i}{\partial V} &= 0 \, , \quad  \frac{\partial^2 a_i}{\partial V^2} = 0 \,  \quad i=1,2
\end{align}
have a non-trivial solution.  This is not easy:  neither the rotating black hole of $5d$ minimal gauged super-gravity~\cite{ChongEtal:2005b} nor those in higher order Lovelock gravity (without hair) admit a non-trivial solution \cite{Hennigar:2016xwd}.
The necessary and sufficient conditions for satisfying \eqref{supercriteqs} for black holes in general remain to be found.

\subsection{Future of black hole chemistry}

Extending the phase space of black hole thermodynamics to include $\Lambda$ as pressure has led us to be able to understand black holes as chemical systems, exhibiting the kind of phenomena found in a broad variety of real-world systems.
The basic results---Van der Waals behavior, reentrant phase transitions, triple points, and Carnot-type cycles---are
extremely robust, having been replicated in a broad variety of settings and contexts
\cite{Grumiller:2014oha, Dolan:2014vba,
Frassino:2014pha,
Zou:2013owa, Hendi:2012um, Hendi:2014kha, Mo:2014qsa,Belhaj:2014tga, Hendi:2015soe, Zeng:2016sei, Gunasekaran:2012dq,
Caceres:2015vsa, Hennigar:2015esa,Zou:2014mha,Sadeghi:2016xal, Belhaj:2012bg, Poshteh:2013pba, Dolan:2014lea, Frassino:2016oom, Tanabe:2016opw,
Belhaj:2013ioa, Ma:2014tka, Cai:2013qga,
Zhang:2014eap,Hristov:2013sya,Zhao:2013oza,
Dehghani:2014caa,Hendi:2015xya,Dutta:2013dca, Chen:2013ce,
Xu:2014kwa, Mo:2013sxa,Suresh:2014pra, Mirza:2014xxa,Xu:2015rfa,Hendi:2015hoa,Hendi:2015pda,Fernando:2016qhq}.
Theories nonlinear in curvature and/or fields also replicate similar behavior. These
include black holes in higher \cite{Gunasekaran:2012dq, Belhaj:2012bg, Poshteh:2013pba, Dolan:2014lea, Frassino:2016oom, Tanabe:2016opw} and lower dimensions \cite{Zou:2014mha, Belhaj:2013ioa, Ma:2014tka}, Lovelock gravity \cite{Cai:2013qga, Xu:2013zea, Mo:2014qsa, Frassino:2014pha, Dolan:2014vba}, nonlinear electrodynamics
\cite{Hendi:2012um, Zou:2013owa,Hendi:2014kha, Mo:2014qsa,Belhaj:2014tga, Hendi:2015soe, Zeng:2016sei, Mo:2014qsa}, Einstein--Yang--Mills gravity
\cite{Zhang:2014eap}, 
black holes with scalar hair \cite{Hristov:2013sya,Zhao:2013oza,Belhaj:2013ioa,Grumiller:2014oha,Dehghani:2014caa,Hendi:2015xya},
dyonic black holes \cite{Dutta:2013dca},
$f(R)$ gravity \cite{Chen:2013ce},
STU black holes \cite{Caceres:2015vsa}, quasi-topological gravity \cite{Hennigar:2015esa},  conformal gravity \cite{Xu:2014kwa}, Horava--Lifshitz black holes \cite{Mo:2013sxa,Suresh:2014pra,Sadeghi:2016xal},
Poincare gauge gravity \cite{Ma:2014tka}, Lifshitz gravity \cite{Gim:2014nba,Brenna:2015pqa},  
massive gravity \cite{Mirza:2014xxa,Xu:2015rfa,Hendi:2015hoa,Hendi:2015pda,Fernando:2016qhq}, and others.\footnote{We refer the reader to citations of \cite{Kastor:2009wy, Kubiznak:2012wp} for further directions and references.}
Most of these articles  obtain the same qualitative phase behavior associated with swallowtails in the Gibbs free energy diagram, indicating that black hole/Van der Waals correspondence is quite robust.   Further reading on these interesting phenomena can be found in \cite{Altamirano:2014tva, Dolan:2014jva, Kubiznak:2014zwa}.

There are a number of new directions to pursue in this subject.   No coexistence line of second-order phase transition and/or second-order triple point\footnote{Typically characterized by 2 order parameters, this is a critical point where a coexistence line of second-order phase transition merges a line of first-order phase transition.},  third-order phase transition, or $n$-tuple  (with $n>3$) critical points (where more than 3 phases meet together), have ever been observed.  It is conceivable that such things may exist in higher-curvature gravity theories, and it would be interesting to either obtain them or to rule out their existence.
While the derivation of the Smarr relation from the first law through the scaling argument has been shown to be very broadly applicable, see e.g. \cite{Brenna:2015pqa}, an assumption of homogeneity is required to obtain it, and it would be preferable to find a more fundamental reason underlying this assumption.  Although its has been shown that the   Lovelock coupling constants can be regarded as thermodynamic variables \cite{Kastor:2010gq, Liberati:2015xcp}, very little is understood \cite{Xu:2013zea} about phase behaviour in the much broader phase space where these quantities are no longer constant.  The relationship between gravitational tension \cite{Armas:2015qsv,El-Menoufi:2013pza} and the approach of black hole chemistry could use further scrutiny and clarification.    Recently it was
shown \cite{Hennigar:1612.06852}
that conformal scalar hair gives rise to a one-parameter family of isolated critical points that occur under much more general conditions than in previous work  \cite{Frassino:2014pha,Dolan:2014vba,Hennigar:2015esa}, clarifying the relationship between isolated critical points and thermodynamically singular points.  However the nature and behaviour of black holes with isolated critical points
  remains to be fully explored.

Modifications that take into account quantum corrections to black holes
  \cite{Smailagic:2012cu, Spallucci:2014kua,Frassino:2015hpa, Frassino:2016oom, Pourhassan:2015cga},
 an extended phase space study of universal horizon products \cite{Pradhan:2016rff}, and the exploration of thermodynamic geometries in the presence of variable $\Lambda$ \cite{Gruber:2016mqb, Dolan:2015xta} are other new frontiers for black hole chemistry.

An intense area of recent activity has been in holography.  This is the subject of the following section.

\section{The AdS/CFT interpretation}\label{Sec:5}
The primary motivation underlying the study of AdS black holes is the AdS/CFT correspondence  \cite{Maldacena:1997re},  which relates a (quantum) gravitational theory in $d$-dimensional AdS space (sometimes called the bulk) to a CFT formulated on its $(d-1)$-dimensional  boundary. Since this correspondence assumes that the cosmological constant $\Lambda$ is fixed,  it is natural to ask what the interpretation of the bulk pressure and volume might be on the boundary  CFT once  $\Lambda$ is treated as a thermodynamic variable.  What do the results from extended thermodynamics in the bulk (first law, Smarr relation, and the various phase transitions) correspond to for the boundary theory? In this section, we will survey recent attempts that address this question.

Recall  that the pressure is defined  in terms of $\Lambda$ (or the AdS curvature radius $l$)
\be\label{eq:pressG}
P = -\frac{\Lambda}{8\pi G_{d}} = \frac{(d-1)(d-2)}{16\pi l^2 G_{d}}\,,
\ee
where we have explicitly included the $d$-dimensional gravitational constant $G_{d}$ in \eqref{eq:press}.
Roughly speaking, $l$ is a measure of the number of degrees of freedom, $N$, of the boundary field theory, the precise correspondence between $l$ and $N$ depending on the family of CFTs being considered.\footnote{As such, the aforementioned examples of heat engines that go around a loop in the $P-V$ plane are, from the viewpoint of the boundary, going around the space of field theories rather than staying within one particular field theory \cite{Johnson:2014yja, Johnson:2016pfa}.}
In particular, in the well-known case of the correspondence between $AdS_{5}\times S^{5}$ and ${\cal N}=4$ $SU(N)$ Yang--Mills theory, the relation is given by \cite{Maldacena:1997re}
\begin{equation}
l^4=\frac{\sqrt{2}\ell_{Pl}^4}{\pi^2}N\,,
\label{eq:NrhL}
\end{equation}
where $\ell_{Pl}$ is the 10-dimensional Planck length.
This relation has its origins in the AdS/CFT correspondence from string theory, where the $AdS_{5}\times S^{5}$ spacetime can be viewed as the near-horizon geometry of $N$ coincident D3 branes in type IIB supergravity.

A few authors \cite{Johnson:2014yja,Dolan:2014cja,Kastor:2014dra} have therefore suggested that varying the pressure, or $\Lambda$, is equivalent to varying the number of colors, $N$, in the boundary Yang--Mills theory. The thermodynamic conjugate of pressure, i.e. the thermodynamic volume, should then be interpreted in the boundary field theory as an associated {\it chemical potential for color}, $\mu$. 
{As shown recently in \cite{Karch:2015rpa} such an interpretation is not entirely correct.  In fact, varying $\Lambda$ in the bulk corresponds both to changing the number of colours $N$ and changing the volume of the CFT. It is possible to hold $N$ fixed, so that we always refer to the same field theory, and thus varying $\Lambda$ in the bulk has a more natural consequence of varying the volume of the field theory. However to stay at fixed $N$ as the volume is varied we have to compensate by varying the Newton constant $G_d$ (see below). 
Let us first recapitulate the first (somewhat simplistic) approach, in which only the number of colors is varied. }

\subsection{First attempts: Chemical potential of the CFT}\label{cp}

From standard thermodynamics, the chemical potential $\mu$ is defined as the thermodynamic variable conjugate to a change in particle number $N_{p}$ and appears in the first law as
\begin{equation}\label{chemfirst}
\delta E = T\delta S +\mu\, \delta N_{p}\,.
\end{equation}
The chemical potential of the CFT can similarly be defined by this first law as the conjugate to its number of degrees of freedom (or colors).

Since varying the pressure means varying $N$, the thermodynamic pressure and volume in the bulk are  respectively dual to the number of colors and the chemical potential in the boundary:
\begin{equation}\label{VdPmudN}
V\delta P  \quad \leftrightarrow \quad  \mu \delta N\,.
\end{equation}
For a general CFT, one can infer from (\ref{eq:pressG}) that $\delta P \sim -\delta N$, likewise implying
a sign difference
\begin{equation}
\mu \sim -V\,,
\end{equation}
from \eqref{VdPmudN} for
the chemical potential and thermodynamic volume. This relation might hint at a meaningful connection between the behavior of a black hole (in the bulk) and that of the holographically dual CFT it describes.

The holographic dictionary maps the black hole mass $M$ to the energy $E$ of the field theory, and the temperature $T$ and thermal entropy $S$ of the black hole to those of the field theory. Taking the specific case of $AdS_{5}\times S^{5}$, we take $\delta N^2$ instead of $\delta N$ to define $\mu$, because in the large $N$ limit, the number of degrees of freedom of the ${\cal N}=4$ $SU(N)$ Yang--Mills theory is proportional to $N^2$.
With this mapping, and the relation between $\Lambda$ and $N$ from (\ref{eq:pressG})  and (\ref{eq:NrhL}), one can define the chemical potential of the field theory, $\mu$, via a first law
\begin{equation}\label{flawN2}
\delta M=T \delta S+\mu \delta N^{2}\,.
\end{equation}
With this definition, the chemical potentials corresponding to  Schwarszchild  \cite{Dolan:2014cja,Zhang:2014uoa} and charged \cite{Zhang:2015ova} black holes in $AdS_{5}\times S^{5}$ have been explicitly calculated.


As an example, we demonstrate how to calculate the chemical potential and relate its behavior to the Hawking--Page transition \cite{Dolan:2014cja,Zhang:2014uoa}. To be specific, consider
the line element \eqref{Schw-Ads} in five dimensions with the corresponding metric function
\begin{equation}
f=1 - \frac{8  G_{5} M }{3 \pi r^2 } + \frac{r^2}{l^2} \, .
\end{equation}
Here $l$ is both the $AdS^{5}$ and the $S^{5}$ radius. Note that because of compactification, $G_{5}=\frac{\ell_{Pl}^8}{\hbar \pi^3 l^5}$ is no longer a constant, but a function of $l$, and $\ell_{Pl}$ is the 10-dimensional Planck length.\footnote{
How does the expression for $G_{5}$ arise? Since we are compactifying a 10-dimensional space into $AdS_5 \times S^5$, we treat  the 10-dimensional Newton constant $G_{10}$ as the true gravitational constant. The corresponding Planck length is defined using the general relation $(\ell_{Pl}^{(d)})^{d-2}=\hbar G_{d}$ for 10 dimensions. To find the effective gravitational constant in 5 dimensions, we divide by the volume of the compactified space, i.e. $G_{5} = G_{10}/V_{S^5}$\,, where $V_{S^5}=\pi^3 l^5$ is the volume of the 5-dimensional sphere with radius $l$.} Because of this extra dependence on $l$, $\mu$ is unfortunately no longer related to $V$ in the simple way described above.

Solving $f(r)=0$ for the event horizon radius, $r_{+}$, we arrive at an expression for $M$ as a function of $r_{+}$ and $l$,
\begin{equation}
M=\frac{3 \pi r_{+}^2(l^2 + r_{+}^2)}{8 G_{5} l^2}\,.
\label{eq:MrhL}
\end{equation}
The next step is to obtain an expression for the black hole mass $M$ as a function of thermodynamic variables in the boundary field theory, namely the entropy $S$ and number of colors $N$.  To this end we employ the Bekenstein--Hawking area formula (\ref{ent}), giving
\begin{equation}
S=\frac{1}{4}\frac{A}{\hbar G_{5}}=
\frac{\pi^5 l^5  r_{+}^3}{2 \ell_{Pl}^8}\,,
\label{eq:SrhL}
\end{equation}
with $A=2\pi^2 r_{+}^3$.
Using (\ref{eq:SrhL}) and (\ref{eq:NrhL}) in (\ref{eq:MrhL}) yields
\begin{equation}
M(S,N)=\frac {3\,\widetilde m_{Pl}}{4}
\Bigl[\Bigl(\frac {S}{\pi} \Bigr)^{\!\frac 2 3} N^{\frac{5}{12}}
+ \Bigl(\frac {S}{\pi} \Bigr)^{\!\frac 4 3} N^{-\frac{11}{12}}\Bigr]\,,
\label{eq:MSN}
\end{equation}
with $\widetilde m_{Pl}=\frac{\sqrt{\pi} \, m_{Pl}}{2^{1/8}}$ and
$m_{Pl}=\frac{\hbar}{\ell_{Pl}}$, the 10-dimensional Planck mass.

We now obtain expressions for the temperature, Gibbs free energy, and chemical potential.
The thermodynamic relation $\delta M=T\delta S+\mu \delta N^2$ implies that
\be
T(S,N)=\left.\frac{\partial M}{\partial S}\right|_N=\frac{\widetilde m_{Pl}}{2 \pi} \left[\left(\frac{S}{\pi} \right)^{-\frac{1}{3}} N^{\frac{5}{12}} + 2\left( \frac{S}{\pi}\right)^{\frac{1}{3}} N^{-\frac{11}{12}} \right]\,,
\ee
and that for fixed $N$, the minimum temperature is
\begin{align}
T_{min}&=\frac {\sqrt{2} \widetilde m_{Pl}}{\pi N^{\frac 1 4}}.
\end{align}
For any $T>T_{min}$, there are two solutions for $S$, corresponding to a large and a small black hole. The heat capacity $C_N=T\left(\frac{\partial S}{\partial T}\right)_N$ diverges at $T_{min}$, and is always negative for small black holes (which are in the $r_{+}^2 < {l^2}/{2}$ regime), rendering them unstable.

To find the conditions for the Hawking--Page transition \cite{Hawking:1982dh} (see also Sec.~\ref{Sec:4}) we compute the Gibbs free energy %
\begin{equation} G(T,N)=M - TS = \frac{\widetilde m_{Pl}}{4}
\Bigl[\Bigl(\frac{S}{\pi}\Bigr)^{\!\frac{2}{3}} N^{\frac{5}{12}} -
\Bigl(\frac{S}{\pi}\Bigr)^{\!\frac{4}{3}} N^{-\frac{11}{12}}\Bigr]\, ,
\end{equation}
which is negative for $N^2 < \frac{S}{\pi}$, corresponding to $r_{+}>l$. Black holes in this regime are more stable than 5-dimensional AdS with thermal radiation at the same temperature, whereas black holes with $r_{+}<l$ are prone to decay into thermal radiation. In terms of $N$,
the Hawking--Page transition temperature, where $G$ changes sign, is
\begin{equation}
T_{\tiny{\mbox{HP}}}=\frac{3}{2\pi}\frac{\widetilde m_{Pl}}{N^{1/4}}\,.
\end{equation}
This transition in the bulk corresponds to a confinement/deconfinement phase transition of the quark-gluon plasma in the boundary field theory \cite{Witten:1998zw}.

\begin{figure*}[]
\centering
\includegraphics[scale=0.7]{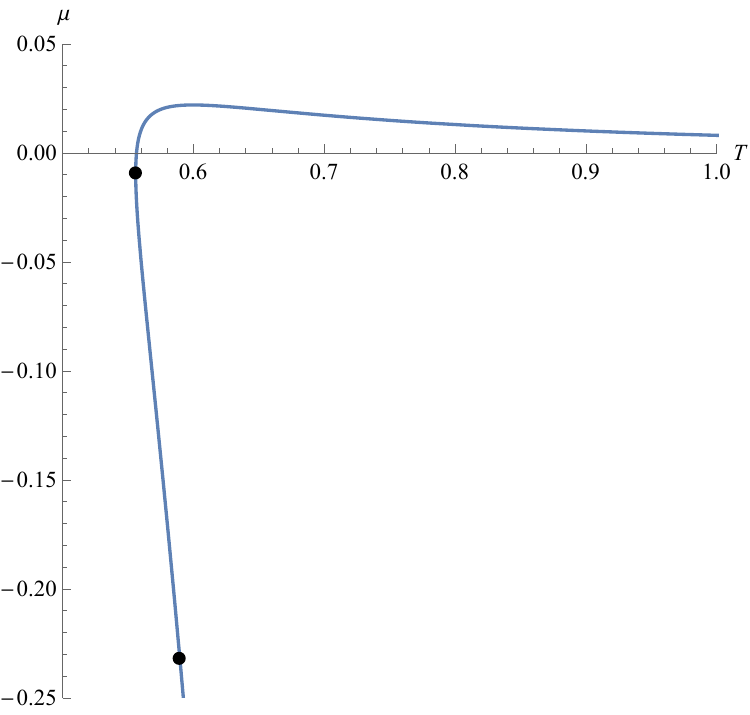}\\
\caption{{\bf
The chemical potential $\mu$ as a function of temperature $T$ at fixed $N=3$ \cite{Zhang:2014uoa}.}
The upper dot denotes the minimal temperature $T_{min}$ and the lower
dot denotes the Hawking--Page transition temperature $T_{\tiny{\mbox{HP}}}\;.$ The curve is parametrized by the horizon radius $r_{+}$; the zero-crossing actually occurs at the small unstable black hole branch.
 }
   \label{fig:mut}
\end{figure*}

Finally, using \eqref{flawN2} and \eqref{eq:MSN} we find  \cite{Johnson:2014yja,Dolan:2014cja,Kastor:2014dra}
\begin{equation}
\mu \equiv \left.\frac{\partial M}{\partial N^2}\right|_S=
\frac {\widetilde m_{Pl}} {32}
\Bigl[5\Bigl(\frac {S}{\pi} \Bigr)^{\!\frac 2 3} N^{-\frac{19}{12}}
-11 \Bigl(\frac {S}{\pi} \Bigr)^{\!\frac 4 3} N^{-\frac{35}{12}}\Bigr]\,,
\label{eq:mu}
\end{equation}
which becomes positive when $N^2 > \left(\frac{11}{5}\right)^{\frac 3 2}\frac{S}{\pi}$, or equivalently, $r_{+}^2 <\frac {5} {11} l^2$. In ordinary chemistry, the chemical potential $\mu$ is negative and large at high temperatures, in the classical regime. When $\mu$ changes sign to become positive, it is an indication that quantum effects are coming into play \cite{Dolan:2014cja}. In terms of temperature, $\mu$ changes sign at
\begin{equation}
T_0=\frac{21}{2\pi \sqrt{55}}
\frac{\widetilde m_{Pl}} {N^{\frac {1} {4} }}\,,
\end{equation}
about 6\% below the Hawking--Page transition temperature $T_{\tiny{\mbox{HP}}}$, and just 0.1\% above the minimum temperature $T_{min}$, as illustrated in  Fig. \ref{fig:mut}. Note, however, a crucial difference: $T_0$ occurs for a black hole in the {\it small} black hole branch, whereas the Hawking--Page is a transition between radiation and a {\it large} black hole.

The high temperature behavior of $\mu$ is consistent with that in ordinary chemistry. In the limit of high temperature (where $\frac{S}{ N^2}\gg1$),
\begin{equation}
\mu \approx -\frac{11 N^{\frac 3 4 }}{2 \widetilde m_{Pl}^3} \left( \frac{\pi T}{2}\right)^4\,
\label{eq:highTmu}
\end{equation}
is indeed negative and a decreasing function of $T$.


An alternate definition of chemical potential makes use of densities of thermodynamic quantities  \cite{Maity:2015ida}.  Dividing all extensive quantities by the {\it field theory volume}
\be\label{VFT}
{\cal V}=\omega_{d-2}l^{d-2}\,,
\ee
the volume of a $(d-2)$-dimensional sphere of radius $l$ (see also Eq.~\ref{VFT2}),
yields
\begin{align}
d\rho = Tds + \Phi dq + \mu dN^{3/2}
\end{align}
for the first law, where $\rho$, $s$ and $q$ are the mass, entropy and charge densities respectively. Using this definition, for the (4+1)-dimensional Schwarszchild and charged black holes, the chemical potential changes sign {\it precisely} at the Hawking--Page transition temperature.
However, this is no longer true for other cases, for example the (3+1)-dimensional case or for rotating black holes.

\subsection{More refined holographic dictionary}
 
{In this subsection we shall discuss a more refined holographic dictionary \cite{Karch:2015rpa}, properly identifying the properties of the holographic fluid with the bulk thermodynamic quantities. In particular, following \cite{Karch:2015rpa} we argue that varying $\Lambda$ in the bulk corresponds to both varying the number of colours and varying the volume of the CFT. The rationale for the latter is simple: varying $\Lambda$ is equivalent to changing the AdS length scale $l$, which in turn changes the curvature radius governing the space on which the field theory resides, varying effectively its volume. We shall also discuss the holographic derivation of the bulk Smarr relation, and the possibility of keeping the number of colours constant, which is only possible if both Newton's constant and $\Lambda$ are  varied.
}

\subsubsection{Pressure and volume of a CFT}
In standard black hole thermodynamics, we write the bulk metric of a generic spherically symmetric black hole in the form \eqref{Schw-Ads}, that is
\begin{equation} \label{BHmetric}
ds^2_b = - f(r) dt^2 + \frac{dr^2}{f(r)} + r^2  d\Omega_k^2\,,
\end{equation}
where
for an asymptotically AdS space, the metric function (or blackening function)
at large $r$ approaches
\begin{equation}
f(r) = \frac{r^2}{l^2} + \ldots .
\end{equation}
To read off the field theory metric, we need to multiply (\ref{BHmetric}) with an overall conformal factor with a double zero at $r=\infty$ and then evaluate the metric at  fixed $r \to \infty$ \cite{Witten:1998qj}. Choosing this factor  to be $l^2/r^2$, we see that the boundary metric is
\begin{equation} \label{bdrymetric}
ds^2 = -dt^2 + l^2 d\Omega_k^2.
\end{equation}
Thus, the {\it volume of the field theory} is defined to be (cf. Eq.~\eqref{VFT})
\be\label{VFT2}
{\cal V}=\omega^{(k)}_{d-2}l^{d-2}\,,
\ee
i.e. the volume of a $(d-2)$-dimensional compact space (a `sphere') of radius $l$.
Denoting by $\Omega$ the {\it free energy of the field theory},
the {\it field theory pressure} is then defined by
\begin{align}\label{pressKarch}
p=-\frac{\partial \Omega}{\partial {\cal V}}\,.
\end{align}

Unfortunately, there is no one-to-one map between bulk and boundary pressures and volumes. To illustrate the issue, note that varying the pressure of the bulk, or equivalently varying only the AdS radius $l$, amounts to varying the following boundary {quantities all at once \cite{Karch:2015rpa}}:
\begin{itemize}
\item
the number of colors $N$, since $l^{d-2} \propto N^{2}$\,,
\item
the volume of the space on which the field theory is formulated, as ${\cal V}\propto l^{d-2}$\,,
\item
the CFT charge $Q$ which is related to the bulk charge $Q_b$ according to $Q=lQ_{b}$\,. \end{itemize}

{Note also that it is possible to
keep constant $N$ while varying $\Lambda$, as follows from the following standard relation: 
\begin{equation}
N^{2} \sim \frac{l^{d-2}}{G_d}\,.
\end{equation}
In this case, the field theory volume ${\cal V}$ changes and keeping $N$ constant means that $G_d$ must vary with $l$.} Here, we see the tension between the description of physics in the bulk and that of the boundary: keeping $N^{2}$ constant is natural for the boundary field theory, but it has the 
consequence that $G_d$ cannot stay fixed in the bulk. On the other hand, varying only the pressure in the bulk (where $G_d$ is kept constant) corresponds to varying both the volume and $N$ of the boundary field theory at the same time.

\subsubsection{The Holographic Smarr relation}

The generalized Smarr relation \eqref{smarrBH} can be equivalently derived by considering the thermodynamics of the dual field theory  \cite{Karch:2015rpa}. Namely, in the limit of a large number of colors, $N$, the free energy of the field theory scales simply as $N^{2}$ (the central charge):
\begin{equation}\label{holosmarr}
\Omega(N,\mu,T,l)= N^{2}\Omega_{0}(\mu,T,l) \,.
\end{equation}
This `{\it holographic Smarr relation}', together with the {\it equation of state}, which for a conformal field theory reads
\be\label{UCFT}
E=(d-2)p{\cal V}\,,
\ee
can be employed for an independent derivation of the Smarr relation \eqref{smarrBH}.
In particular, the
consistency of the
equations \eqref{holosmarr} and \eqref{UCFT} with the Smarr relation \eqref{smarrBH} has been explicitly demonstrated for the charged-AdS black hole spacetime in 
any dimension in \cite{Karch:2015rpa}.
In the same paper, a modified equation of state was also found for a non-trivial example of a large $N$ gauge theory with hyperscaling violation. Here, a different Smarr relation for the bulk would be expected and would be interesting to  find (see however \cite{Brenna:2015pqa}).

The holographic Smarr relation (\ref{holosmarr}) gives a simple but important insight into the behavior of the boundary field theory:
no non-trivial phase transition can occur by varying $N$ alone.
In fact, from a holographic perspective, non-trivial phase transitions happen to black holes in the bulk as we vary the bulk pressure because we are inevitably varying both $N$ and the volume of the boundary field theory.

Beyond the leading large $N$ limit, the free energy of the boundary field theory is no longer  proportional to $N^{2}$ and will instead depend nontrivially on $N$.
Such theories would correspond in the bulk to gravitational theories that include higher curvature terms \cite{Karch:2015rpa}. These exhibit exotic behavior such as reentrant phase transitions and non-mean field critical points as seen in Sec. \ref{Sec:4}.  Although less straightforward, it would be interesting to use holography to rederive  the generalized Smarr relations \eqref{SmarrLovelock} for these systems.

\subsection{$p-{\cal V}$ criticality of boundary CFT}

{Using the pressure and volume of a CFT as defined in the previous subsection, one can study the corresponding $p-{\cal V}$ criticality \cite{Dolan:2016jjc}.}
Consider for example a charged AdS black hole in the 5-dimensional bulk, dual to a 4-dimensional $\mathcal{N} = 4$ SUSY Yang-Mills on the boundary, and define  pressure and volume via \eqref{pressKarch} and \eqref{VFT2}.  Since charged AdS black holes exhibit a line of first order phase transitions terminating at a critical point with mean field exponents
\cite{Kubiznak:2012wp}, from the CFT viewpoint a critical point with mean field exponents appears in its  $p-{\cal V}$ plane, keeping $N$ constant.

However, the $p-{\cal V}$ behavior of the boundary field theory differs from that of the black hole in the bulk in a number of ways.
There is a single phase at low temperatures and two phases above the critical temperature.
It is the pressure, rather than the volume, that jumps across the phase transition. Hence, it is more appropriate to let $p$, instead of ${\cal V}$, be the order parameter.  Furthermore with ${\cal V}$ and $Q$ fixed, the critical exponents turn out to be \cite{Chamblin:1999tk,Wu:2000id,Dolan:2016jjc}
\begin{align}
\alpha=\frac{2}{3}, \quad \beta=1, \quad \gamma=-\frac{2}{3}, \quad \delta= \frac{1}{3} \, ,
\end{align}
which are not mean field, though the equation of state of the CFT is still analogous to the Van der Waals equation of state.
However, if $\Phi$ instead of $Q$ is taken as the order parameter, the critical exponents become mean field. Finally, since the boundary field theory is conformal, ${\cal V}$ and $T$ are not independent, suggesting that $Q$ rather than $T$ should be used as the control parameter
\cite{Dolan:2016jjc}.

\subsection{Holographic entanglement entropy} \label{hee}

The AdS/CFT correspondence has enabled us to study the entanglement entropy of a boundary CFT via the geometry of the bulk \cite{Ryu:2006bv}. There have also been several developments regarding the behavior of entanglement entropy by considering extended thermodynamics in the bulk \cite{Kastor:2014dra,Kastor:2016bph,Caceres:2016xjz,Johnson:2013dka,Caceres:2015vsa,Zeng:2015wtt}. We start off by defining entanglement entropy and stating its relation to the bulk geometry.

\begin{figure*}
\centering
\rotatebox{0}{
\includegraphics[scale=0.7]{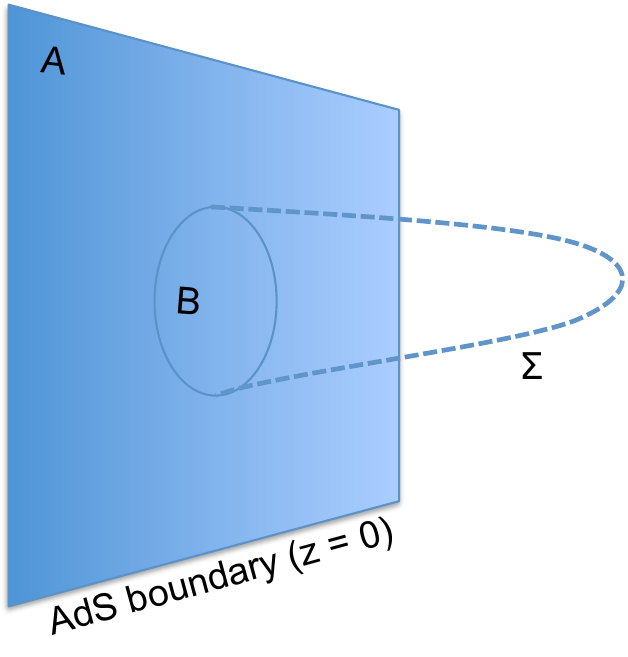}}\\
\caption{
The boundary $z = 0$ divided into two complementary regions $A$ and $B$. The minimal surface $\Sigma$, which lives in the AdS bulk, has a boundary at spatial infinity that matches the boundary between $A$ and $B$.
 }
   \label{fig:bulkboundary}
\end{figure*}

In quantum theory, a system can be partitioned into a subsystem $A$ and its complement $B$ by splitting its Hilbert space. For example, $A$ and $B$ can be complementary spatial volumes in a quantum field theory on a given constant time slice, separated by an ``entangling surface''. If the full system is in the pure state $\qket{\psi}$, the state of subsystem $A$ is described by a reduced density matrix
$ \rho_{A}=\text{Tr}_{B}\qket{\psi}\qbra{\psi} \,, $
where the degrees of freedom in $B$ are traced over. The entanglement between $A$ and $B$ can be quantified by the entanglement entropy, defined as the von Neumann entropy of the reduced density matrix $\rho_{A}$:
\begin{equation}\label{entShan}
S_{E}= - \text{Tr} \rho_{A} \log \rho_{A} \,.
\end{equation}
The entanglement entropy in a CFT living on an AdS boundary is encoded in the AdS bulk by virtue of the Ryu--Takayanagi proposal \cite{Ryu:2006bv}, which states that the entanglement entropy $S_{E}$ between two complementary regions $A$ and $B$ in the CFT is given by  a generalization of the Bekenstein--Hawking formula
\begin{equation} \label{eqn:ryu}
S_{E} = \frac{A_{\Sigma}}{4G_d}\,,
\end{equation}
applied to a bulk minimal surface $\Sigma$ (with area $A_{\Sigma}$)\footnote{Since the minimal surface area in an asymptotically AdS bulk (and thus also the entanglement entropy) is formally divergent, to obtain a finite result one needs to employ a due regularization procedure.
}
whose boundary at spatial infinity matches the entangling surface in the CFT
(see Fig.~\ref{fig:bulkboundary}).

Expressing $\rho_{A}$ in the form of an effective thermal system
\begin{equation} \label{eqn:mod}
\rho_{A}= \frac{e^{-H_{A}/T_{0}}}{\text{Tr}(e^{-H_{A}/T_{0}})}\, ,
\end{equation}
where $H_{A}$ is known as the {\it modular Hamiltonian}, yields, upon employing \eqref{entShan}
\begin{equation}
\label{eqn:ee}
T_{0} \delta S_{E} = \text{Tr}\bigl(H_{A} \delta \rho_{A}\bigr) \equiv \delta \left< H_{A}\right>\,,
\end{equation}
showing that entanglement entropy in CFTs satisfy a first law \cite{Blanco:2013joa,Wong:2013gua}.
In the case of a spherical entangling surface (with radius $r_{0}$), the first law (\ref{eqn:ee}) has been shown \cite{Faulkner:2013ica} to follow from the bulk gravitational first law associated with the minimal surface $\Sigma$. The bulk gravitational first law applies in our non-black hole context because $\Sigma$ in this spherical case is a 
Killing horizon (with Killing vector $\xi$), and is given by
\begin{equation} \label{eqn:bulklaw}
\frac{\kappa \delta A_\Sigma} {8\pi G_d}=\delta E_\xi\,,
\end{equation}
where $\kappa$ is the surface gravity of the horizon, while $\delta A_\Sigma$ is, in this case, the change in area of the minimal surface $\Sigma$ under a perturbation to a nearby solution of the bulk equations of motion which keeps $\partial B$ fixed, and $\delta E_\xi$ is the change in the conserved quantity associated with the Killing vector $\xi$.

Similar to the case for black holes, using the Hamiltonian arguments, the bulk gravitational first law can be extended to include variations in $\Lambda$, and is given by
\begin{equation}\label{extended}
\delta E_\xi = T_{0}{\delta A_\Sigma\over 4G_d} -{V\delta\Lambda\over 8\pi G_d}\,,
\end{equation}
where the ``thermodynamic volume'' is now
\begin{equation}\label{Theta}
V = {l^2\over (d-1)r_{0}}\, A_\Sigma \,,
\end{equation}
and the area $A_{\Sigma}$ was found by imposing a cutoff  \cite{Kastor:2014dra}.

Using (\ref{eqn:ryu}) and (\ref{eq:pressG}),
the extended bulk first law can be rewritten as
\begin{equation}\label{extended2}
\delta E_\xi = T_{0}\delta S_E - (d-2)T_{0}S_E{\delta l\over l} \,.
\end{equation}
Identifying a bulk quantity $E_{\xi}$ with $\left< H_{A}\right>$ of the CFT, and using the precise correspondence between $l$ and $N$ (which depends on the CFT in consideration), one will get an extended first law for the CFT entanglement entropy. For example, for a 2-dimensional CFT with a 3-dimensional bulk, this relation was shown to be \cite{Kastor:2014dra}
\begin{equation}\label{extended3}
\delta \left< H_{A}\right> = T_{0}\delta S_E - T_{0}{S_E\over N}\delta N \,.
\end{equation}
An extended first law for entanglement entropy was also derived for more general scenarios that include variable $\Lambda$, $G_d$, and Lovelock couplings \cite{Kastor:2016bph,Caceres:2016xjz}.


Another development regarding holographic entanglement entropy was considered in \cite{Johnson:2013dka}.  For a charged black hole in $AdS_{4}$, it was found (at fixed $\Lambda$) that as temperature increases, the entanglement entropy in the CFT
undergoes a discontinuous jump as the black hole undergoes a Van der Waals phase transition, similar to what happens for the black hole thermal entropy as discussed in Sec. \ref{Sec:4}.

It has been further  argued  \cite{Caceres:2015vsa} that such behavior of the entanglement entropy can diagnose the $P-V$ phase structure of the bulk. Computing the entanglement entropy of a circular region on the boundary for the field theory dual to a (3+1)-dimensional STU black hole with given charges, it was shown that for charge configurations where a Van der Waals phase transition is present (i.e. for 3 or 4 charges turned on), the entanglement entropy\footnote{To obtain a finite entanglement entropy, the area of the minimal surface in pure AdS was subtracted from the black hole configuration.} exhibits a 
jump at the same critical temperature. This was done numerically for various temperatures and charges to obtain $T-S$ curves which can indicate the presence of a phase transition
\cite{Caceres:2015vsa}.

The behavior of the entanglement entropy and {\it two-point correlation function} have also been studied numerically for the quintessence\footnote{Quintessence is a scalar field which can vary in space and time. It was proposed to explain dark energy, and is an alternative to the cosmological constant model.} charged AdS black hole \cite{Zeng:2015wtt},  further hinting that these quantities might be good indicators of phase transitions in the bulk.

\subsection{Other directions}

Among other directions related to the AdS/CFT correspondence and black hole chemistry, let us mention the following two.

{\it Holographic superconductors.} A new venue for the black hole chemistry is the study of $P-T$ phase diagrams for spacetimes with multiple fields (going so beyond a simple case of negative cosmological constant) that are used to model holographic superconductors \cite{Hartnoll:2008vx}.
The first such study \cite{Nie:2015zia}
constructed  phase diagrams for the ``$s+p$ model'' in Gauss--Bonnet gravity by employing both the
bulk \eqref{eq:press} and the CFT (through the boundary stress energy tensor) definitions of pressure. Curiously, it was the bulk definition that gave rise to a $P-T$ diagram reminiscent of the actual $P-T$ diagram of Helium-3 \cite{Nie:2015zia}.

{\it Kerr/CFT correspondence.}
Another related study concerns the Kerr/CFT correspondence \cite{Guica:2008mu} for the super-entropic holes discussed in Sec.~\ref{Sec:3}.  This correspondence posits a duality between the horizon of a Kerr black hole (bulk) and a  2-dimensional CFT.
For super-entropic black holes  it is not a-priori obvious that the correspondence exists, given the
non-compactness of their event horizons, and indeed it was recently shown \cite{Sinamuli:2015drn} that  some but not all super-entropic black holes exhibit this correspondence.  For example $d=4$ singly-spinning super-entropic black holes
exhibit the correspondence, but their higher-dimensional counterparts do not since  extremal super-entropic black holes no longer exist in $d>4$.
 A sufficient condition for the applicability of the Cardy formula (relating the CFT central charge to the entropy)
is that  the electric charge of such  black holes is large relative to the AdS length $(q \gg l)$   in contrast to
 Kerr--Newman-AdS black holes in which  small rotation parameters are additionally required.

\section{Beyond $\Lambda<0$}\label{Sec:6}

We devote this section to extending of some of the ideas presented in this review to more general settings.
Specifically we shall consider a positive cosmological constant $\Lambda >0$ (extending the concepts  presented in Sec.~\ref{Sec:1} and Sec.~\ref{Sec:2}) \cite{Dolan:2013ft, Kubiznak:2015bya}, asymptotically Lifshitz spacetimes \cite{Brenna:2015pqa},
and connections with so-called horizon thermodynamics \cite{Padmanabhan:2002sha}.

\subsection{Thermodynamics of de Sitter black holes}

The thermodynamics of asymptotically de Sitter (dS) black holes is much more complex than that of their asymptotically flat or AdS cousins for two basic reasons. First, the existence of a cosmological horizon in addition to a black hole horizon means that the system associated with an observer located between these horizons is in a non-equilibrium state---such an observer would find herself
in a thermodynamic system characterized by two temperatures.
Second, the absence of a Killing
vector that is timelike everywhere outside the black hole horizon prevents one from
defining a good notion of the asymptotic mass\footnote{See \cite{Ashtekar:2014zfa, Ashtekar:2015ooa} for recent developments on other peculiar features of asymptotically dS spacetimes.}.
This is quite unfortunate as dS black holes are of direct interest in cosmology.

There have been only a few investigations of the thermodynamics with variable $\Lambda$ in cosmological settings, e.g.
\cite{Li:2014ixn, Tian:2014ila,Azreg-Ainou:2014lua, Kastor:2016bnm}. In this subsection we  discuss which features of
black hole chemistry  can be carried over to the dS black hole case.

\subsubsection{Multiple horizons and their first laws}

One way to deal with the thermodynamics of spacetimes with multiple horizons is to formulate several separate thermodynamic first laws, one for {each `physical' horizon present in the spacetime}.
Specifically, let us consider a general rotating dS black hole with several $U(1)$ charges in $d$-dimensions. Such a black hole typically admits three horizons, located at real positive radii $r$, determined from the horizon condition, say $f(r)=0$.
The {\it cosmological horizon} (denoted with subscript $c$) is located at the largest positive root $r_c$,
the {\it black hole horizon} (denoted with subscript $b$) corresponds to the second largest positive root $r_b$, and the {\it inner horizon} (if it exists) corresponds to the third largest positive $r_i$
(denoted with subscript $i$).

 The arguments in App.~\ref{FirstLaw} can be recapitulated for $\Lambda > 0$  \cite{Dolan:2013ft}
(see also \cite{Cai:2001sn, Cai:2001tv, Sekiwa:2006qj}), and indicate that the following first laws:
\ba
\delta M&=&T_{b} \delta S_{b}+\sum_k \Omega^k_{b}\delta J^k+ \sum_j\Phi_{b}^j\delta Q^j\, {+\, V_{b}\delta P\,,}\label{firstBHb}\\
\delta M&=&-T_{c} \delta S_{c}+\sum_k \Omega^k_{c}\delta J^k+ \sum_j\Phi_{c}^j\delta Q^j
{\, +\, V_{c}\delta P}\,,\label{firstBHc}\\
\delta M&=&-T_{i} \delta S_{i}+\sum_k \Omega^k_{i}\delta J^k+ \sum_j\Phi_{i}^j\delta Q^j
{\, +\, V_{i}\delta P}\,,
\label{firstBHi}
\ea
hold for these horizons\footnote{
See e.g. \cite{Cvetic:2010mn, Castro:2012av, Page:2015gia} for a discussion of why a rather formal thermodynamics of inner horizons may play a role in understanding black hole microscopics.}.
Here, $M$ represents a quantity that {\it would be} the {\it ADM mass} in asymptotically AdS and flat cases\footnote{In the dS case such a quantity is ``conserved in space'' (rather than in time) due to the spacelike character of the Killing field $\partial_t$ in the region near infinity \cite{Ghezelbash:2001vs}.} and the horizon temperatures $T_b$, $T_c$ and $T_i$ are all
defined to be proportional to the magnitudes of their respective surface gravities and so are all
positive.   $S_b$, $S_c$ and $S_i$ denote the horizon entropies, the $\Omega$'s and $J$'s denote the respective angular velocities and momenta, the $\Phi$'s and $Q$'s stand for the respective electric potentials and charges, and the quantity $P$ is related to the positive cosmological constant $\Lambda$ according to the same relation \eqref{eq:press} as for the AdS case
\be\label{P}
P=-\frac{\Lambda}{8\pi}=-\frac{(d-1)(d-2)}{16\pi l^2}<0\,,
\ee
commensurate with the form of a perfect fluid stress-energy tensor.  Since it is now negative, $P$ is perhaps best understood as a {\it tension} rather than a pressure, though we shall continue to call $P$ pressure throughout this section.
The quantities
$V_c$, $V_b$ and $V_i$ are the
thermodynamic volumes, that is, the quantities thermodynamically conjugate to $P$:
\be
{
V_c=\Bigl(\frac{\partial M}{\partial P}\Bigr)_{S_c, J^1,Q^1\dots }\,,\quad
V_b=\Bigl(\frac{\partial M}{\partial P}\Bigr)_{S_b, J^1,Q^1\dots }\,,\quad
V_i=\Bigl(\frac{\partial M}{\partial P}\Bigr)_{S_i, J^1,Q^1\dots }\,.
}
\ee

Starting from \eqref{firstBHb}--\eqref{firstBHi}, it is also possible to formulate  `subtracted' first laws. For example, for an observer in between the cosmological and
the black hole horizon, the difference between \eqref{firstBHb} and \eqref{firstBHc} gives
\be
0=T_{b} \delta S_{b}+T_{c}\delta S_{c} +\sum_i (\Omega^i_{b}-\Omega^i_{c})\delta J^i
+\sum_j(\Phi_{b}^j-\Phi_c^j)\delta Q^j
-V \delta P\,,\label{first2}
\ee
 where $V$ stands for the net volume of the `observable universe',
\be
 V=V_c-V_b\geq 0\,,
\ee
and for standard  examples  equals the naive geometric volume \cite{Dolan:2013ft}.

Various isoperimetric inequalities for the different thermodynamic volumes associated with dS black hole spacetimes
have also been formulated   \cite{Dolan:2013ft}.  For example, the entropy is increased by adding black holes
for fixed
volume of the observable universe for all considered examples.

 The above three laws are accompanied by the corresponding {Smarr--Gibbs--Duhem} formulae
\ba
\frac{d-3}{d-2}M&=&T_{b}S_{b}+\frac{d-3}{d-2}\sum_j\Phi_{b}^jQ^j
+\sum_{k}\Omega_{b}^kJ^k-\frac{2}{d-2}V_{b}P\,,\quad \label{Smarr1}\\
\frac{d-3}{d-2}M&=&-T_{c}S_{c}+\frac{d-3}{d-2}\sum_j\Phi_{c}^jQ^j +\sum_{k}\Omega_{c}^k
J^k-\frac{2}{d-2}V_{c}P\,,\quad \label{Smarr2}\\
\frac{d-3}{d-2}M&=&-T_{i}S_{i}+\frac{d-3}{d-2}\sum_j\Phi_{i}^jQ^j +\sum_{k}\Omega_{i}^kJ^k-\frac{2}{d-2}V_{i}P\,,\quad \label{Smarr2}
\ea
which can be derived from the corresponding first laws \eqref{firstBHb}--\eqref{firstBHi}
via  the dimensional scaling argument  \cite{Kastor:2009wy}.

\subsubsection{Effective thermodynamic description and phase transitions}

Having formulated the first law for each physical horizon a natural question arises:
do dS black holes admit phase transitions similar to their AdS cousins? Surprisingly, at the moment there is no consensus
in the literature for how to do consistent thermodynamics for these black holes. Let us discuss several recent proposals.

Perhaps the most `naive' and straightforward proposal is to study the thermodynamics of all three dS horizons \cite{Kubiznak:2015bya}, treating them as if they were {\it independent thermodynamic systems}, characterized by their own temperature and thermodynamic behavior. Technically the behavior of all such systems is captured by a single `thermodynamic potential' that corresponds in some sense to a continuation of the Gibbs free energy of AdS black holes to negative pressures (to describe positive cosmological constant) and ``negative temperatures'' (to be able to describe inner and cosmological horizons as well). A phase transition of {\it any} of these systems is then taken as a sign of a phase transition of the whole dS black hole spacetime. Using this criterion, it was shown that a  6-dimensional doubly spinning rotating black hole in dS space admits a {\it reentrant phase transition}, similar to the one observed for their AdS cousins \cite{Kubiznak:2015bya}.

Another approach is that of the {\it effective equilibrium thermodynamic description}.
The key ingredient for such a description is to
concentrate on an observer who is located in an `observable part of the universe', in between the black hole horizon and the cosmological horizon, and assign to the system an ``{\it effective temperature}'' $T_{\mbox{\tiny eff}}$,
through a postulated effective thermodynamic first law. Depending on the
interpretation of the mass parameter $M$, there are various versions of the effective approach, each giving rise to a different effective thermodynamic description and different $T_{\mbox{\tiny eff}}$.

In the first version,  originated by Urano et al. \cite{Urano:2009xn} and elaborated upon in
\cite{Ma:2013aqa, Zhao:2014raa, Zhang:2014jfa, Ma:2014hna, Guo:2015waa, Guo:2016eie} 
the mass parameter $M$ is treated as the {\it internal energy} $E$ of the system.
The system is assigned a {\it `total entropy'} $S$, given by the  sum of the black hole horizon and the cosmological horizon entropies \cite{Kastor:1992nn, Bhattacharya:2015mja}, and an effective volume $V$ that equals the volume of the observable universe,
\be\label{Stotal}
S=S_b+S_c\,,\quad V=V_c-V_b\,,\quad E=M\,.
\ee
By recasting \eqref{firstBHb} and \eqref{firstBHc}, the effective temperature $T_{\mbox{\tiny eff}}$ and the effective pressure $P_{\mbox{\tiny eff}}$ are then determined from the following first law:
\be\label{US}
\delta E=T_{\mbox{\tiny eff}} \delta S-P_{\mbox{\tiny eff}} \delta V
+\sum_i \Omega^i_{\mbox{\tiny eff}}\delta J^i+ \sum_j \Phi_{\mbox{\tiny eff}}^j\delta Q^j\,,
\ee
where the new thermodynamic quantities $\Omega^i_{\mbox{\tiny eff}}$ and $\Phi_{\mbox{\tiny eff}}^j$ are
defined as quantities thermodynamically conjugate to $J^i$ and $Q^j$.
Interestingly, this identification leads to a ``Schwarzschild-dS black hole analogue'' of the {\it Hawking--Page transition}.
However, as discussed in \cite{Kubiznak:2016prep}, physically unclear results emerge for more complicated black hole spacetimes.
For example, already for the spherically symmetric charged-dS black hole, up to four branches of admissible black holes appear for a given $T_{\mbox{\tiny eff}}$, leading to a very complicated phase structure. 
Moreover, neither the effective temperature nor the effective pressure, as defined by \eqref{US}, are manifestly positive.

\begin{figure*}
\centering
\begin{tabular}{cc}
\rotatebox{-90}{
\includegraphics[width=0.47\textwidth,height=0.27\textheight]{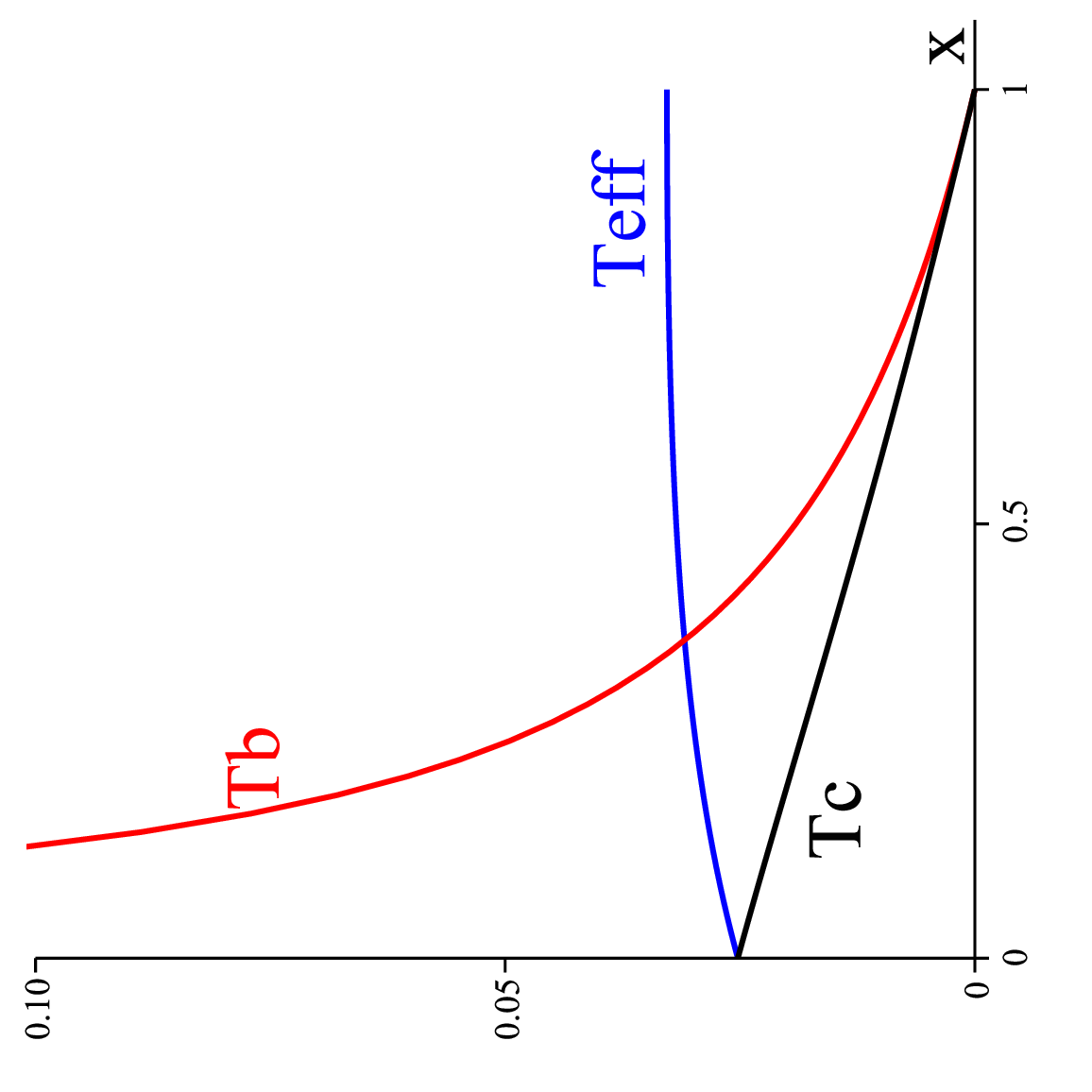}}&
\rotatebox{-90}{
\includegraphics[width=0.47\textwidth,height=0.27\textheight]{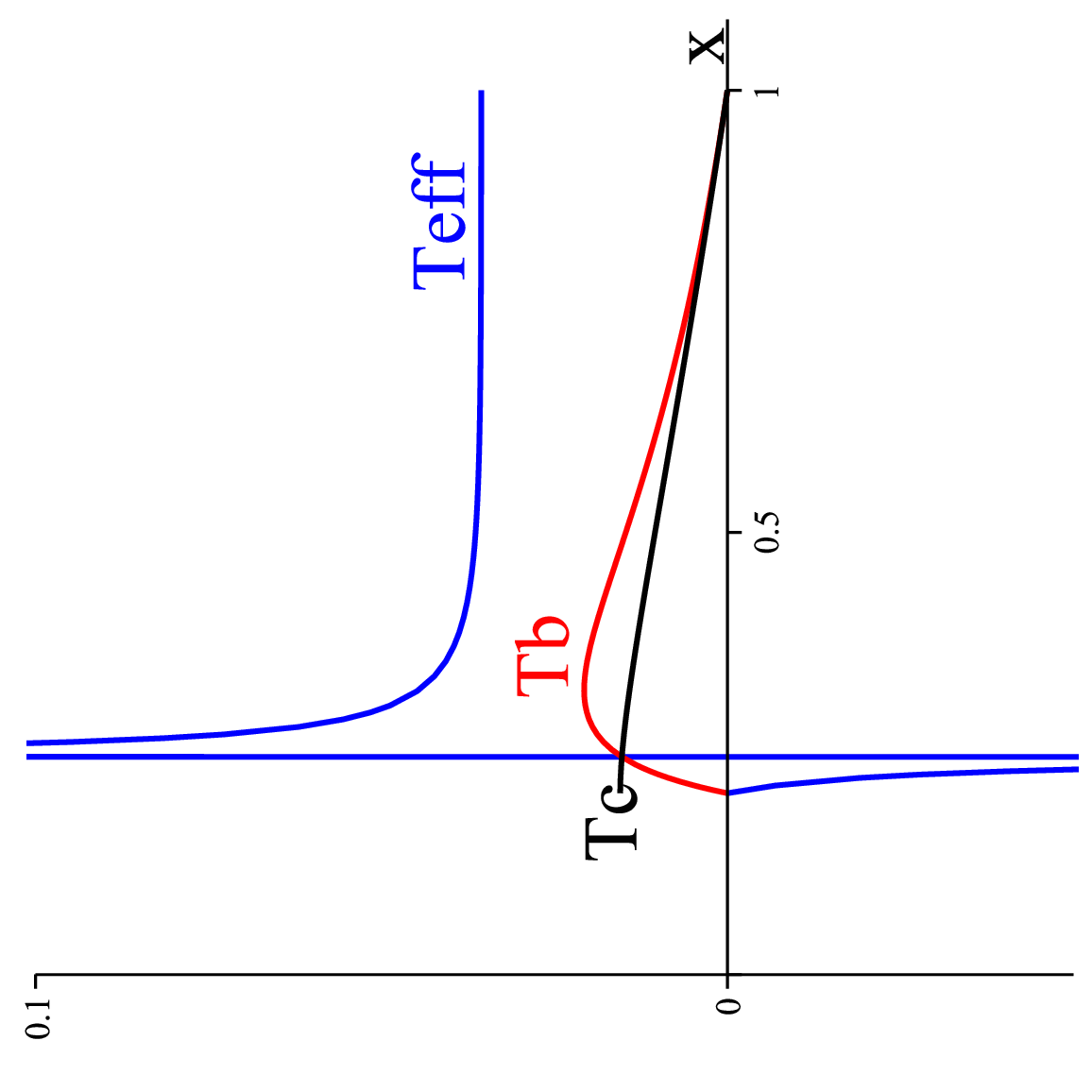}}
\end{tabular}
\caption{{\bf $M$ as enthalpy: Effective temperature (sum of entropies).}
The effective temperature as predicted by the `sum entropy rule', $S=S_c+S_b$, is displayed for ${\cal P}=0.003$ for the
Schwarzschild-dS case (left) and the charged-dS case with $Q=1$ (right) as a function of $x=r_b/r_c$. The black curve displays the cosmological horizon temperature $T_c$, the red curve corresponds to black hole horizon temperature $T_b$, and the blue curve to the effective temperature $T_{\mbox{\tiny eff}}$. Note that in the charged case, at $x=x_{\tiny L}\approx 0.205$ (the lukewarm solution $T_b=T_c$) the effective temperature suffers from an infinite jump while it is negative for $x\in(x_{\tiny \mbox{min}}, x_{\tiny L})$.
}
\label{figTeff1}
\end{figure*}
In the second version of the effective approach, the parameter $M$ is, similar to the AdS case, treated as {\it gravitational enthalpy} \cite{Li:2016zca, Kubiznak:2016prep}.
Namely, the following effective first law is imposed:
\be\label{effective4}
\delta H=T_{\mbox{\tiny eff}}\delta S+V_{\mbox{\tiny eff}} \delta {\cal P}
+\sum_i \Omega^i_{\mbox{\tiny eff}}\delta J^i+ \sum_j \Phi_{\mbox{\tiny eff}}^j\delta Q^j\,,
\ee
where $H=-M$ and ${\cal P}=-P$.  Note that the volume $V=V_c-V_b$ is no longer treated as fundamental and is replaced by $V_{\mbox{\tiny eff}}$.
Starting again from the two first laws \eqref{firstBHb} and \eqref{firstBHc}, and to write \eqref{effective4}, one needs to identify the entropy $S$ of the effective system. It is not very hard to see that provided we take the sum of the entropies as in \cite{Li:2016zca}, $S=S_c+S_b$, we get
\be\label{Teff1}
S=S_c+S_b\quad \Rightarrow\quad T_{\mbox{\tiny eff}}=\Bigl(\frac{1}{T_c}-\frac{1}{T_b}\Bigr)^{-1}\,.
\ee
Although this formula for $T_{\mbox{\tiny eff}}$ often appears in the literature, one can easily see its unphysical properties. Considering
for example the charged-dS black hole solution,  we observe, as illustrated in Fig.~\ref{figTeff1}, that the black hole horizon and cosmological horizon temperatures are not ordered: we can have $T_b<T_c$ (close to the extremal limit), $T_b>T_c$ (close to the Nariai limit), or even $T_b=T_c$ (in the case of the lukewarm solution). Consequently, the effective temperature as defined by formula \eqref{Teff1} is not necessarily positive and becomes ill defined (suffers from an `infinite jump') when $T_b=T_c$. This unphysical behavior of $T_{\mbox{\tiny eff}}$ \eqref{Teff1} is likely to prevail for any solutions with inner horizons and is thence generic.

An `ad hoc solution' of this problem would be to identify  the effective entropy as
$S=S_c-S_b\geq 0\,$.\footnote{This ad hoc postulate   is not justified by anything apart from simplicity and the fact that it does not produce pathologies at least at first sight. For example, both effective temperature $T_{\mbox{\tiny eff}}$ and effective volume $V_{\mbox{\tiny eff}}$ are now manifestly positive.}
The effective first law \eqref{effective4} then implies
\ba
T_{\mbox{\tiny eff}}&=&\Bigl(\frac{1}{T_c}+\frac{1}{T_b}\Bigr)^{-1}\geq 0\,,\quad
V_{\mbox{\tiny eff}}=T_{\mbox{\tiny eff}}\Bigl(\frac{V_c}{T_c}+\frac{V_b}{T_b}\Bigr)\geq 0\,,\quad\nonumber\\
\Omega^i_{\mbox{\tiny eff}}&=&-T_{\mbox{\tiny eff}}\Bigl(\frac{\Omega^i_b}{T_b}+\frac{\Omega^i_c}{T_c}\Bigr)\,,\quad
\Phi^j_{\mbox{\tiny eff}}=-T_{\mbox{\tiny eff}}\Bigl(\frac{\Phi^j_b}{T_b}+\frac{\Phi^j_c}{T_c}\Bigr)\,.\qquad
\ea
Consequences of this proposal are currently under investigation \cite{Kubiznak:2016prep}.

In summary, at the moment there is no accepted formalism describing thermodynamics of asymptotically dS black holes and their possible phase transition, irrespective of whether the cosmological constant is allowed to vary or not. Perhaps an approach where one allows a temperature gradient, $T=T(r)$  \cite{York:1986it} might offer some solution. See also   \cite{McInerney:2015xwa} for a recent alternative study.

\subsection{Pressure and volume in horizon thermodynamics}

 The concepts of pressure and volume as well as that of the equation of state $P=P(V,T)$ applied to black holes (with a modified first law)  appeared in the literature prior to the idea of extended phase space thermodynamics.
A particular manifestation that has received much attention  is  {\it horizon thermodynamics} \cite{Padmanabhan:2002sha, Padmanabhan:2009vy}.

After Hawking's seminal paper \cite{Hawking:1974sw} elevated  the laws of black hole mechanics to the laws of thermodynamics, identifying
the geometric concept of surface gravity $\kappa$ with the quantum mechanical temperature of the black hole, $\kappa\hbar\propto T$, two important questions arose: How do the completely classical Einstein equations know about quantum effects? Can we understand gravity from a thermodynamic  viewpoint?

Several answers, dependent on the nature of the horizon in consideration, subsequently appeared \cite{Cai:2014bwa}: considerations about the {\it local Rindler horizon} led Jacobson to `re-derive' the Einstein equations as a thermodynamic equation of state from the Clausius relation \cite{Jacobson:1995ab} ii) the study of the Einstein equations evaluated on the {\it black hole horizon} led Padmanabhan to formulate horizon thermodynamics \cite{Padmanabhan:2002sha}, and similarly iii) the Friedmann equations at the {\it apparent cosmological horizon} can be re-cast in the form of the first law of thermodynamics \cite{Akbar:2006kj, Cai:2006rs}, though understanding mass as enthalpy is problematic in this setting \cite{Tian:2014ila}. See \cite{Sakharov:1967pk, Hayward:1997jp, Padmanabhan:2003gd} for further results and references.  In this section we concentrate on horizon thermodynamics, which explicitly works with notions of pressure and black hole volume.

The basic idea of horizon thermodynamics is as follows.
Consider a static spherically symmetric black hole spacetime, written in
standard coordinates
\be\label{metricHT}
ds^2=-f(r) dt^2+\frac{dr^2}{g(r)}+r^2d\Omega^2\,,
\ee
with a non-extremal horizon located at $r_+$ given by $f(r_+)=0$,
and identify the total pressure $P$ with the $T^r{}_r$ component of the energy-momentum tensor
of all the matter fields, including the cosmological constant, if present. The radial Einstein equation evaluated on the black
hole horizon can then be regarded as an Horizon Equation of State (HES)
\be\label{HES}
P = P(V, T)\,,
\ee
which, upon a virtual displacement of the horizon, gives rise to the
Horizon First Law (HFL)
\be\label{HFL}
\delta E=T\delta S-P\delta V\,,
\ee
both relations being reminiscent of the extended phase space thermodynamics studied  in Sec.~\ref{Sec:4}.

For example, in  4-dimensional Einstein gravity minimally coupled to matter, the pressure is identified as
\be\label{joj}
P\equiv T^r{}_r|_{r_+} =
\frac{T}{2r_+}-\frac{1}{8\pi r_+^2}=P(V,T)
\ee
from the radial  Einstein equation in the spherically symmetric case, using the familiar relation $T=\frac{\kappa}{2\pi}$ for the temperature.  Equation
\eqref{joj} is the HES \eqref{HES}, upon identifying the volume $V$ with the geometric volume
\be\label{VHTDs}
V=\frac{4}{3}\pi r_+^3\,.
\ee
Employing the Bekenstein relation $S=\frac{A}{4}=\pi r_+^2$ for the entropy yields
\be
T\delta S=\underbrace{4\pi r_+^2\delta r_+}_{\delta V}P+\underbrace{\frac{\delta r_+}{2}}_{\delta E}\quad \Rightarrow \quad \delta E=T\delta S-P\delta V\,,
\ee
which is the HFL.

Note that to obtain the HES and HFL  it was necessary not only to use the identification \eqref{joj} for the pressure but
also to determine the thermodynamic quantities $T,S,$ and $V$ by some other criteria. The output is the identification of
the horizon `quasi-local' energy
\be\label{HE}
E=\frac{r_+}{2}\,,
\ee
which is nothing other than a Misner--Sharp energy\footnote{This is a meaningful concept of quasi-local energy in spherically symmetric spacetimes \cite{Hayward:1994bu}, evaluated on the black hole horizon.} and a  {\it universal} HES \eqref{joj} whose form depends only on the type of gravitational theory considered (Einstein's gravity in our case)  and whose dependence on matter content is `entirely captured' by the notion of pressure $P$.
We also note that the resultant HFL  \eqref{HFL} is of  `{\it cohomogeneity-one}',  contrary to expectations (from the presence of two terms on the  right-hand side of \eqref{HFL}) that it should be a cohomogeneity-two relation.  The only independent variation $\delta r_+$ corresponds to the virtual displacement of the horizon.

\begin{figure}
\begin{center}
\includegraphics[width=0.6\textwidth,height=0.33\textheight]{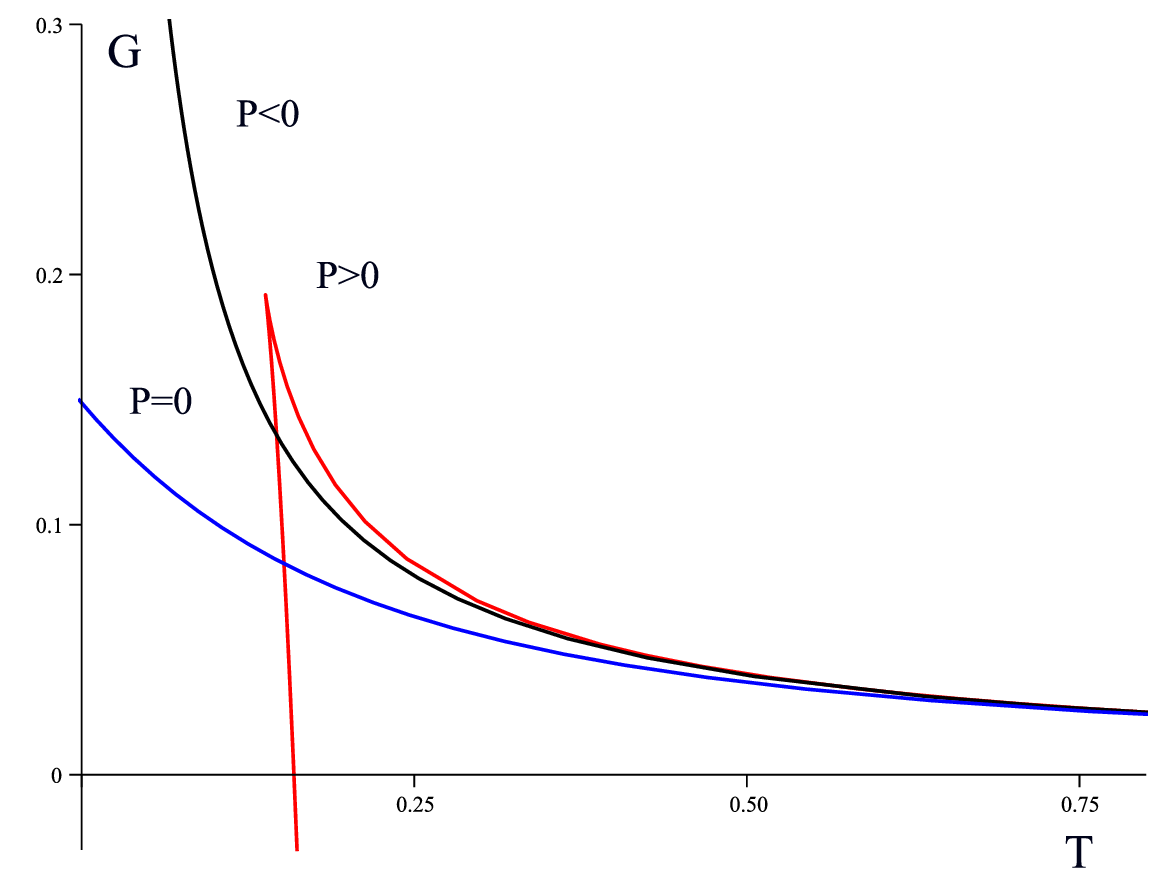}
\caption{\textbf{Gibbs free energy in horizon thermodynamics: Einstein black holes.}
The $G-T$ diagram is displayed for $P=0.03$ (red curve), $P=0$ (black curve) and $P=-0.2$ (blue curve). For positive pressures we observe a characteristic shape reminiscent of the Hawking--Page behavior.
}
\label{FigHTDs}
\end{center}
\end{figure}
 Defining the horizon Gibbs free energy $ G=E-TS+PV=G(T,P)$ from the quantities in  \eqref{HFL} yields three qualitatively distinct behaviours, plotted in Fig.~\ref{FigHTDs}. This indicates that horizon thermodynamics is `universal':  it depends only on the gravitational theory under consideration and not the matter content. For example, in Einstein gravity one always recovers the same HES \eqref{joj} and all possible phase diagrams are of the type displayed in Fig.~\ref{FigHTDs}.
 However the actual interpretation of the phase diagram requires some care, since it  depends on the {\it actual matter content} and is inherently degenerate: different ``points'' on the curves display black holes that not only differ by their size, but also may have different charges, or even be in a different environment   \cite{Hansen:2016ayo}. The resemblance between the $P>0$ curve and  the Hawking--Page transition \cite{Hawking:1982dh} is only superficial because the horizon thermodynamics diagram corresponds to a  thermodynamic ensemble in which one fixes the total pressure of all matter fields.
\begin{figure}
\begin{center}
\includegraphics[width=0.6\textwidth,height=0.33\textheight]{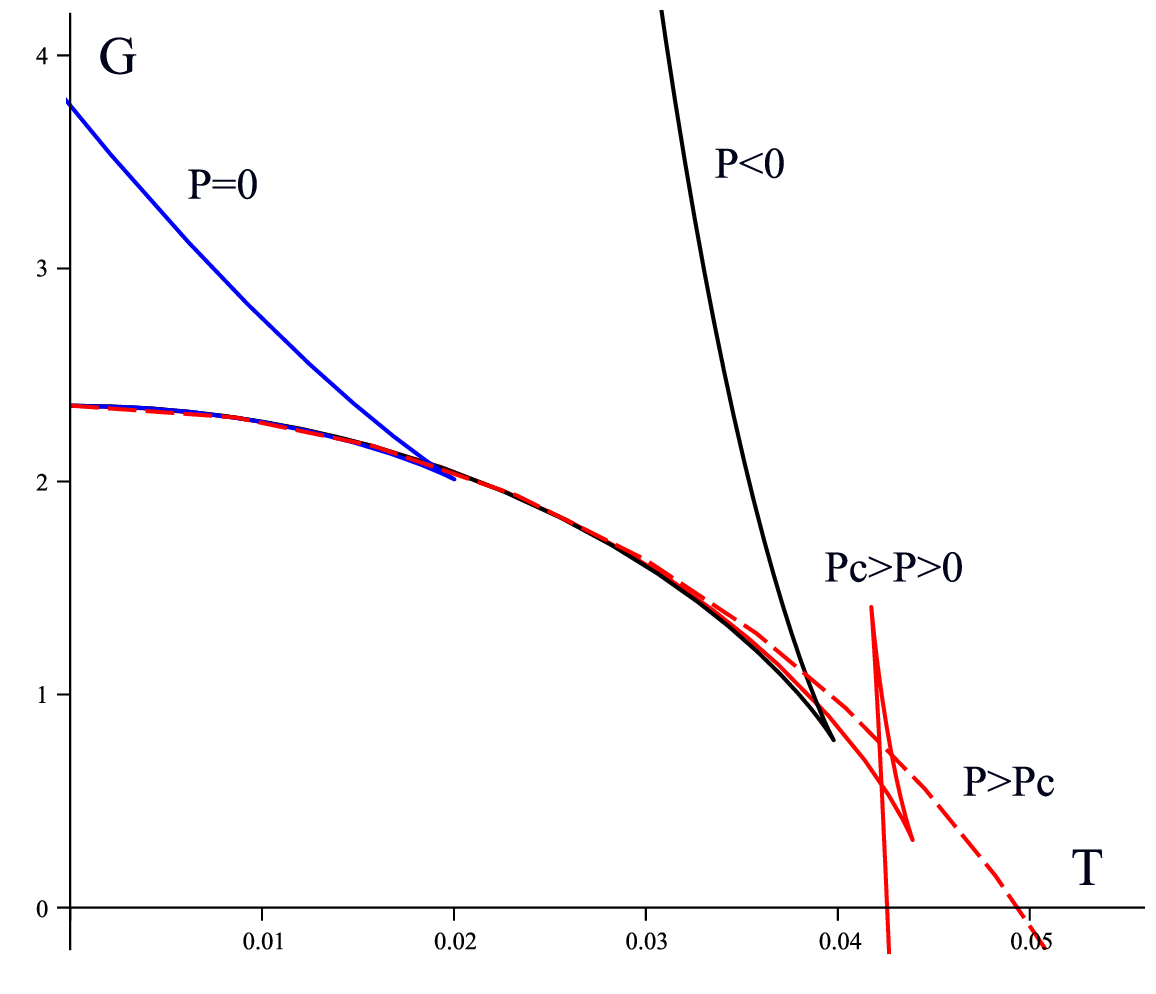}
\caption{\textbf{Gibbs free energy in horizon thermodynamics: $d=5$ spherical Gauss--Bonnet black holes.}
The $G-T$ diagram is displayed for $P=0.01$ (red dash curve), $P=0.0025$ (red solid curve), $P=0$ (black curve), and $P=-0.05$ (blue curve) and $\alpha_2=1$. For small positive pressures we observe a characteristic swallow tail  reminiscent of the Van der Waals-like phase transition.
}
\label{FigHTDs2}
\end{center}
\end{figure}

To obtain more complicated diagrams  other theories of gravity must be considered \cite{Chakraborty:2015wma}. For example Fig.~\ref{FigHTDs2} displays the horizon Gibbs free energy for the five-dimensional Gauss--Bonnet black hole.
Various gravitational theories can be classified based on their corresponding phase diagrams, which corresponds to a classification of `vacuum' black hole solutions in a given theory \cite{Ma:2015llh, Hansen:2016ayo}.  For example,
if the stress-energy vanishes then the thermodynamics reads
\be\label{vacuumHFL}
\delta E=T_0\delta S\,,\quad \mbox{where}\quad E=\frac{r_+}{2}\,,\quad T_0=\frac{1}{4\pi r_+}\,,\quad S=\pi r_+^2\,
\ee
for the vacuum Schwarzschild black hole. Upon adding matter whilst retaining spherical symmetry, the HES can be written as \be\label{joj3}
T=2r_+P+\frac{1}{4\pi r_+}=2r_+P+T_0\,,
\ee
where $T$ corresponds to the `true' Hawking temperature via \eqref{temp}; the latter term can be written in terms of the `vacuum black hole temperature' $T_0$. The HFL  is given by the `vacuum black hole first law' \eqref{vacuumHFL} described from the viewpoint of an  observer able to measure the true Hawking temperature $T$. The matter contribution
\be
\delta E=\underbrace{T_0}_{T-2r_+P}\delta S=T\delta S-P\underbrace{2r_+\delta S}_{\delta V}\,
\ee
can be interpreted as a work term.  This explains the true meaning of the HFL and the origin of the universality of horizon thermodynamics:  the horizon equations represent an `equivalence class of vacuum relations' described from a point of view of observers who can identify
the true matter content of the spacetime.

It is possible to extend these considerations to rotating black holes \cite{Hansen:2016wdg}.  The HFL becomes
a  {\it cohomogeneity-two}  relation

\be
\delta E=T\delta S+\Omega\delta J-\sigma \delta A\,,
\ee
where $\sigma$ stands for the horizon `{\it surface tension}' of all the matter fields, and
the horizon radius $r_+$ and the rotation parameter $a$ are now allowed to vary independently.
One can show that only in very special cases, the tension term can be re-expressed as a pressure--volume term, obtaining so a more specialized HFL
$\delta E=T\delta S+\Omega\delta J-P\delta V$, see \cite{Hansen:2016wdg} for more details.

\subsection{Lifshitz Spacetimes}

Lifshitz spacetimes have attracted much attention in connection with a generalized AdS/CFT correspondence, where they play the role of gravity duals to condensed matter systems with anisotropic scaling \cite{Kachru:2008yh}. Namely, the field theory is characterized by a dynamical critical exponent $z$, which governs the anisotropy between spatial and temporal directions,
\be\label{scalingL}
t\to \lambda^zt\,,\quad x^i\to \lambda x^i\,,\quad z\neq 1\,,
\ee
a toy model being the Lifshitz field theory with $z=2$.

In order to describe the field theory at a finite temperature one considers asymptotically Lifshitz black hole spacetimes, e.g. \cite{Danielsson:2009gi, Mann:2009yx, Bertoldi:2009vn,Balasubramanian:2009rx} (see also \cite{Copsey:2010ya, Horowitz:2011gh} for comments on their pathologies).
Employing the metric ansatz
\be\label{LL}
ds^2=-\Bigl(\frac{r}{l}\Bigr)^{2z}f(r)dt^2+\frac{l^2dr^2}{r^2g(r)}+r^2d\Omega_k^2\,,
\ee
one recovers the required scaling \eqref{scalingL}, accompanied by $r\to \lambda^{-1}r$, provided that $f$ and $g$ approach unity at large $r$. The metric is typically recovered as a solution to field equations modified from general relativity by adding a Proca field or specially tuned higher-curvature terms; the AdS asymptotics is recovered upon setting $z=1$.

Contrary to the asymptotically AdS spacetimes where the asymptotic mass plays the role of an enthalpy and can be uniquely defined through e.g. the method of conformal completion \cite{Ashtekar:1984zz, Ashtekar:1999jx, Das:2000cu}, the unusual asymptotics of spacetime \eqref{LL} makes the concept of mass difficult to define, leading to a question of  whether or not consistent thermodynamics can be for such black holes formulated.
In particular, there exists some disagreement over the correct mass and several proposals for a generalized Smarr relation
\begin{equation}\label{eqn:othersmarr}
 (d + z - 2) M = (d - 2) T S\,,
\end{equation}
paired with a first law
\begin{equation}
\delta M = T \delta S\,,
\end{equation}
appeared in the literature
\cite{Bertoldi:2009dt,Bertoldi:2009vn, Dehghani:2010kd,Liu:2014dva, Berglund:2011cp,Dehghani:2011hf,Dehghani:2013mba, Way:2012gr}
and were found valid for concrete examples of Lifshitz black holes. The relation \eqref{eqn:othersmarr} apparently differs from the standard formula \eqref{smarrBH}
and is not  consistent with the first law through the Eulerian scaling  \cite{Bertoldi:2009vn, Liu:2014dva}; for example the mass term $M$ evidently
 would have to scale as $L^{d+z-2}$.

The situation was recently resolved in \cite{Brenna:2015pqa} where the authors proposed a procedure where the `standard AdS-type' Smarr relation \eqref{smarrBH} and first law were imposed and exploited to derive the thermodynamic mass  and volume for various kinds of Lifshitz black holes. This is possible in general since notions of entropy (horizon area), temperature (surface gravity), pressure (cosmological constant), and charge are each well-defined.  This approach for computing mass agrees in all cases where mass can be computed by independent means.  It can furthermore be extended to cases where there is disagreement on how to compute mass, and in general adjudicates between ambiguities that occur for particular kinds of black holes that have degeneracies in their parameter space.
The  Smarr relation  \eqref{eqn:othersmarr}  (and others that previously appeared in the literature)  were shown to be unified into the standard AdS one
upon exploiting additional identities valid for  particular varieties of  Lifshitz black holes. More concretely,  one can for a given solution \eqref{LL} algorithmically construct the mass $M$ and thermodynamic volume $V$, starting from a knowledge of $T$ and $S$ \cite{Brenna:2015pqa}. Curiously, the computed volume $V$ often violates (at least in some range of parameters) the AdS version of reverse isoperimeric inequality discussed in Sec.~\ref{Sec:3}, suggesting that in the asymptotically Lifshitz case such an inequality has to be accordingly modified.

For example, an exact solution to a higher-curvature gravity theory \cite{Gim:2014nba}
\be
\label{eqn:lifd5hc}
f(r) = g(r) = 1 - \frac{ m l^{5/2}}{r^{5/2}}\,,
\ee
describing a $k=0$ Lifshitz black hole with $\Lambda = \frac{-2197}{551 l^2}$, was found to have
\be\label{MVGim}
M =   \frac{297}{1102} \frac{r_+^5 \omega_{3}^{(0)}}{l^3}\,,	\quad
V =  \frac{1782}{2197} \frac{r_+^5 \pi \omega_{3}^{(0)}}{l}\,,
\ee
where $\omega_{3}^{(0)}$ is the surface area of the constant $(t,r)$ toroidal section,
consistent with the mass obtained via other methods \cite{Gim:2014nba}.  The other thermodynamic parameters are
\be
T = \frac{ 5 r_+^2}{ 8 \pi l^3}\,, \quad	S = \frac{396 \pi r_+^3 {\omega_3^{(0)}}}{551}\,, \quad P = \frac{2197}{4408 \pi l^2}\,,
\ee
for the temperature, entropy, and pressure, respectively.  The ratio \eqref{ratio} is easily computed to be \cite{Brenna:2015pqa}
\be
	\mathcal{R} = 3 \left( \frac{88 \pi }{13^{3}} \frac{r_+}{l} \right)^{\frac{1}{4}}\,,
\ee
and so for sufficiently small $r_+$ we will have $\mathcal{R}  < 1$ violating the  AdS version of the reverse isoperimetric inequality.

\subsection{Symmetry breaking vacua}

Lovelock gravity theories have several  maximally symmetric vacua with different values of the curvature.  There can
be reduced symmetry vacua that are separated by critical surfaces in the space of Lovelock couplings, with a
variety of possibilities for vacuum symmetry breaking \cite{Kastor:2015sxa}.  The potential physical relevance of transitions across such critical surfaces   could happen if the cosmological constant were dynamical.  A form of such transition from
AdS to de Sitter space with a black hole was recently considered \cite{Camanho:2013uda} and shown to have  an interpretation within the context of black hole chemistry \cite{Hennigar:2015mco}.

\section{Summary \& future outlook}

 Black hole thermodynamics has undergone a renaissance in recent years, as many theorists have explored the implications of
understanding $\Lambda$ as  thermodynamic pressure $P$.  This is an exciting and fruitful area of research, with many new developments over the past few years.   The basic picture of black holes as Van der Waals fluids with the associated critical exponents is now firmly established. Furthermore, it is quite robust, being replicated in pretty much any gravitational theory, with pretty much any couplings to matter, and with unconventional asymptotics.  New phenomena familiar from everyday thermodynamics, such as enthalpy, reentrant phase transitions, triple points, and Carnot cycles have all now entered the language and structure of the subject, broadening it to what is called Black Hole Chemistry.

Despite this,  many problems and open questions remain to be explored.  Although the thermodynamic correspondence with Van der Waals fluids is well established, the reasons for this are still somewhat puzzling.  Furthermore there are noteworthy exceptions. Black holes in higher curvature gravity can exhibit both multiple reentrant phase transitions and novel behaviour at their critical points under the right circumstances.  Super-entropic black holes violate the reverse isoperimetric inequality conjecture that is otherwise satisfied by the vast majority of AdS black holes.  For certain hairy black holes the entropy can even become negative for certain values of the parameters.  A better understanding of these unusual cases should help us to better understand the thermodynamic character of all black holes.

To this end, perhaps the most important quantity to understand  is volume.  It is straightforwardly defined as the thermodynamic conjugate to pressure, but its physical meaning and interpretation are rather mysterious, particularly since this notion survives in the
flat-space limit where the pressure vanishes.   This is perhaps to be expected since physical systems can have volume even in the
absence of pressure, but its relevance for asymptotically flat black holes is largely unexplored.
The significance of the reverse isoperimetric inequality conjecture in this context is also not clear, particularly in view of the super-entropic cases.
If the asymptotics are broadened to include Lifshitz black holes, then the conjecture also does not hold. Since the pressure  differs
in this case there is no a-priori reason to expect it  to hold; whether some generalized version of the conjecture can be formulated is
unknown.  Furthermore, in de Sitter spacetime there are several notions of volume, and in NUT-charged spacetimes the volume can even be negative.   The diversity of the situation suggests that some deeper understanding of volume and/or reverse isoperimetric inequality remains to be found.

While this review has primarily concentrated on regarding $\Lambda$ as a thermodynamic quantity,  we have noted that any dimensionful coupling, such as occurs in Lovelock gravity, Quasi-topological gravity, or Born--Infeld electrodynamics can be regarded as a thermodynamic variable, and associated conjugates exist.  This greatly extends the thermodynamic phase space.
However there has been almost no investigation of the implications of this finding.  Perhaps even more exotic phenomena, such as
quadruple or $n$-tuple critical points (generalizing the triple point), remain to be found.  Do  third and higher-order black hole phase transitions exist?  Are there new kinds of isolated critical points?  Do any of these phenomena require a larger extended phase space or is variable $\Lambda$ sufficient?

Black hole chemistry should likewise have implications for understanding the microscopic degrees of freedom of black holes.  While a concept of molecular density has been developed for black holes, the implications of this idea have yet to be fully explored.  In more general terms, what do phenomena such as reentrant phase transitions and isolated critical points tell us about the
``microscopic structure'' of AdS black holes?  And how transferable are these notions to black holes with differing   asymptotic structures?

An important related question is that of gauge/gravity duality in the context of extended phase space.   In one interpretation, varying $\Lambda$ in the AdS bulk is equivalent to varying the number of degrees of freedom, $N$, in the boundary CFT, leading to the definition of a chemical potential associated with $N$. Using this identification and previous results shown in holography, for pure AdS spacetimes (dual to pure states of CFTs on the AdS boundary),  one can derive an extended form of the first law for holographic entanglement entropy (which includes a chemical potential term) from the extended gravitational first law. The chemical potential of the CFTs behaves similarly to that of an ideal gas.  However much remains to be done. While it has been suggested that the chemical potential of the CFT and the black hole volume in the bulk are closely related ($\mu \propto -V$), this is the case only if the compactified dimensions (if there is compactification) stay fixed. However in the particular case of  AdS$_5$ $\times$ $S^5$, the size of the $S^5$ also varies,
rendering this relationship between $\mu$ and $V$ suspect. More generally, what is the interpretation of the chemical potential $\mu$ of the gauge theory to the bulk spacetime?    Another  interpretation is that varying $\Lambda$ means varying the volume the CFT resides on; this yields an extended thermodynamic phase space for the CFT since pressure and volume can be defined. The full implications of this approach remain to be explored.  Computation of higher-order corrections beyond the leading large $N$ limit have yet to be carried out, and their implications for black hole chemistry should prove most interesting.

Finally there is considerable work to be done in extending Black Hole Chemistry beyond $\Lambda<0$. While in principle
one can incorporate variable $\Lambda$ into black hole spacetimes with any asymptotic structure, physically interpreting
these cases remains a considerable challenge, particularly in asymptotically de Sitter spacetimes.   Yet transitions between
spacetimes with $\Lambda<0$ and $\Lambda>0$ exist, so it is essential that we obtain a better understanding of this latter case.

Chemistry, the interplay of matter, is a vast subject with many diverse applications.
Black hole chemistry may prove to be just as multi-faceted.

\section*{Acknowledgements}

This research was supported in part by Perimeter Institute for Theoretical Physics and in part by the Natural Sciences and Engineering Research Council of Canada. Research at Perimeter Institute is supported by the Government of Canada through the Department of Innovation, Science and Economic Development Canada and by the Province of Ontario through the Ministry of Research, Innovation and Science.

\appendix

\section{Generalized first law of black hole mechanics}\label{FirstLaw}
In this appendix we reproduce the Hamiltonian derivation \cite{Kastor:2009wy} of the extended first law \eqref{law1mech}.\footnote{ We refer to \cite{Wu:2016auq} for a covariant treatment in more general setting of variable background fields.}

Consider a solution to Einstein's equations in $d$  spacetime dimensions that describes a black hole with a Killing field.  Decompose the metric
\begin{equation}\label{metricsplit}
g_{ab} = h_{ab}-n_a n_b\,,
\end{equation}
where  $n^a$ is  the unit timelike normal ($n\cdot n=-1$) to a hypersurface $\Sigma$, whose  induced metric $h_{ab}$  satisfies $h_a{}^b n_b =0$.  Foliating spacetime by a family of such hypersurfaces, the system can
be taken to evolve along the vector field
\begin{equation}\label{xidecomp}
\xi^a=N n^a +N^a\,,
\end{equation}
where $N=-\xi\cdot n$ is the lapse function and $N^a$ the shift vector, which is  tangential to $ {\Sigma}$.
The dynamical variables in the phase space are comprised of the metric  $h_{ab}$ and its conjugate momentum
$\pi ^{ab} =-\sqrt{h} (K^{ab}-K h ^{ab} )$, where  {$K_{ab} =h_a{}^c \nabla _c n_b$} is the extrinsic curvature  of ${\Sigma}$.  Here  we denote $K=K^a{}_a$ and $\pi=\pi^a{}_a$  as the traces
of these respective tensors, but  $h$ is the determinant of the metric $h_{ab}$ restricted to ${\Sigma}$.

The full gravitational  Hamiltonian  is given by ${\cal H} =NH+N^a H_a$\,, where
\ba
H&\equiv& -2G_{ab} n^a n^b =  -\, R^{(d-1)}  + {1 \over |h|}  \Bigl({\pi ^2 \over d-2 } - \pi^{ab} \pi_{ab} \Bigr)\,,\label{hamconstraint}\nonumber\\
H_b&\equiv& -2G_{ac} n^a h^c _b=-2\,  D_a (|h|^{-{1 \over 2}} \pi^{ab} )\,.\label{momconstraint}
\ea
Here $D_a$ is the covariant derivative operator with respect to $h_{ab}$  on $ {\Sigma}$, and   $R^{(d-1)}$ its scalar curvature. Setting ${8\pi} T^a _b =-\Lambda g^a _b $ then yields
\begin{equation}\label{Hconstraint}
H =- {2} \Lambda\,,  \quad H_b =0\,,
\end{equation}
for the constraint equations.

Consider a solution  $g_{ab}$  of the field equations with Killing vector $\xi^a$
and cosmological constant $\Lambda$.  Let $\tilde g_{ab} =g_{ab} +\delta g_{ab}$ be an `infinitesimally close' solution (not necessarily admitting any Killing vector)
with $\tilde\Lambda = \Lambda +\delta\Lambda$, and correspondingly
$\tilde h_{ab}= h_{ab} + \gamma_{ab}$,  $\tilde \pi_{ab}= \pi_{ab} + p_{ab}$, with  $h_{ab}$ and $\pi^{ab}$ regarded as the initial data for the original (background) solution $g_{ab}$, and $\gamma_{ab} = \delta h_{ab}$\,, $p_{ab} = \delta\pi_{ab}$.  Incorporating this into
\eqref{Hconstraint} gives $D_a B^a = N\delta H + N^a\delta H_a = - {2}N \delta\Lambda\,,$ or alternatively \cite{Traschen:1985,SudarskyWald:1992,TraschenFox:2004, Kastor:2009wy}
\be\label{altgauss}
 D_a ( B^a   -{2}\delta \Lambda \omega ^{ab} n_b ) =0\,,
 \ee
where  $N= - \xi^a n_a=-D_c (\omega ^{cb} n_b )$,
\be\label{gaussvector}
 B^a [\xi ] =    N(D^a \gamma^c_c \!-\! D_b \gamma^{ab})  - \gamma^c_c D^a N + \gamma^{ab} D_b N
+ |h|^{-\frac{1}{2}} N^b\bigl(\pi^{cd} \gamma_{cd} h^a{}_b \!-\! 2 \pi^{ac} \gamma_{bc} \!-\!2p^a{}_b \bigr)\,,
\ee
and  $\omega^{ab} = -\omega^{ba} $, referred to as the {\it Killing co-potential} from section \ref{3aa},
satisfies \cite{Kastor:2008xb,Kastor:2009wy}:
 \begin{equation}\label{omegadef}
 \nabla _c \omega ^{cb} = \xi ^b\,,
 \end{equation}
and is not unique; it is only defined up to a divergence-less term. If $ \omega_{ab}$ solves
$ \nabla ^a \omega_{ab} = \xi_b$\,, then so does $\omega^\prime_{ab}=  \omega_{ab} + \zeta_{ab}$\,, where
$ \nabla ^a \zeta_{ab} =0$.

Equation \eqref{altgauss} is  a Gauss' law relation.
Integrating  it  over a volume $\hat V$ contained in $\Sigma$ gives
\be \label{gaussint}
\int_{\partial \hat V_{out}} d\mathcal{S} r_c \left( B^c [\xi ]  - {2}\delta \Lambda \omega ^{cb} n_b \right)
 = \int_{\partial \hat V_{in}} d\mathcal{S} r_c \left( B^c [\xi ]   - {2} \delta \Lambda \omega ^{cb} n_b \right)\,,
\ee
where  $r^c$ is the unit normal respectively pointing into and out of the inner and outer boundaries   $\partial \hat V_{in, out}$ of  $\hat V$.
The ambiguity in  $ \omega_{ab}$ implies that in general the values of the integrals on the outer and inner boundaries cannot be given separate interpretations; it is only their difference that is meaningful. Writing
$\omega ^{cb}= \omega ^{cb} - \omega_{AdS}^{cb} + \omega_{AdS}^{cb}$ for the $\partial \hat V_{out}$ integral
yields
\ba \label{gaussint2}
\int_{\partial \hat V_{out}} d\mathcal{S} r_c \left( B^c [\xi ]  - {2}\delta \Lambda  \omega_{AdS}^{cb}  n_b \right)
&=& \int_{\partial \hat V_{out}} d\mathcal{S} r_c \left({2}\delta \Lambda  (\omega ^{cb} - \omega_{AdS}^{cb}) n_b \right)\nonumber\\
&&\qquad + \int_{\partial \hat V_{in}} d\mathcal{S} r_c \left( B^c [\xi ]   - {2} \delta \Lambda \omega ^{cb} n_b \right)\,,\qquad
\ea
where $\omega_{AdS}^{ab}$  is the Killing co-potential of the `background AdS spacetime'.

Setting the outer boundary at spatial infinity,
the respective variations in the total mass  $M$ and angular momentum $J$ of the space-time   are defined as
\begin{eqnarray}
 16\pi \delta M  &=&
 -\int_\infty d\mathcal{S} r_c \left(B^c  [{\partial_t}]  - {2}\delta \Lambda \omega_{AdS}^{cb} n_b\right)\,,
   \label{flatdm} \\
  16\pi \delta J  &=&  \int_\infty d\mathcal{S} r_c  B^c  [{\partial_\varphi} ]\,, \label{flatdj}
 \end{eqnarray}
and are obtained by respectively setting $\xi^a = (\partial_t)^a$ (time translations) and $ \xi^a = (\partial_\varphi)^a$
(rotations).  The $ \omega_{AdS}^{cb}$ term ensures  $\delta M$  is finite \cite{Kastor:2009wy}.

Taking the inner boundary to be the event horizon $H$ of a black hole generated by the Killing vector
$\xi^a = ({\partial_t+ \Omega \partial_\varphi})^a$ yields
\be
   2\kappa \delta A  =  -\int_{H} d\mathcal{S} r_c  B^c  [ {\partial_t+ \Omega \partial_\varphi} ]\,,  \label{bhhor}
 \ee
provided the horizon is a  bifurcate Killing horizon of area $A$ on which $\xi$ vanishes,
and $\kappa=\sqrt{ -\half  \nabla^a\xi^b \nabla_a\xi_b }\;\bigr\vert_{r=r_+}$ is its surface gravity.

Since $\delta \Lambda$ is spacetime-independent we can define
\begin{equation}\label{vdef}
V=   \int _{\infty} d\mathcal{S} r_c n_b \left(\omega ^{cb} - \omega_{AdS}^{cb}\right) -\int _{H} d\mathcal{S} r_c n_b \omega ^{cb} \,,
\end{equation}
and interpret the remaining terms in \eqref{gaussint2} as $V\delta P$;
note that $V$ is finite because of the presence of the $\omega_{AdS}^{cb}$ term.
So we have recovered
\be
\delta M =T\delta S  + V \delta P+\Omega \delta J\,, 
 \ee
 which (upon including the electromagnetic terms) becomes \eqref{law1mech}.

\section{The $d$-dimensional Kerr-AdS Metric}
\label{AppB}

In this appendix we review the general Kerr-AdS black hole spacetimes \cite{Gibbons:2004uw, Gibbons:2004js}
and their basic characteristics.
These are $d$-dimensional metrics that solve the Einstein equations with  cosmological constant
\be
R_{ab} =\frac{2\Lambda}{(d-2)}  g_{ab}\,,
\ee
and generalize the $d$-dimensional asymptotically-flat rotating black hole spacetimes of Myers and Perry \cite{Myers:1986un}.
In   Boyer--Lindquist  coordinates the metric takes the form
\ba \label{metric}
ds^2&=&-\frac{W\rho^2}{l^2}d\tau ^2+\frac{2m}{U} \Bigl(W d\tau -\sum_{i=1}^{N} \frac{a_i \mu_i ^2 d\varphi _i}{\Xi _i}\Bigr)^2+\frac{U dr^2}{F-2m}\nonumber\\
&+&\sum_{i=1}^{N} \frac{r^2+a_i^2}{\Xi _i} \mu_i ^2 d\varphi _i^2+\sum_{i=1}^{N+\varepsilon}\frac{r^2+a_i ^2}{\Xi _i} d\mu _i ^2
-\frac{1}{W \rho^2}\Bigl(\sum_{i=1}^{N+\varepsilon}\frac{r^2+a_i ^2}{\Xi _i} \mu_i d\mu_i\Bigr)^2\,,
\ea
where $\rho^2=r^2+l^2$\,,
\ba\label{metrcifunctions}
W&=&\sum_{i=1}^{N+\varepsilon}\frac{\mu _i^2}{\Xi _i}\,,\quad U=r^\varepsilon \sum_{i=1}^{N+\varepsilon} \frac{\mu _i^2}{r^2+a_i^2} \prod _j ^N (r^2+a_j^2)\,,\nonumber\\
F&=&r^ {\varepsilon -2} \frac{\rho^2}{l^2}\prod_{i=1}^N (r^2+a_i^2)\,,\quad \Xi_i=1-\frac{a_i^2}{l^2}\,.\quad
\ea
To treat even ($\varepsilon=1)$  odd ($\varepsilon=0)$ spacetime dimensionality $d$ simultaneously, we have parameterized
\be
d=2N + 1 + \varepsilon\,,
\ee
and in even dimensions set for convenience $a_{N+1}=0$.
The  coordinates $\mu_i$ are not independent, but obey the constraint
\begin{equation}\label{constraint}
\sum_{i=1}^{N+\varepsilon}\mu_i^2=1\,,
\end{equation}
in addition to $0 \le \mu_i \le 1$ for $1 \le i \le N$ and $-1 \le \mu_{N+1} \le 1$ (in even dimensions).

The spacetime admits up to $N$ independent angular momenta $J_i$, described by
$N$ rotation parameters $a_i$, and generalizes the previously known singly-spinning case \cite{Hawking:1998kw}.
In $d=4$ it reduces to the four-dimensional Kerr-AdS metric. In any dimension, the general rotating Kerr-AdS geometry admits a hidden symmetry
of the Killing--Yano tensor \cite{KubiznakFrolov:2007} that is responsible for integrability of geodesic motion and various test field equations in these spacetimes \cite{Frolov:2008jr}.

The thermodynamic quantities associated with Kerr-AdS black holes were first calculated in \cite{Gibbons:2004ai}.
The mass $M$, the angular momenta $J_i$, and the angular velocities $\Omega_i$ read
\be \label{TD}
M=\frac{m \omega _{d-2}}{4\pi (\prod_j \Xi_j)}\Bigl(\sum_{i=1}^{N}{\frac{1}{\Xi_i}-\frac{1-\varepsilon }{2}}\Bigr)\,,\quad
J_i=\frac{a_i m \omega _{d-2}}{4\pi \Xi_i (\prod_j \Xi_j)}\,,\quad \Omega_i=\frac{a_i (1+\frac{r_+^2}{l^2})}{r_+^2+a_i^2}\,,
\ee
while the temperature $T$, and the entropy $S$ are given by
\ba\label{TS}
T&=&\frac{1}{2\pi }\Bigr[r_+\Bigl(\frac{r_+^2}{l^2}+1\Bigr)
\sum_{i=1}^{N} \frac{1}{a_i^2+r_+^2}-\frac{1}{r_+}
\Bigl(\frac{1}{2}-\frac{r_+^2}{2l^2}\Bigr)^{\!\varepsilon}\,\Bigr]\,,\nonumber\\
S&=&\frac{A}{4}=\frac{\omega _{d-2}}{r_+^{1-\varepsilon}}\prod_{i=1}^N
\frac{a_i^2+r_+^2}{4\Xi_i}\,.
\ea
The horizon radius $r_+$ is determined as the largest root of $F-2m=0$ and $\omega_{d}$ is given by \eqref{omeg-d}.
Finally, the thermodynamic volume reads \cite{Cvetic:2010jb}:
\begin{eqnarray} \label{VBHKerr}
V=\frac{r_+A}{d-1}\Bigl(1+\frac{1+r_+^2/l^2}{(d-2)r_+^2}\sum_i\frac{a_i^2}{\Xi_i}\Bigr)=\frac{r_+A}{d-1}+\frac{8\pi}{(d-1)(d-2)}\sum_i a_i J_i\,.
\end{eqnarray}


\providecommand{\href}[2]{#2}\begingroup\raggedright\endgroup

\end{document}